\documentclass[12pt]{article}
\usepackage{epsf}
\catcode`\@=11
\newcount\hour
\newcount\minute
\newtoks\amorpm
\hour=\time\divide\hour by60
\minute=\time{\multiply\hour by60 \global\advance\minute by-\hour}
\edef\standardtime{{\ifnum\hour<12 \global\amorpm={am}%
        \else\global\amorpm={pm}\advance\hour by-12 \fi
        \ifnum\hour=0 \hour=12 \fi
        \number\hour:\ifnum\minute<10 0\fi\number\minute\the\amorpm}}
\edef\militarytime{\number\hour:\ifnum\minute<10 0\fi\number\minute}
\def\draftlabel#1{{\@bsphack\if@filesw {\let\thepage\relax
   \xdef\@gtempa{\write\@auxout{\string
      \newlabel{#1}{{\@currentlabel}{\thepage}}}}}\@gtempa
   \if@nobreak \ifvmode\nobreak\fi\fi\fi\@esphack}
        \gdef\@eqnlabel{#1}}
\def\@eqnlabel{}
\def\@vacuum{}
\def\draftmarginnote#1{\marginpar{\raggedright\scriptsize\tt#1}}
\def\draft{\oddsidemargin -.2truein
        \def\@oddfoot{\sl preliminary draft \hfil
        \rm\thepage\hfil\sl\today\quad\militarytime}
        \let\@evenfoot\@oddfoot \overfullrule 3pt
        \let\label=\draftlabel
        \let\marginnote=\draftmarginnote
   \def\@eqnnum{(\theequation)\rlap{\kern\marginparsep\tt\@eqnlabel}%
\global\let\@eqnlabel\@vacuum}  }
\relax
\def\sqr#1#2{{\vcenter{\vbox{\hrule height.#2pt
        \hbox{\vrule width.#2pt height#1pt \kern#1pt
           \vrule width.#2pt}
        \hrule height.#2pt}}}}

\def\lsim{{\displaystyle
{{\raise-8pt\hbox{$ <$}}
\atop{\raise5pt\hbox{$\sim$}}}}}
\def\gsim{{\displaystyle
{{\raise-8pt\hbox{$ >$}}
\atop{\raise5pt\hbox{$\sim$}}}}}
\def\slsim{{\displaystyle
{{\raise-8pt\hbox{$\scriptstyle <$}}
\atop{\raise5pt\hbox{$\scriptstyle \sim$}}}}}
\def\sgsim{{\displaystyle
{{\raise-8pt\hbox{$\scriptstyle  >$}}
\atop{\raise5pt\hbox{$\scriptstyle \sim$}}}}}
\newcommand{\ar}[2]{{#1\atopwithdelims[]#2}}
\newcommand{\sump}[0]{\sum_{(h,g)}\!{\raise 4pt \hbox{$'$}}\,}
\def\pa{\partial}
\def\bpa{\bar{\partial}}
\def\ve{\vert}

\def\ra{\to}
\def\lra{\leftrightarrow}
\def\ti{\times}
\def\ss{\ar{a}{b}}
\def\bss{\ar{\ba}{\bb}}
\def\hg{\ar{h}{g}}

\def\na{\nabla}
\def\w{\wedge}

\def\fig#1#2#3#4{
\begin{figure}
\begin{center}
\mbox{\epsfysize #1 \epsffile{#2}}
\end{center}
\caption{#3}
\label{#4}
\end{figure}}
\def\he#1{{\rm HET}_{#1}}
\def\tn#1{{\rm II}_{#1}^{(2,2)}}
\def\tbs{{\overline{\rm II}}_{6}^{(2,2)}}
\def\tfs{{\rm II}_6^{(4,0)}}
\def\ihe#1{{\it HET}_{#1}}
\def\itn#1{{\it II}_{#1}^{(2,2)}}
\def\itbs{{\overline{\it II}}_{6}^{(2,2)}}
\def\itfs{{\it II}_6^{(4,0)}}

\def\Tr{\,{\rm Tr}\, }
\def\det{\,{\rm det}\, }
\def\Str{\,{\rm Str}\, }

\def\Im{\,{\rm Im}\, }
\def\Re{\,{\rm Re}\, }

\def\iT{T_2}
\def\iU{U_2}
\def\I{\rm I}
\def\II{\rm II}
\def\III{\rm III}

\def\ifd{\int_{\cal F}\frac{{\rm d}^2\tau}{\t_2}}
\def\ifdd{\int_{\cal F}\frac{{\rm d}^2\tau}{\t_2^2}}
\def\ph{\vphantom}
\def\phl{\vphantom{l}}
\def\a{\alpha}
\def\b{\beta}
\def\c{\gamma}
\def\g{\gamma}
\def\d{\delta}
\def\e{\epsilon}
\def\vep{\varepsilon}
\def\m{\mu}
\def\n{\nu}
\def\t{\tau}
\def\p{\pi}
\def\ps{\psi}
\def\Ps{\Psi}
\def\r{\rho}
\def\th{\vartheta}
\def\Th{\Theta}
\def\s{\sigma}
\def\l{\lambda}
\def\k{\kappa}

\def\f{\phi}
\def\x{\xi}

\def\et{\eta}

\def\Ga{\Gamma}
\def\Fi{\Phi}
\def\O{\Omega}
\def\D{\Delta}

\def\thb{\bar{\th}}
\def\etb{\bar{\eta}}
\def\bz{\bar{z}}

\def\bv{\bar{v}}
\def\bq{\bar{q}}

\def\ba{\bar{a}}
\def\bb{\bar{b}}

\def\bi{\bar{\imath}}

\def\bps{\bar{\psi}}
\def\bPs{\bar{\Psi}}

\def\bT{\overline{T}}
\def\bU{\overline{U}}

\def\Z{Z}

\def\tm{\tilde m}
\def\tnn{\tilde n}

\def\tX{\widetilde{X}}
\def\tPs{\widetilde{\Psi}}

\def\T{{\cal T}}

\def\N{{\cal N}}

\def\sp{\, , \; \;}

\def\nl{\hfil\break}

\def\vs{\vskip}

\def\rd{{\rm d}}
\def\re{{\rm e}}

\def\ni{\noindent}
\def\nl{\newline}
\def\ed{\end{document}}
\newtoks\@stequation
\def\subequations{\refstepcounter{equation}%
  \edef\@savedequation{\the\c@equation}%
  \@stequation=\expandafter{\theequation}
  \edef\@savedtheequation{\the\@stequation}
  \edef\oldtheequation{\theequation}%
  \setcounter{equation}{0}%
  \def\theequation{\oldtheequation\alph{equation}}}

\def\endsubequations{\setcounter{equation}{\@savedequation}%
  \@stequation=\expandafter{\@savedtheequation}%
  \edef\theequation{\the\@stequation}\global\@ignoretrue
  \vspace*{-12pt} \\}
\def\thefootnote{\fnsymbol{footnote}}
\def\bea{\begin{eqnarray}}
\def\eea{\end{eqnarray}}
\def\be{\begin{equation}}
\def\ee{\end{equation}}
\def\bs{\begin{subequations}}
\def\es{\end{subequations}}
\newskip\humongous \humongous=0pt plus 1000pt minus 1000pt
\def\caja{\mathsurround=0pt}
\def\eqalign#1{\,\vcenter{\openup1\jot \caja
        \ialign{\strut \hfil$\displaystyle{##}$&$
        \displaystyle{{}##}$\hfil\crcr#1\crcr}}\,}
\newif\ifdtup

\def\limit#1#2{\smash { \mathop{#1} \limits_{#2} }  }

\def\np#1#2#3{Nucl. Phys. {\bf{B#1}} (#2) #3}
\def\pl#1#2#3{Phys. Lett. {\bf{B#1}} (#2) #3}

\def\thebibliography#1{%
\vskip 0.5cm \centerline{\bf References}
\list{%
[\arabic{enumi}]}{\settowidth\labelwidth{[#1]}
\leftmargin\labelwidth
\advance\leftmargin\labelsep
\usecounter{enumi}}
\def\newblock{\hskip .11em plus .33em minus .07em}
\sloppy\clubpenalty4000\widowpenalty4000
\sfcode`\.=1000\relax}

\renewcommand{\theequation}{\arabic{section}.\arabic{equation}}

\renewcommand{\section}{\setcounter{equation}{0}\@startsection%
{section}{1}{0mm}{-\baselineskip}{0.5\baselineskip}%
{\normalfont\normalsize\bfseries}}

\renewcommand{\subsection}{\@startsection%
{subsection}{2}{0mm}{-\baselineskip}{0.5\baselineskip}%
{\normalfont\normalsize\itshape}}

\begin{document}

\begin{titlepage}
\begin{flushright}
CERN-TH/97-103, NEIP-97-006\\
LPTENS/97/24, CPTH-S507.0597 \\
hep-th/9708062 \\
\end{flushright}
\begin{centering}
\vspace{.3in}
\boldmath
{\bf $R^2$ CORRECTIONS AND NON-PERTURBATIVE DUALITIES \\ OF
$N=4$ STRING GROUND STATES}\\
\unboldmath
\vspace{0.4 cm}
A. GREGORI$^{\ 1}$,
E. KIRITSIS$^{\ 1}$,
C. KOUNNAS$^{\ 1,\, \ast}$,
N.A. OBERS$^{\ 1}$,
P.M.~PETROPOULOS$^{\ 1,\, 2,\, \diamond}$
and
B. PIOLINE$^{\ 1,\, \diamond}$\\
\vskip 1cm
{\it $^1 $ Theory Division, CERN$^{\ \dagger}$}\\
{\it 1211 Geneva 23, Switzerland}\\
\medskip
{\it and}\\
\medskip
{\it $^2 $ Institut de Physique Th\'eorique, Universit\'e de
Neuch\^atel}\\
{\it 2000 Neuch\^atel, Switzerland}\\
\vspace{0.4cm}
{\bf Abstract}\\
\end{centering}
\vspace{.1in}
We compute and analyse a variety of four-derivative gravitational terms
in the
effective action of six- and four-dimensional type II string
ground states with $N=4$ supersymmetry.
In six dimensions, we compute the relevant perturbative corrections for
the
type II string compactified on $K3$.
In four dimensions we do analogous computations for several
models with ($4,0$) and ($2,2$) supersymmetry.
Such ground states are related by heterotic--type II duality or type
II--type II
$U$-duality.
Perturbative computations in one member of a dual pair give a
non-perturbative
result in the other member. In particular, the exact CP-even $R^2$
coupling on the $(2,2)$ side reproduces the tree-level term plus NS
5-brane instanton contributions on the $(4,0)$ side. On the other hand,
the exact CP-odd
coupling yields the one-loop axionic interaction $a R\w R$ together
with a similar instanton sum.
In a subset of models, the expected breaking of the $SL(2,Z)_S$
$S$-duality
symmetry to a $\Ga(2)_S$ subgroup is observed on the
non-perturbative thresholds.
Moreover, we present a duality chain that provides evidence for the
existence of heterotic $N=4$ models
in which $N=8$ supersymmetry appears at strong coupling.

\vspace{.3cm}
\begin{flushleft}
CERN-TH/97-103, NEIP-97-006\\
LPTENS/97/24, CPTH-S507.0597 \\
August 1997 \\
\end{flushleft}
\hrule width 6.7cm \vskip.1mm{\small \small \small
$^\ast$\ On leave from {\it Laboratoire de Physique Th\'eorique de
l'Ecole Normale Sup\'erieure,}
{\it CNRS,} 24 rue Lhomond, 75231 Paris Cedex 05, France.\\
$^\diamond$ On leave from {\it Centre de Physique Th\'eorique,
Ecole Polytechnique,}
{\it CNRS,} 91128 Palaiseau Cedex, France.\\
$^\dagger$ e-mail addresses: agregori, kiritsis, kounnas, obers,
petro, bpioline@mail.cern.ch.}
\end{titlepage}
\newpage
\setcounter{footnote}{0}
\renewcommand{\thefootnote}{\arabic{footnote}}

\setcounter{section}{0}
\section{Introduction}

There has been intriguing evidence that different perturbative string
theories might be non-perturbatively equivalent \cite{d,ht,w1,pw}.
In six dimensions, there is a conjectured duality between the heterotic
string
compactified on $T^4$ and the type IIA string compactified on $K3$
\cite{d,ht}.
Both theories have $N=2$ supersymmetry and 20 massless vector
multiplets
in six dimensions. Several arguments support this duality:

\romannumeral1)
 The tree-level two-derivative actions of the two theories
(in the Einstein frame) are related by a duality transformation.
In particular, the field strength of the heterotic antisymmetric tensor
(with the gauge Chern--Simons form included) is dual to that of the
type
II string.

\romannumeral2) The relation described above implies that the heterotic
string is a magnetic or solitonic string of the type II theory and
vice versa.
This is also supported by the following facts~\cite{s1}.
There is a singular string solution of the heterotic theory,
electrically
charged under the antisymmetric tensor, which can be identified with
the perturbative heterotic string \cite{hs}.
There is also a magnetically charged solitonic (regular at the core)
string
solution, which has the correct zero-mode structure to be identified
with the type II string.
Upon the duality map, their role is interchanged in the type II
theory \cite{s1}.

\romannumeral3) Anomaly cancellation of the heterotic string implies
that
there should be a one-loop $R^2$ term in the type II theory. Such a
term was found
by direct calculation in \cite{vw2}.
Its one-loop threshold correction upon compactification to four
dimensions
\cite{hm1} implies instanton corrections on the heterotic side due to
5-branes wrapped on the six-torus.

\romannumeral4) Upon toroidal compactification to four dimensions,
heterotic--type II
duality translates into $S\leftrightarrow T$ interchange \cite{d}.
As a consequence, perturbative $T$-duality of the type II string
implies
$S$-duality \cite{s} of the heterotic string (and vice versa).
Electrically charged states are interchanged with magnetically charged
states.

\romannumeral5) The six-dimensional heterotic--type II duality implies
by the adiabatic
argument \cite{vw1} non-perturbat\-ive
dualities in lower-dimensional models obtained as freely-acting
orbifolds of
the original pair.

\romannumeral6) More general (non-free) symmetric orbifolds still give
rise
to $N=2$ heterotic--type~II dual pairs in four dimensions
\cite{kv,fhsv,re}. On the heterotic side they can be viewed as
compactifications on $K3$, while on the type IIA side they correspond
to
compactifications  on $K3$-fibred Calabi--Yau manifolds.
On the heterotic side the dilaton is in a vector multiplet and the
vector moduli space receives both perturbative and non-perturbative
corrections.
On the other hand, the hypermultiplet moduli space does not receive
perturbative corrections, or non-perturbative ones, if $N=2$ is assumed
to be unbroken.
On the type II side the dilaton is in a hypermultiplet and the
prepotential
for the vector multiplets comes only from the tree level.
This fact provides a quantitative test of duality; this was shown in
Refs. \cite{kv,re}, where the
tree-level type II prepotential was computed and shown to give the
correct one-loop heterotic result and to predict the non-perturbative
corrections on the heterotic side.
This quantitative test is not applicable to $N=4$ string duality.

There is another class of non-perturbative duality symmetries
known as
$U$-duality \cite{ht}, which relates type II vacua with maximal
supersymmetry.
They are obtained from the convolution of the $SL(2,Z)$ symmetry of
type
IIB in ten dimensions and $O(d,d,Z)$ duality upon compactification.
Using freely-acting orbifolds, the supersymmetry can be reduced but
there
should still be a $U$-duality symmetry \cite{vw1}. In Ref. \cite{sv} a
class
of models with $N=4$ and $N=2$ supersymmetry was discussed; these are
related by $U$-duality.
Again, there are several arguments in favour of $U$-duality, but no
quantitative
test to our knowledge.

In this paper we shall focus on the implications of heterotic--type
IIA
and $U$-dualities for higher-derivative gravitational
terms in the effective action, namely $R^2$ couplings and variations
thereof. These terms have the property that in vacua with
16 supercharges ($N=4$ in four dimensions) they only receive
contribution
from short representations of the supersymmetry algebra, the
so-called BPS multiplets. This property becomes obvious once
these terms are written
in terms of helicity supertraces \cite{bk}, which are known
to count only BPS states. Therefore,
$R^2$ terms in $N=4$ vacua are very similar
to the terms in the two-derivative action for vacua with 8
supercharges
($N=2$ in four dimensions). In fact, the two-derivative action
can be shown to be uncorrected both perturbatively and
non-perturbatively
in $N=4$ vacua, so that these couplings are the first terms where
quantum corrections manifest themselves, in a still controllable way,
though. The $F^4$ \cite{bk}
and $R^4$ terms in vacua with ($4,0$) supersymmetry
also belong to this class of BPS-saturated couplings, together
with higher-derivative terms constructed out of the Riemann
tensor and the graviphoton field strengths~\cite{fg}.

Contributions to $R^2$ couplings depend on the type of $N=4$ vacua
we are considering: ($2,2$) vacua, where two supersymmetries come
from the left-movers and two from the right-movers, or ($4,0$) vacua,
where all four supersymmetries come from the left-movers only.
All heterotic ground states with $N=4$ supersymmetry are of the ($4,0$)
type, but ($4,0$) type~II vacua  can also be constructed
\cite{ht,fk}. In that case, the axion--dilaton corresponds to the
complex scalar in the gravitational multiplet in four dimensions
and, as such,
takes values in an $SU(1,1)/U(1)$ coset space,
while the other scalars
form an $SO(6,N_V)\Big/\Big(SO(6)\times SO(N_V)\Big)$ manifold, where
$N_V$ is the
number of vector multiplets in four dimensions.
On the other hand
($2,2$) models only exist in type II and have a different
structure: the dilaton is now part of the $SO(6,N_V)\Big/\Big(SO(6)\ti
SO(N_V)\Big)$
manifold, while the $SU(1,1)/U(1)$ coset is spanned by a perturbative
modulus. Duality always maps a ($2,2$) ground state to a ($4,0$) ground
state
\cite{ht}.
We shall argue that $R^2$ couplings are exactly given
by their one-loop result in ($2,2$) vacua. Translated
into the dual ($4,0$) theory, the exact $R^2$ coupling
now appears to arise from non-perturbative effects,
identified with NS 5-brane instantons in Ref. \cite{hm1}.
Here we shall carry the work of \cite{hm1} further,
and extend it to more general gravitational couplings and
to other $N=4$ models, obtained as freely-acting
orbifolds of the usual type IIA on $K3\ti T^2$ and heterotic on $T^6$
theories.  These exotic ground states possess a number of vector
multiplets smaller than that of their parents ($N_V=22$), and we shall
generically
refer to them as {\it reduced-rank $N=4$ models}.
They reduce to standard $N=4$ or $N=8$ models in proper
decompactification limits, and are invariant under reduced groups
of $T$- or $S$-dualities.

To be specific, we will consider the following $N=4$ models:

a) Type II theory compactified on $K3\times T^2$ with ($2,2$)
supersymmetry and 22 vector multiplets. We will denote this ground
state
by $\tn{22} $. It is conjectured to be dual to
the heterotic string
compactified on $T^6$ (denoted henceforth by $\he{22}$) via
$S\leftrightarrow T$ interchange~\cite{ht}.
The $R^2$ coupling has already been considered in the case \cite{hm1}.
We will reconsider it here in order to compute also the thresholds of
other four-derivative terms, as well as to compare it with the
six-dimensional
thresholds once we decompactify the $T^2$.

b) Type II theory compactified on a six-dimensional manifold
with $SU(2)$ holonomy, which is locally but {\it not globally}
$K3\times
T^2$.
The supersymmetry is still ($2,2$). We present examples with
$N_{V}=6,10,14$. The class of models with $N_V$ reduced  was initially
constructed in Ref. \cite{fk}, using a fermionic construction
\cite{abk,klt}.
Here we shall  construct them by starting with the $K3\times
T^2$ model,
going to
a subspace of $K3$ with a $Z_2$ (non-freely-acting) symmetry and
orbifolding with this symmetry, accompanied by a lattice shift $w$ on
the two-torus
($w$ is a four-dimensional vector of mod(2) integers with
$w^2=0$).
In practice, we consider the $T^4/Z_2$ orbifold limit of $K3$.
The $Z_2$ symmetry we use is a subgroup of the $(D_4)^4$ symmetry of
$T^4/Z_2$.
Appropriately choosing this $Z_2$ subgroup
allows the construction of ($2,2$) ground states
with $N_{V}=6,10,14$ vector multiplets.
Such a $Z_2$ symmetry has the property that if we orbifold by it,
without a $T^2$ shift, it reproduces the $K3\times T^2$ models at
a different point in the $K3$ moduli space.
Moreover, because of the shift on the $T^2$, the $SL(2,Z)_T$ duality
symmetry is broken to a $\Gamma(2)_T$ subgroup.
We will denote these ground states by $\tn{N_V} (w)$.

Since the orbifold acts freely, by the adiabatic argument, the new
model should be dual to a corresponding orbifold of the heterotic
string on $T^6$ with reduced rank, which we will denote by
$\he{N_V}(w)$.
Duality will then imply that such $N=4$ ground states have a reduced
$S$-duality group, $\Gamma(2)_S \subset SL(2,Z)_S$.
This property is reflected in the non-invariance of the $R^2$ threshold
under the full $SL(2,Z)_S$ group.
When the shift vector $w$ involves projections on momenta only on the
type
II side, then its action is perturbatively visible on the heterotic
side,
since it acts again on momenta.
If, on the other hand, it contains projections on the winding numbers
on
the type II side, then in heterotic language the projection is on
non-perturbative states carrying magnetic charges.

c) A ($2,2$) type II model obtained by orbifolding the type
II string
compactified on $T^6$ (maximal $N=8$ supersymmetry).
We split $T^6=T^2\times T^4$ and the $Z_2$ orbifold action is an
inversion
on $T^4$ and a shift $w$ on $T^2$.
This is a ground state, where $N=8$ supersymmetry is {\it
spontaneously}
broken
to $N=4$.
It has $N_V=6$ vector multiplets and will be denoted by~$\tbs (w)$.

d) A ($4,0$) type II model, constructed by
freely orbifolding
by $(-1)^{F_{\rm L}}$ times a $Z_2$ lattice shift $w$ on $T^6$
($(-1)^{F_{\rm L}}$ is the left-moving fermion number).
Such a ground state has $N_V=6$  and we will denote it by $\tfs (w)$.
Here again $N=8$ supersymmetry is spontaneously broken to $N=4$.
It was argued in \cite{sv} to be
$U$-dual
to the $\tbs (w)$ ground state of the previous
paragraph via
$S\leftrightarrow T$ interchange.
There is a map of the two-torus electric and magnetic charges
similar to the case of string--string duality.

String--string duality and $U$-duality  imply that the aforementioned
models are related through
\bs
\be
\he{N_V} (w)(S,T) = \tn{N_V} (w') (T,S)\, ,
\label{int1}\ee
\be
\tfs (w)(S,T) = \tbs (w')(T,S)\, ,
\label{int2}\ee
\es
where the lattice shift $w'$ is obtained from $w$ through the
duality map. For the particular case of a shift vector $w^*$
acting on the momenta only, ${w^{*}}'=w^*$.

Moreover, we shall prove that, at least in the weak-coupling
regime $S_2 \to \infty$, the two models
$\tbs(w)$ and $\tn{6}(w)$ are actually identical, up to relabelling
of perturbative moduli. In particular, for $w^*=w_{\I}\equiv
(0,0,1,0)$:
\be
\tbs (w_{\I})(S,T,U)=
\tn{6} (w_{\I})(S,-2/T,-1/2U)\, .
\label{int3}
\ee
We also have the following decompactification limits, at least
in the perturbative regime:
\be
\tn{6}(w_{\I})(T_2\to \infty)= \tbs (w_{\I})(T_2 \to 0)
= \, {\rm type \ IIA \ on \ } K3 \, ,
\label{int4}\ee
\be
\tn{6} (w_{\I})(T_2\to 0)= \tbs(w_{\I})(T_2\to \infty)
= \, {\rm type \ IIA \ on \ } T^4 \, .
\label{int5}\ee
Now making use of the string--string duality (\ref{int1}) we obtain
that, at least in the {\it large-radius} limit,
\be
\he{6} (w^*)(S_2\to \infty)= \, {\rm heterotic \ on \ } T^4\, ,
\label{int6}\ee
\be
\he{6}(w^*)(S_2\to 0) = \, {\rm type \ IIA \ on \ } T^4\, .
\label{int7}\ee
Thus we find that
at weak coupling the $\he{6}(w^*)$ has $N=4$ supersymmetry
while at strong coupling $N=8$ supersymmetry is restored.
It is known that the heterotic string can be viewed as a
(non-freely-acting) $Z_2$ orbifold
of M-theory \cite{hw}.
Here, however, we find a ground state of the heterotic string in which
$N=8$ supersymmetry is {\it spontaneously} broken to $N=4$. The extra
gravitinos are magnetic solitons, with masses scaling as the
inverse of the heterotic coupling constant. Therefore, $N=8$
supersymmetry is
restored in the strong-coupling limit where
these gravitinos become massless.
This is similar to the situation described in \cite{decoa, kk, decol}.

The structure of this paper is as follows:
in Section 2 we describe the potential perturbative and
non-perturbative contributions
to the $R^2$ couplings in the various string models we analyse further.
In Section 3 we consider the type IIA, B string compactified to six
dimensions
on $K3$ and compute the perturbative corrections to the four-derivative
couplings involving the metric, the NS--NS antisymmetric tensor
and the dilaton.
In Section 4 we describe the calculation of the one-loop $R^2$
threshold in generic type II orbifold models with $N=4$ supersymmetry.
In Section 5 we reconsider the $R^2$ thresholds of type II string on
$K3\times T^2$ and its dual heterotic theory.
In Section 6 we analyse the BPS spectrum and $R^2$ thresholds of the
various models with $N=4$ supersymmetry and reduced rank.
Section~7 contains our conclusions.
In Appendix A we describe the kinematics of on-shell string vertices
relevant for our threshold calculations, and in Appendix B the
calculation of helicity supertraces that count BPS states in string
ground states
with $N=4$ supersymmetry and some associated $\vartheta$-function
identities.
In Appendix C we calculate the relevant fundamental-domain integrals
appearing in the one-loop calculation of the thresholds.
Details of one-loop string calculations are left to Appendix D.

\boldmath
\section{Perturbative and instanton corrections to $R^2$ couplings}
\unboldmath

In this paper we shall be interested in four-derivative gravitational
couplings in the low-energy effective action of superstring vacua with
16 supercharges ($N=4$ in four dimensions, or $N=2$ in six dimensions).
The prototype of these terms is $R^2 \equiv R_{\a\b\c\d} R^{\a\b\c\d}$,
but we shall also consider couplings involving the NS antisymmetric
tensor $B_{\m\n}$ and the dilaton $\Phi$.

At tree level, such terms can be obtained directly from the relevant
ten-dimensional calculations (see \cite{sloan}) upon compactification
on the appropriate manifold, $K3$, $K3\times T^2$ or $T^6$.
They turn out to be non-zero in ($4,0$) ground states  (heterotic or
type
II)
and zero for ($2,2$) ground states. They may a priori also receive
higher-loop
perturbative corrections, but
($4,0$) ground states appear to have no perturbative corrections at
all,
while the perturbative corrections in ($2,2$) vacua are expected
to come only
from one loop owing to the presence of extended
supersymmetry.

These terms are related by
supersymmetry to eight-fermion couplings.
As such they may receive non-perturbative corrections
from instantons having not more than 8 fermionic zero-modes.
This rules out generic instanton configurations, which break
all of the 16 supersymmetric charges and therefore
possess at least 16 zero-modes. However, there exist
particular configurations that preserve one half of the
supersymmetries (this is the only possibility in six dimensions
where the supersymmetry is $N=2$), thereby possessing 8
fermionic zero-modes\footnote{Instantons with less than 8
zero-modes do not exist, in agreement with the absence of
corrections to the two-derivative or four-fermion action.}.
These configurations correspond to
the various $p$-brane configurations of the original ten-dimensional
theory:
a Euclidean $p$-brane can generate an instanton when its
($p+1$)-dimensional
world-volume wraps around some appropriate submanifold of the
compactification
manifold ($K3$).
All superstrings in ten dimensions have in common the NS
5-brane that couples to the dual of the NS--NS antisymmetric tensor and
breaks
half of the ten-dimensional supersymmetry.
Type II superstrings also have D $p$-branes that are charged under the
various R--R forms
and their duals:
$p=0,2,4,6,8$ for type IIA theory,
$p=-1,1,3,5,7$ for type IIB. Obviously,
D-branes are absent from heterotic ground states, but also
from the $($4,0$)$ type II model we shall consider, since the latter
has no massless R--R fields. We conjecture that this is in fact true
for any ($4,0$) vacuum. The only instanton configuration for such vacua
is therefore the NS 5-brane, which only starts to contribute for
dimensions
less than or equal to four.

In ($2,2$) models the situation is a bit more involved.
Let us consider first the type IIA, B string compactified on $K3$ to
six
dimensions.
Since $K3$ is four-dimensional, only branes with $p+1\leq 4$ need be
considered as instantons.
Wrapped in a generic fashion around  submanifolds of $K3$
they
break all
supersymmetries and thus do not contribute, in our calculation.
There are, however, supersymmetric $0, 2$ and 4 cycles in $K3$.
The relevant instantons will then have $p+1=0,2,4$, found only in type
IIB.
Thus in type IIA theory we do not expect any instanton corrections.
In type IIB theory, all scalar fields span an
$SO(5,21)\Big/\Big(SO(5)\times SO(21)\Big)$ coset
space.
The perturbative $T$-duality symmetry $O(4,20,Z)$ combines with the
$SL(2,Z)$ symmetry in ten dimensions into an $O(5,21,Z)$
$U$-duality symmetry group.
The exact non-perturbative threshold should therefore be an
$O(5,21,Z)$-invariant function of the moduli and, as argued in
\cite{kp}, it
can be written as linear
combinations
of the Eisenstein--Poincar\'e series. However, all such series
have distinct and non-zero perturbative terms when expanded in terms of
any
modulus, in disagreement with the fact that all perturbative
corrections
should vanish.
We thus conclude that the $R^2$ threshold is  non-perturbatively zero
also in type IIB on $K3$.

There is an independent argument pointing to the same result.
Consider compactifying type IIA, B on $K3 \times S^1$.
Then IIA and IIB are related by
inverting the circle radius.
{}From the type IIA point of view there are now potential instanton
corrections
from the $p=0,2,4$-branes wrapping around a $0,2,4$ $K3$ cycle
times $S^1$.
However, on the heterotic side we are still in a dimension larger than
four
so we still have no perturbative or non-perturbative corrections.
This implies that the contribution of the IIA instantons still
vanishes,
as it does for the IIB instantons, which are just the same as the
six-dimensional ones. The instanton contributions in six dimensions
thus also have to vanish.

Compactifying further to four dimensions on an extra circle,
the scalar manifold becomes
$SU(1,1)/U(1) \times  SO(6,22)\Big/\Big(SO(6)\times SO(22)\Big)$
and the duality group
$SL(2,Z)\times O(6,22,Z)$.
The instanton contributions
can come from 5-branes wrapped around $ K3\times T^2$ as well as,
in type IIB, from D 3-branes wrapped around $T^2$ times a
$K3$ 2-cycle, and $(p,q)$
D 1-branes wrapped around $T^2$.
The 1-brane contribution is zero since it is related via $SL(2,Z)$
duality to that of the fundamental string world-sheet
instantons, which vanish
from the one-loop
result\footnote{This is equivalent to the statement that in IIB
the one-loop threshold only depends on the complex structure $U$
of the torus. This no longer holds for other
thresholds such as $\na H \na H$ and, accordingly, we shall
find that those are non-perturbatively corrected even in type II.}.
All other instanton corrections depend non-trivially on the $O(6,22)$
moduli.
Again, it is  expected that an $O(6,22)$-invariant result would imply
perturbative corrections depending on the $O(6,22) $ moduli, which are
absent, as we
will show. Therefore,  we again obtain that the
non-perturbative corrections vanish in IIB, and also in IIA.
This can also be argued via type II--heterotic--type I triality.
On the heterotic and type I side these corrections come from the
5-brane
wrapped on $T^6$. The world-volume action of the D 5-brane in type II
theory
is known and wrapping it around $T^6$ and translating to heterotic
variables produces a result depending only on the $S$ field.
Thus on the heterotic side we do not expect $O(6,22)$-dependent
corrections, and therefore no instanton contributions in type II.

The upshot of the above discussion is that, in $(2,2)$ models, various
dualities imply that
on the type II side instanton corrections to $R^2$ terms are absent in
six, five and four dimensions.
Similar arguments apply to the other reduced-rank ($2,2$) models
considered in this paper, since their instanton corrections are related
to the ones above by applying some selection rules on the possible
brane-wrappings.
We can therefore
restrict ourselves to
a one-loop computation on both type IIA and type IIB ($2,2$)
models.

\setcounter{equation}{0}
\section{One-loop corrections in six-dimensional
type IIA and IIB theories\label{6d}}

In this section, we compute the one-loop four-derivative terms in the
effective action for type IIA and IIB theory compactified to six
dimensions on
the $K3$ manifold. We will work in the $Z_2$ orbifold limit of $K3$ in
order to be explicit but,  as we will show, the result will be valid
for
all values of the $K3$ moduli.

\subsection{Type II superstring on $K3$: a reminder \label{remin}}

The one-loop partition function of type IIA, B theory compactified on
the $T^4/Z_2$ orbifold is
\bea
Z_{\; \II}^{\rm six \ dim} &=& {1 \over \t_2^2 \, |\et|^{24}}
\frac{1}{2}
\sum_{a,b=0}^1
(-1)^{a+b+ab} \th^2 \ss
\frac{1}{2}
\sum_{\bar{a}, \bar{b} =0}^1
(-1)^{\ba + \bb + \m \ba \bb }
\thb^2\ar{\ba}{\bb }\cr
&&\ti \,
\frac{1}{2}
\sum_{h,g=0}^1
 \th \ar{a +h}{ b +g }
 \th \ar{a -h}{ b -g }
\thb \ar{\ba +h }{\bb +g}
\thb \ar{\ba -h }{\bb -g}
\Gamma_{4,4}\hg \, ,
\label{61}
\eea
where the $T^4$ orbifold blocks $\Gamma_{4,4} \hg$ are given by
\be
\Gamma_{4,4} \ar{0}{0} =  \Ga_{4,4}
\sp
\Gamma_{4,4} \hg  =16 {\vert \et \vert^{12}  \over
\left\vert  \th \ar{1 +h}{ 1 +g } \right\vert^4} \sp (h,g) \neq (0,0)\,
,
\label{z44}
\ee
where $\Ga_{4,4}$ is the ($4,4$) lattice sum.
The parameter $\m$  takes the value $0$ or $1$ for type IIA or IIB
superstrings, respectively, as it determines the sign of the
antiholomorphic Ramond sector and hence the space-time chirality of the
fermions. We shall also use the notations $\vep=(-1)^{\m}$ and denote
by
left and right the holomorphic and antiholomorphic side, respectively.
On each side the sum over spin structures splits into three even
structures $(a,b)\in \{(0,0),(0,1),(1,0)\}$ and one odd $(a,b)=(1,1)$
according to the number of fermionic zero-modes on the world-sheet.

\vs .1cm
To compute the massless spectrum we need the following geometric data
of $K3$: the Einstein metric on $K3$ is parametrized by 58 scalars, and
the non-zero Betti numbers are $b_0=b_4=1$ and $b_2=22$.
 Out of the 22 two-forms, 3 are
self-dual, while the remaining 19 are anti-self-dual. At the $T^4/Z_2$
orbifold point of $K3$, those correspond to the $3+3$ $Z_2$-even
two-forms $\rd x^i \w \rd x^j$ and to 16 anti-self-dual two-forms
supported by
the two-sphere that blows up each of 16 fixed points.
With this in mind, it is easy to derive the massless spectrum:

{\sl Type IIA.} The ten-dimensional bosonic massless spectrum
consists of the  NS--NS fields
$G_{MN}$, $B_{MN}$, $\Phi$ and of the R--R three-form and one-form
potentials
$A_{MNR}$ and $A_{M}$. Compactification on $K3$ then gives in the
NS--NS
sector
$G_{\m \n}$ and 58 scalars,  $B_{\m \n}$ and 22 scalars, and the
dilaton
$\Phi$; in the R--R sector we have $A_{\m \n \r}$ and 22 vectors in
addition
to $A_{\m}$.
In six dimensions, $A_{\m \n \r}$ can be dualized into a vector, so
all in all the bosonic fields comprise a graviton, 1 antisymmetric
two-form tensor, 24 $U(1)$ vectors
and 81 scalars.  Hence, we end up with the following supermultiplets of
six-dimensional ($1,1$) (non-chiral) supersymmetry:
\be
\mbox{1 supergravity multiplet} \sp
\mbox{20 vector multiplets}\, ,
\ee
where we recall that:
\nl
-- the ($1,1$) supergravity multiplet comprises a  graviton, 2 Weyl
gravitinos of
opposite chirality, 4 vectors, 4 Weyl spinors of opposite chirality,
1 antisymmetric tensor, 1 real scalar;
\nl
-- a vector multiplet comprises 1 vector,  2 Weyl spinors of opposite
chirality,
4 scalars. \nl
The scalars parametrize $R \ti SO(4,20)\Big/\Big(SO(4) \ti
SO(20)\Big)$, where the first
factor corresponds to the dilaton up to a global  $O(4,20,Z)$
$T$-duality
identification.

{\sl Type IIB.} The ten-dimensional massless bosonic spectrum
consists of the NS--NS fields $G_{MN}$, $B_{MN}$, $\Phi$, and the
self-dual
four-form $A^+_{MNRS}$, the two-form $A_{MN}$ and the zero-form  $A$
from the
R--R sector. Compactification on $K3$
then gives in the NS--NS sector the same as for type IIA. In the R--R
sector, we obtain respectively $A^+_{\m \n \r\s}$ (which is not
physical),
22 $B^{\rm R-R}_{\m \n}$
(of which 19 anti-self-dual and 3 self-dual) and 1 scalar,
$A_{\m \n}$ and 22 scalars, and the scalar $A$ itself.
If we decompose both $B_{\m \n}$ and $A_{\m \n}$ into a self-dual and
an
anti-self-dual part, the bosonic content comprises a graviton,
5 self-dual and 21 anti-self-dual antisymmetric tensors and 105
scalars.
Hence, we end up with the following six-dimensional ($2,0$) (chiral)
supermultiplets:
\be
\mbox{1 supergravity multiplet} \sp
\mbox{21  tensor multiplets}\, ,
\ee
where we recall that:
\nl
-- the ($2,0$) supergravity mulitplet comprises a graviton, 5 self-dual
antisymmetric
tensors, 2 left Weyl gravitinos, 2 Weyl fermions;
\nl
-- a ($2,0$) tensor multiplet comprises  1 anti-self-dual antisymmetric
tensor, 5
scalars, 2 Weyl fermions of chirality opposite to that of the
gravitinos.
\nl
The scalars including the dilaton parametrize the coset space
$SO(5,21)\Big/\Big(SO(5) \ti SO(21)\Big)$, and the
low-energy supergravity has  a global $O(5,21,R)$ symmetry
\cite{romans}.

\subsection{One-loop three-graviton scattering amplitude in six
dimensions
\label{1l6d} }

We must  consider the piece quartic in momenta of the
one-loop three-point function:
\be
{\cal I} = \e_{1\m\n} \e_{2\k \l}
\e_{3\r \s}
\ifdd \int
\prod_{i=1}^3  {\rd^2 z_i  \over \p}
\left\langle V^{\m \n}(p_1,\bz_1,z_1)
 V^{\k \l}(p_2,\bz_2,z_2)
V^{\r \s} (p_3,\bz_3,z_3)
\right\rangle\, .
\label{64}
\ee
Here the space-time indices run over $\m = 0, \ldots ,5$
(see Appendix A for conventions), and the
vertex operators in the 0-picture are
\be
V^{\m \n} (p,\bz,z) =
\left(\bpa X^\m (\bz,z) + i p \cdot \bps (\bz) \bps^\m (\bz) \right)
\Big(\pa X^\n (\bz,z) + i p \cdot \ps (z) \ps^\n (z)\Big)
{\rm e}^{i p \cdot X (\bz,z) }\, ,
\label{vop}
\ee
where the polarization tensor $\e_{\m \n}$ is symmetric
traceless for
a graviton ($\r\equiv 1$) and antisymmetric for an antisymmetric
two-form
gauge field
($\r\equiv -1$).

Altogether the physical conditions are
\be
\e_{\m \n} = \r \e_{\n \m} \sp
 p^{\m} \e_{\m \n} = 0 \sp
 p^{\m} p_{\m} =0 \sp
 p_1+p_2+p_3 =0\, .
\label{66}
\ee
Note that they imply $p_i \cdot p_j=0$ for all $i,j$. Were the $p_i$'s
real
and
the metric Minkowskian, this would indicate that the momenta are in
fact
collinear, and all three-point amplitudes would vanish due to
kinematics.  This can be
evaded by going to complex momenta in Euclidean space.

The expression (\ref{vop}) gives the form for all the vertex operators
when we take the even spin structure both on the left and the right.
When one spin structure (say left) is odd, though,
the presence of a conformal Killing
spinor together with a world-sheet gravitino zero-mode requires
one of the vertex operators (say the last one) be converted to the
$-1$-picture on the left
\be
V^{\m \n} (p,\bz,z) =
\left(\bpa X^\m (\bz,z) + i p \cdot \bps (\bz) \bps^\m (\bz) \right)
  \ps^\n (z) {\rm e}^{i p \cdot X (\bz,z) }\, ,
\label{mop} \ee
and a left-moving supercurrent
\be
G_F^{\vphantom{int}} =  \pa X^\g \ps_\g + G_F^{\rm int}
\label{68}
\ee
be  inserted at an arbitrary point on the world-sheet \cite{gm}.

There are four possible spin-structure combinations to consider,
which can be grouped in two pairs according to whether they describe
CP-even or CP-odd couplings,
\be
\mbox{CP-even:}\, \cases{\bar{e}{\rm-}e \cr \bar{o}{\rm-}o \cr }
\quad
\mbox{CP-odd:}\,  \cases{\bar{e}{\rm-}o \cr \bar{o}{\rm-}e \, ,\cr }
\label{cpc}
\ee
where we denote $e$ ($o$) the even (odd) spin structure on the left
and the barred analogues for those on the right.

Because of the physical conditions (\ref{66}), the only kinematic
structures
that can appear at four-derivative order are, in an index-free notation
(see Appendix A):
\be
(p_1 \e_2 p_3) (p_2 \e_1 \e_3 p_2) \quad
\mbox{ and } \quad p_1 \w p_2 \w p_1 \, \e_2 \w p_2 \, \e_1 \w \e_3
\label{authkin}
\ee
up to permutations of $(1,2,3)$.

The low-energy action can then be determined by finding
Lorentz-invariant terms that yield the same vertices on shell.
Depending on the polarization of the incoming particles, the
string amplitude can be reproduced by the following terms in
the effective action (see Appendix A for more details):
\be
\label{R2vert}
\eqalign{
R^2
\equiv&
R_{\m\n\r\s} R^{\m\n\r\s}
=
\Big(p_1 h(p_2) p_3\Big) \Big(p_2 h(p_1) h(p_3) p_2\Big)
\cr
\na H \na H
\equiv&
\na_\m H_{\n\r\s} \na^\m H^{\n\r\s}
=
6 \Big(p_1 h(p_2) p_3\Big) \Big(p_2 b(p_1) b(p_3) p_2\Big)
\cr
B \w R \w R
\equiv&
\e^{\m\n\k\l\r\s}_{\phl} B_{\m \n}^{\phl}
R_{\k\l}^{\hphantom{\k\l}\a\b}
R_{\r\s\a\b}^{\phl}
=
- 2 p_1 \w p_2\w p_1 \, h(p_2)\w p_2 \, h(p_1)\w  b(p_3)
\cr
B \w \na H \w \na H
\equiv&
\e^{\m \n \k \l \r \s}_{\phl} B_{\m \n}^{\phl} \na_{\k}^{\phl}
H_{\l}^{\hphantom{\l}\a \b}
 \na_{\r}^{\phl} H_{ \s \a \b}^{\phl}
=
-2 p_1\w p_2\w p_1 \, b(p_2)\w p_2 \, b(p_1)\w  b(p_3)
\cr
H \w H \w R
\equiv&
\e^{\m \n \k \l \r \s}_{\phl} H_{\m \n \k}^{\phl}
H_{\l}^{\hphantom{\l}\a \b}
 R_{ \r \s \a \b}^{\phl}
=
6 p_2\w p_3\w p_2 \, h(p_3)\w p_3 \, b(p_2)\w  b(p_1)\, .
\cr}
\ee
In these expressions, $h(p_i)$ and
$b(p_i)$ denote the Fourier components
of the graviton and antisymmetric tensor, which we identify with the
polarizations $\e_i$ in the string calculation,
$H_{\m\n\r}=\pa_{\m} B_{\n\r} + \pa_{\r} B_{\m\n} + \pa_{\n} B_{\r\m}$
is the field strength of the two-form potential, and the left-hand side
defines a short-hand notation for the corresponding term (in agreement
with standard notation up to factors of $\sqrt{-g}$).

The precise
meaning to be attributed to Eq. (\ref{R2vert}) is, for instance:
\be
\int \rd^6 x \sqrt{-g} R_{\m\n\r\s} R^{\m\n\r\s}
\stackrel{\rm on \ shell}{=}
  \int { \rd^6 p_1 \, \rd^6 p_2 \, \rd^6 p_3 \over (2\pi)^{12} }
  \d^{(6)}(p_1+p_2+p_3)
  \Big(p_1 h(p_2) p_3\Big) \Big(p_2 h(p_1) h(p_3) p_2\Big)\, .
\ee
Note that other four-derivative terms such as squared Ricci tensor
or squared scalar curvature do not contribute at three-graviton
scattering in traceless gauge, so that their coefficient cannot
be fixed at this order. That this remains true at four-graviton
scattering was proved in~\cite{fotw}; it can be seen as a consequence
of the field redefinition freedom
$g_{\m\n} \rightarrow g_{\m\n} + a R_{\m\n} + b R g_{\m\n}$,
which generates $R^2$ and $R_{\m\n}R^{\m\n}$ couplings from
the variation of the Einstein term.
Similarly, the coupling of two antisymmetric tensors and one graviton
could as well be reproduced by a variety of $RHH$ terms, equivalent
under field redefinitions.

We now defer the interested reader to Appendix D for the
actual detailed evaluation of the string amplitude,
and merely state the salient results:

\begin{itemize}
\item{}
The ${\bar{e}{\rm-}e}$ sector manifestly receives $O(p^4)$
contributions from contractions of four fermi- ons on both sides,
and the resulting terms in the effective action are
\be
{\cal I}_{\rm eff}^{\bar{e}{\rm-}e}
 = 32 \p^3  \int \rd^6 x \sqrt{-g} \left( R^2 + {1\over 6 }
\na H \na H \right)\, .
\label{621b}
\ee
\item{}
In the ${\bar{o}{\rm-}o}$ sector we find the same result, but with
 an overall minus sign
depending on whether we consider type IIA or IIB:
\be
{\cal I}_{\rm eff}^{\bar{o}{\rm-}o}
 =  32 \p^3 \vep \int \rd^6 x \sqrt{-g}
\left(  R^2 + {1\over 6}  \na H \na H\right)\, .
\label{628b}
\ee
Therefore,  one-loop string corrections generate $R^2$ and
$\na H \na H$ terms in the effective action of type IIA
superstring
on $K3$, while no such terms appear in the type IIB superstring.
\item{}
The CP-odd sectors ${\bar{e}{\rm-}o}$ and ${\bar{o}{\rm-}e}$ again lead
to
the same vertices up to a sign depending on type IIA, B but also
on the nature of the particles involved. This leaves
\bs
\be
{\cal I}_{\rm eff, \ IIA}^{\rm CP-odd} =
 32\p^3 \int \rd^6 x \sqrt{-g}\,
{1 \over 2}\, ( B\w R \w R + B\w \na H \w \na H )\, ,
\ee
\be
{\cal I}_{\rm eff, \ IIB}^{\rm CP-odd} =
 - 32\p^3 \int \rd^6 x \sqrt{-g}\,
{1 \over 6}\,   H \w H \w R \, .
\ee
\label{637}
\es
\end{itemize}
Summarizing, we can put the results (\ref{621b}), (\ref{628b})
for the CP-even
terms and (\ref{637}) for the CP-odd terms together, and we record
the one-loop four-derivative terms in the six-dimensional effective
action for type IIA and IIB:
\bs
\be
{\cal I}_{\rm eff, \  IIA} = 32\p^3 \N_6 \int \rd^6 x \sqrt{-g}
\left( 2  R^2  + {1\over 3} \na H \na H + {1\over 2}
 B \wedge ( R \wedge R + \na H \wedge \na H) \right)\, ,
\label{638}
\ee
\be
{\cal I}_{\rm eff, \ IIB} =  - 32\p^3 \N_6 \int \rd^6 x \sqrt{-g}
              \,  {1 \over 6}\, H \wedge  H \wedge R\, ,
\label{639}
\ee
\es
where we introduced
an overall normalization constant~$\N_6$.

As a check note that the type IIA theory should be invariant
under a combined space-time ($P$) and world-sheet parity ($\O$).
Since the Levi--Civita $\e$ tensor changes sign under $P$ while the $B$
field
changes sign under $\O$, we verify the correct invariance under $P\O$.
On the other hand, the type IIB theory is correctly invariant under the
world-sheet parity $\O$, since the interactions contain an even number
of antisymmetric tensor fields.

We should stress here that these thresholds, although they were
computed at
the $T^4/Z_2$ orbifold point of $K3$ are valid for any value of the
$K3$
moduli.
The reason is that the threshold is proportional to the elliptic genus
of $K3$ (which in this case is equal to the $K3$ Euler number) and thus
is moduli-independent.
It can also be seen directly in the $T^4/Z_2$ calculation as follows.
The result is obviously independent of the ($4,4$) orbifold moduli.
All the other moduli have vertex operators that are proportional to the
twist fields
of the orbifold. The correlator of three gravitons or antisymmetric
tensors
and one of the extra moduli is identically zero, since the symmetry
changes the sign of twist fields. Thus, the derivatives of the
threshold with respect to the extra moduli are zero.

\section{One-loop gravitational corrections in four-dimensional
type II models\label{4d}}
\label{susy4D}

Further compactification of six-dimensional $N=2$ type IIA, B string
theory
on a two-torus yields $N=4$ string theories in four dimensions.
Six-dimensional duality between heterotic string on $T^4$ and type IIA
string on $K3$ is expected to
descend to a duality between the corresponding four-dimensional $N=4$
compactified
theories. Moreover, the two compact flat dimensions make it possible
to construct more exotic compactifications,
preserving $N=4$ in four dimensions \cite{fk,kk}, via the
fermionic construction or constructions based on freely-acting
asymmetric
orbifolds. As we will see in Section \ref{rrm}, the models obtained in
this
way may have heterotic $S$-duals
or type II $U$-duals. In the following we shall be interested in
computing the four-dimensional counterparts of the six-dimensional
four-derivative gravitational terms for generic $N=4$ ground states.
Before that, however, we shall briefly recall some features
of $N=4$ supersymmetry.

\subsection{Four-dimensional $N=4$ supersymmetry and its BPS states
\label{4db}}

Massless multiplets of $N=4$ four-dimensional supersymmetry, with
helicity
less than or equal to 2,
are the gravity multiplet (1 graviton, 4 gravitinos, 6 graviphotons,
4 fermions, 1 complex scalar) and the vector multiplet (1 photon,
4 fermions and 6 real scalars).
In particular, the six-dimensional $N=(1,1)$ gravity multiplet
decomposes under reduction into the four-dimensional $N=4$ gravity
multiplet plus two $N=4$ vector multiplets, while the six-dimensional
chiral
$N=(2,0)$
gravity multiplet yields one four-dimensional $N=4$ gravity multiplet
plus one
$N=4$ vector multiplet (upon dualization of four-dimensional two-form
potentials into scalars). On the other hand, both the six-dimensional
$N=(1,1)$
vector and $N=(2,0)$ tensor multiplets reduce to one $N=4$ vector
multiplet
each.

The generic massive $L^j$
representation of $N=4$ supersymmetry contains 128 bosonic
plus 128 fermionic states generated by the action of eight fermionic
raising operators on a spin $j\in\Z/2$ vacuum ($j$ denotes the
representation of the $SO(3)$ little group of massive representations).
However, when the central charge matrix degenerates, only 6 or 4
of the raising operators survive, respectively yielding
intermediate BPS representations $I^j$ of dimension 64 or short
BPS representations\footnote{The massless
representations are always short representations.} $S^j$ of dimension
16.
Such BPS states can be traced by using helicity supertraces, which
behave
as ``indices''
counting unpaired BPS multiplets \cite{bk}. More details about
the actual computation of helicity supertraces can be found in Appendix
B.

\subsection{Gravitational thresholds in four dimensions \label{4dt} }

The two-derivative low-energy effective action for $N=4$ theories is
believed to be exact at tree level, but higher-derivative terms
can receive perturbative and non-perturbative one-loop corrections.
We will be interested in computing the moduli dependence of
the four-derivative terms involving the graviton, antisymmetric tensor
and dilaton, more generally called gravitational thresholds. The
terms of interest are therefore:
\bea
{\cal I}_{\rm eff}
= \int \rd^4 x \sqrt{-g} & \Big( &\! \! \! \!
\D_{\rm gr} (T,U) R_{\m\n\r\s} R^{\m\n\r\s} +
\Theta_{\rm gr} (T,U) \e^{\m\n\r\s}_{\phl} R_{\m\n\a\b}^{\phl}
R_{\r\s}^{\hphantom{\r\s}\a\b}
\cr
&+& \! \! \! \!
\D_{\rm as} (T,U) \na_\m H_{\n\r\s}\na^\m H^{\n\r\s} +
\Theta_{\rm as} (T,U) \e^{\m\n\r\s}_{\phl}
\na_\m^{\phl}H_{\n\a\b}^{\phl} \na_{\r}^{\phl}
H_\s^{\hphantom{\r}\a\b} \cr
&+&\! \! \! \!
\D_{\rm dil} (T,U) \na_\m \na_\n \Fi \na^\m \na^\n \Fi +
\Theta_{\rm dil-as}(T,U) \e^{\m\n\r\s} \na_\m \na_\a \Fi \na^\a
H^{\n\r\s}
\cr
&+&\! \! \! \!
\Theta_{\rm gr-as} (T,U) \e^{\m\n\r\s}_{\phl} R_{\m\n\a\b}^{\phl}
\na_{\r}^{\phl} H_\s^{\hphantom{\r}\a\b}\,
\Big)\, .
\label{eac}
\eea
Again, we shall use a short-hand notation for each term appearing in
the above expression:
$R^2$, $R\w R$, $\na H \na H$,
$\na H \w \na H$, $\na\na\Fi\na\na\Fi$, $\na\na\Fi\w\na H$, $R \w \na
H$. Note
that there is no non-vanishing on-shell  $ R H $-coupling between one
graviton and one two-form, nor any $\na\na\Phi\w\na\na\Phi$ or
$\na\na\Phi\w R$ couplings.
The various terms in Eq.  (\ref{eac}) will turn out to be expressible
in terms of
helicity supertraces and, as such, will receive contributions from
BPS states only. They therefore offer a reliable window into the
strong-coupling regime.

We will now concentrate on the derivation of general
formulas for gravitational
thresholds in four-dimensional
type II models descending from type II six-dimensional vacua
compactified on $K3$.
At first, one might think that such thresholds could be evaluated by
computing two-graviton scattering. Such an amplitude, however,
vanishes on shell, and is potentially infrared-divergent.
A rigorous and unambiguous way to deal with this problem
was described in \cite{KK,KK1} and further analysed in
\cite{PR};
this amounts to regularizing the infrared
by turning on background fields that provide the theory with
a mass gap.
This method preserves some of the original supersymmetries of the
theory: up to $N=2$ for heterotic ground states, and up to
$(p,q)=(2,2)$
for type II ground states, where we denote by $p$ and $q$ the number
of supersymmetries coming from the left and the right. However, this
procedure does not allow us to discriminate the various interaction
terms appearing in (\ref{eac}), by lack of a sufficient number of
marginal operators that could be turned on as background fields.

Here, however, we shall only be interested in the
$(T,U)$ moduli dependence of the four-derivative
gravitational couplings in the effective action;
it will therefore be sufficient to compute the
scattering amplitude between two gravitons (or two two-forms or two
dilatons)
and moduli fields.
This will give access to $\pa_\phi \D$ and $\pa_\phi \Theta$,
which are infrared-finite.

The same comments as in the six-dimensional case apply to the choice
of vertices in Eq.  (\ref{eac}) for describing the string amplitude.
In particular, one may add to this expression terms
such as $R_{\m\n}R^{\m\n}$ or $R^2$ without changing the S-matrix,
and for instance choose instead of $R^{\m\n\r\s}R_{\m\n\r\s}$
the Gauss--Bonnet combination
$(R_{\m\n\r\s}R^{\m\n\r\s} - 4 R_{\m\n} R^{\m\n} + R^2)$,
which has the advantage of being a total derivative at second order
in $h$
and therefore does not correct the graviton
propagator\footnote{The Gauss--Bonnet combination in four dimensions is
a
total
derivative
to any order in $h$, so one might wonder how it could describe vertices
at all. The answer is that the vertices derived from it only vanish
when taking into account all kinematical restrictions on momenta and
polarizations special to four dimensions. Those no longer exist
when going to Euclidean complex momenta. We thank R. Woodard for
discussions
on this point.}.
This would be useful
if one were to look at four-particle scattering, where field theory
subtraction enters into play \cite{sloan}. Also,
it will turn out that the naive $R\wedge R$, $\na H \w \na H$ and
$\na\na\Fi\w \na H$
terms, chosen to represent the CP-odd interaction of gravitons with
moduli,
are inadequate
and have to  be supplemented by Chern--Simons couplings.
With these provisos, the kinematical structures contributing to
gravitational thresholds read:
\be
\label{4Dvert}
\eqalign{
R^2
=&\,
\Big(p_1 h(p_2) p_1\Big)\Big(p_2 h(p_1) p_2\Big)
-2 (p_1 p_2) \Big(p_2 h(p_1) h(p_2) p_1\Big) + (p_1 p_2)^2 \Big(h(p_1)
h(p_2)\Big)
\cr
\na H \na H
=&\,
 6 (p_1 p_2) \Big(p_2 b(p_1) b(p_2) p_1\Big) -3 (p_1 p_2)^2 \Big(b(p_1)
b(p_2)\Big)
\cr
R \w R
=&
- 2 h(p_2)\,  p_1 \w h(p_1)\,  p_2 \w p_1\w p_2
  - 2(p_1 p_2) \, h(p_1) \, h(p_2) \w p_1\w p_2
\cr
\na H \w \na H
=&
- 2b(p_2)\,  p_1\w b(p_1)\,  p_2\w p_1\w p_2
  + 2(p_1 p_2) \, b(p_1)\, b(p_2)\w p_1\w p_2
\cr
R \w \na H
=&
- 2b(p_2)\,  p_1\w h(p_1)\,  p_2\w p_1\w p_2
  + 2(p_1 p_2)\,  h(p_1)\, b(p_2)\w p_1\w p_2
\cr
\na\na\Fi\na\na\Fi
=&\,
(p_1 p_2)^2
\cr
\na\na\Fi \w \na H
=&\,
3 (p_1 p_2)\,  p_1 \w p_2 \w b(p_2)
\cr
}
\ee
and it is readily checked that these expressions are consistent with
gauge
invariance $\e \to \e + p\otimes k + \r\,k\otimes p$ with
$k\cdot p=0$.

The second equation in (\ref{4Dvert}) shows that the $\na H \na H$
coupling cannot be revealed by a three-particle amplitude.
This forces us to look at scattering amplitudes involving at least
two gravitons (or two two-forms or two dilatons) and two moduli.
In fact, the insertion of any number of moduli remains tractable
as long as two complex-conjugated moduli are not simultaneously
present, and we shall therefore
keep with the general case of $N$ moduli.

\subsection{Two-graviton--$N$-moduli scattering amplitude
\label{2gnm} }

The class of ($2,2$) supersymmetric models descending from
six-dimensional type II string on $K3$ can be generically described at
the $Z_2$ orbifold point of $K3$ by the following partition function:
\begin{eqnarray}
 Z_{\; \II}^{\rm four \ dim} &=&
{1 \over \t_2\,  \vert\eta \vert^{12}}
 {1\over 2}\sum_{a,b=0}^1 (-1)^{a+b+ab}
 \vartheta^2{a\atopwithdelims[]b}
{1\over 2}\sum_{\ba,\bb=0}^1 (-1)^{\ba+\bb+\mu \ba \bb}
 \thb^2{\ba\atopwithdelims[]\bb}
\cr
&& \ti \, {1\over 2}\sum_{h,g=0}^1
\vartheta{a+h\atopwithdelims[]b+g}
\vartheta{a-h\atopwithdelims[]b-g}
\thb{\ba+h\atopwithdelims[]\bb+g}
\thb{\ba-h\atopwithdelims[]\bb-g}
 Z_{6,6} \ar{h}{g}
\cr
&\equiv &
 \sum_{a,b=0,1} \sum_{\ba,\bb=0,1}
Z\ar{\ba\ a}{\bb\ b}\, ,
\label{gp}
\end{eqnarray}
where $Z_{6,6} {h\atopwithdelims[]g}$ are
generic\footnote{
In particular, they are not necessarily lattice partition functions,
and may carry dependence on several untwisted or twisted moduli.}
orbifold blocks whose structure
depends on the specific way the $Z_2$ group acts on the various states
of the spectrum.

For such a vacuum, we shall need to extract the four-momenta part
from the following amplitude:
\be
{\cal I}_{\f} = \e_{1\m\n} \e_{2\k \l}
\ifdd \int \prod_{i=1}^{N+2} {\rd^2 z_i  \over
\p}
\left\langle V^{\m \n}(p_1,\bz_1,z_1)
 V^{\k \l}(p_2,\bz_2,z_2)
\prod_{j=3}^{N+2} V_{\f_j}(p_j,\bz_j,z_j)
\right\rangle\, ,
\label{tpa}
\ee
containing two gravitons or antisymmetric tensor fields (depending on
the
polarization tensors $\e_{i \m \n}$) and $N$ two-torus
moduli fields.
In contrast to Section \ref{6d}, the space-time
indices now run
over $\m=0,\ldots , 3$, but the vertex operators of the space-time
fields
are identical to those given in Eq. (\ref{vop}) for the 0-picture,
Eq. (\ref{mop}) for the $-1$-picture on the left, etc.
In close analogy, in the 0-picture the vertex operators of the moduli
fields
are
given by
\be
V_\f(p,\bz,z) =
v_{IJ}(\f)  \left(\bpa X^I (\bz,z) + i p \cdot \bps (\bz) \bps^I
(\bz)\right)
\Big( \pa X^J (\bz,z) + i p \cdot \ps (z) \ps^J (z) \Big)
{\rm e}^{i p \cdot X (\bz,z) }\, ,
\ee
where
\be
v_{IJ}(\f) = \pa_\f \left(G_{IJ} + B_{IJ} \right)
\sp I,J=1,2\, .
\ee
In particular, in the standard $(T,U)$ parametrization recalled in
Appendix C, we have
\be
v_{IJ}(T) \bpa X^I \pa X^J ={ 1 \over 2i \iU} \bpa X \pa \tX
\sp
v_{IJ}(U) \bpa X^I \pa X^J ={ i \iT \over 2 \iU^2} \bpa \tX \pa \tX
\ee
with $X=X^4+U X^5$, $\tX = X^4 + \bU X^5$ (and similarly
$\Ps = \ps^4 + U \ps^5$), while the vertices for $\bT,\bU$ are
obtained
by complex conjugation. Note that chiral moduli $(T,U)$ have $\pa \tX$
as left-moving part, while the antichiral ones $(\bT,\bU)$ have
$\pa X$ instead.

The modifications for $-1$-picture on the right and/or left
are as described in Section \ref{6d}, so that for example for the
$-1$-picture
on the left we have
\be
V_T(p,\bz,z) = {1\over 2i \iU} \left(\bpa X + i p\cdot \bps \bPs\right)
\tPs
               {\rm e}^{i p \cdot X (\bz,z) }
\ee
together with an insertion of the left-moving supercurrent
\be
G_F =  \pa X^\m \ps_\m + G_{KL} \pa X^K  \ps^L =
\pa X^\m \ps_\m + \pa X \tPs + \pa \tX \Ps\, ,
\ee
where we omitted the $K3$ internal part of $G_F$.

We will again defer the details of the computation to Appendix D, and
simply
outline the calculation here. A drastic simplification occurs thanks
to a selection rule that forbids contractions not conserving the $U(1)$
charge
of the $T^2$ superconformal theory:
\be
\langle X X \rangle = \left\langle \tX \tX \right\rangle
=\langle \Ps \Ps \rangle = \left\langle \tPs \tPs \right\rangle =0\, .
\ee
Except when a pair of complex-conjugated moduli occurs, only the
zero-mode
of the bosonic part of the moduli vertices contributes and generates
for
each insertion a derivative with respect to the corresponding modulus
(together,
in the odd structure, with a sign depending on the nature of the last
modulus). Supersymmetry
then demands that the fermionic part of the two gravitons be contracted
together, yielding the four powers of momenta as desired. The
$\bar{e}{\rm-}e$ and $\bar{o}{\rm-}o$ kinematics
turn out to be equal in the two-graviton case  and opposite in the
two-antisymmetric-tensor case (zero in the graviton--two-form case).

Our final result for the one-loop moduli dependence of the
four-derivative gravitational couplings in Eq. (\ref{eac}) is
summarized by
\bs
\label{gfo}
\be
\pa_\f \D_{\rm gr} (T,U) = \,
{\N_4\over \p^4}  \int_{\cal F} \rd^2 \t \,  {1\over 2}\, \pa_\f \left(
\k^{\bar{e}e} Z^{\bar{e}e}
-\k^{\bar{o}o} Z^{\bar{o}o}\right)
\label{dfo}
\ee
\be
\pa_\f \D_{\rm as} (T,U) = \,
{\N_4\over \p^4}  \int_{\cal F} \rd^2 \t \,  {1\over 12}\, \pa_\f
\left(
\k^{\bar{e}e} Z^{\bar{e}e}
+\k^{\bar{o}o} Z^{\bar{o}o}\right)
\label{dfo2}
\ee
\be
\pa_\f\D_{\rm dil} (T,U) = \,
{\N_4\over \p^4}  \int_{\cal F} \rd^2 \t \,  {1\over 2}\, \pa_\f \left(
\k^{\bar{e}e} Z^{\bar{e}e}
+\k^{\bar{o}o} Z^{\bar{o}o}\right)
\label{dfo3}
\ee
\be
\pa_\f \Th_{\rm gr} (T,U) = -
{\N_4 \over \p^4} \int_{\cal F} \rd^2 \t \,  {1 \over 4}\, \pa_\f
\left(
 \k^{\bar{e}o} Z^{\bar{e}o}
+ \k^{\bar{o}e} Z^{\bar{o}e}\right)
\label{tfo}
\ee
\be
\label{tfo2}
\pa_\f \Th_{\rm as} (T,U) = -
{\N_4 \over \p^4} \int_{\cal F} \rd^2 \t \,  {1 \over 4}\, \pa_\f
\left(
 \k^{\bar{e}o} Z^{\bar{e}o}
+ \k^{\bar{o}e} Z^{\bar{o}e}\right)
\ee
\be
\label{tfo3}
\pa_\f \Th_{\rm gr-as} (T,U) = -
{\N_4 \over \p^4}  \int_{\cal F} \rd^2 \t \,  {1 \over 2}\, \pa_\f
\left(
 \k^{\bar{e}o} Z^{\bar{e}o}
- \k^{\bar{o}e} Z^{\bar{o}e}\right)
\ee
\be
\label{tfo4}
\pa_\f \Th_{\rm dil-as} (T,U) = -
{\N_4 \over \p^4}  \int_{\cal F} \rd^2 \t \,  {1 \over 3} \, \pa_\f
\left(
 \k^{\bar{e}o} Z^{\bar{e}o}
- \k^{\bar{o}e} Z^{\bar{o}e}\right)\, ,
\ee
\es
where $\N_4$ is a normalization constant that we will fix later.
The derivative $\pa_\f$ stands for the product $\prod_{j=3}^N
\pa_{\f_j}$.
The $\k^{\bi j}$ are numerical coefficients that depend on the choice
of type II string as well as on the choice of moduli:
\be
\k^{\bi j} = \cases{
1
\sp \bi,j=\bar{e},e \cr
-\s_\f \vep
\sp \bi,j=\bar{o},o \cr
i \chi_\f
\sp \bi,j=\bar{e},o \cr
i \s_\f \chi_\f \vep
\sp \bi,j=\bar{o},e\, , \cr }
\label{kad}
\ee
where $(\chi_\f,\s_\f)$ specifies the nature of the last modulus
(see Eqs. (\ref{mid},b))
and the conformal blocks $Z^{\bi j}$ are expressed in terms of
the blocks $Z \ar{\ba \  a}{\bb \  b}$ appearing in the
four-dimensional partition function (\ref{gp}):
\be
Z^{\bi j} = \cases{
16 \p^2
\sum_{(a,b),\ (\ba,\bb) {\rm \ even}}
Z \ar{\ba \ a}{\bb \ b}
\pa_{\bar{\t}} \log \left( { \thb \bss  \over \etb }
\right)
\pa_\t \log \left( { \th \ss  \over \et }
\right)
  \sp \bi,j=\bar{e},e \cr
{'\!Z'} \ar{1 \ 1}{1 \ 1}
  \sp \bi,j=\bar{o},o \cr
-4 \p i
\sum_{(\ba,\bb) {\rm \ even}}
\pa_{\bar{\t}} \log \left ({\thb \bss  \over \etb }\right)
Z' \ar{\ba \ 1}{\bb \ 1}  \sp \bi,j=\bar{e},o \cr
{\hphantom -}4 \p i
\sum_{(a,b){\rm \ even}}
{'\!Z} \ar{1 \ a}{1 \ b}
\pa_\t \log \left ({\th \ss  \over \et} \right)
  \sp \bi,j=\bar{o},e \, .\cr}
\label{zij}
\ee
In the previous expression, a prime on the left and/or the right
stands for the operation in Eq.  (\ref{611}).

\subsection{Gravitational thresholds and  helicity supertraces}

Using Riemann identity
and ($2,2$) supersymmetry, it is readily seen that the four blocks
$Z^{\bi j}$ are equal to $Z^{\bar{o}e}$.
Moreover, identity (\ref{pv2}) allows us
to convert the $\pa_\t$ derivative in $Z^{\bar{e}e}$ into
a second-order derivative with respect
to the variable $v$ conjugate to the left helicity $\l_{\rm L}$,
as described in Appendix B. A similar statement applies to the right
side,
yielding:
\be
Z^{\bi j} = {16\p^4\over\t_2} \left\langle \l_{\rm L}^2 \l_{\rm R}^2
\right\rangle
  = {8\p^4 \over 3\t_2} \left\langle (\l_{\rm L} + \l_{\rm R})^4
\right\rangle
  = {8\p^4 \over 3\t_2} B_4\, .
\ee
Substituting in Eq.  (\ref{gfo}), we obtain for instance
\be
\pa_\f \D_{\rm gr} (T,U) = {8 \over 3} \N_4
{1 + \vep \s_\f \over 2}
\int_{\cal F} {\rd^2 \t\over\t_2} \pa_\f B_4\, ,
\label{db4}
\ee
and similar relations for the other thresholds.
This makes it obvious that only short BPS states contribute to
the one-loop four-derivative gravitational corrections. From now on,
it will be convenient to fix the normalization constant
to
\be
\N_4={3\over 8}\, .
\ee
We note that the four different spin structures contribute in the same
way to $\pa_\f \Delta_{\rm gr}$, but for signs depending on the type A
or B of
superstring and the modulus $\f$ we are considering.
As a result of this interference:
\bs
\label{tdc}
\be
\mbox{type IIA:}\, \cases{
\pa_T \D_{\rm gr} =  \int_{\cal F} {\rd^2 \t \over \t_2}  \pa_T B_4
\cr
\pa_U \D_{\rm gr} = 0
\cr}
\label{da1}
\ee
\be
\mbox{type IIB:}\, \cases{
\pa_T \D_{\rm gr} = 0
\cr
\pa_U \D_{\rm gr} = \int_{\cal F} {\rd^2 \t \over \t_2}  \pa_U  B_4\, .
\cr}
\label{db1 tdc}
\ee
\es
We recover in this way the well-known result that $\Delta_{\rm
gr}$
only depends on the K\"{a}hler moduli $T$ and not on the
complex-structure
moduli $U$ in type IIA, while the reverse is true in type IIB
\cite{bcov}.
Similar interferences occur for all thresholds and yield the
following moduli dependences:
\bs
\be
{\rm IIA}: \; \;
\Delta_{\rm gr}(T)\sp
\Delta_{\rm as}(U)\sp
\Delta_{\rm dil}(U)\sp
\Theta_{\rm gr}(T)\sp
\Theta_{\rm as}(T)\sp
\Theta_{\rm gr-as}(U)\sp
\Theta_{\rm dil-as}(U)\, ,
\ee
\be
{\rm IIB}: \; \;
\Delta_{\rm gr}(U)\sp
\Delta_{\rm as}(T)\sp
\Delta_{\rm dil}(T)\sp
\Theta_{\rm gr}(U)\sp
\Theta_{\rm as}(U)\sp
\Theta_{\rm gr-as}(T)\sp
\Theta_{\rm dil-as}(T)\, .
\ee
\es
The dependence of $\Delta_{\rm gr}(T)$ is consistent with
our argument that the $R^2$ term does not get corrections
beyond one loop.
However, there exists a subgroup of $SO(6,N_V,Z)$ that exchanges
the (type IIA) $U$-modulus with the dilaton $S$-modulus,
so that $SO(6,N_V,Z)$ duality implies that $\Delta_{\rm as},\Delta_{\rm
dil},
\Theta_{\rm gr-as},\Theta_{\rm dil-as}$ are also $S$-dependent, i.e.
are
perturbatively and non-perturbatively corrected. The loophole
in the argument of Section 2 is that, for these couplings, the
world-sheet instantons of the type IIB string are non-zero
(since they depend on the type IIB $T$-modulus), and therefore
the $(p,q)$ D 1-branes do contribute to instanton corrections.
{}From now on we shall restrict ourselves to $R^2$ thresholds,
for which the type II one-loop result is exact \cite{kp, bk2, apt}.

\boldmath
\section{Gravitational thresholds in ordinary type II on $K3 \ti T^2$}
\label{m1}
\unboldmath

We now apply the previous formalism to the trivial reduction to four
dimensions
of type II string theory on $K3$.  Using the considerations in
Subsection
\ref{4db} and the six-dimensional spectrum, it follows that
both type IIA and type IIB on $K3\ti T^2$ have
\be
\mbox{1 supergravity multiplet} \sp \mbox{22 vector multiplets}\, .
\label{ordsp}
\ee
The two theories are indeed exchanged by
$T$-duality on one circle of $T^2$, which corresponds to the exchange
of
$T$ and $U$ moduli. The scalars therefore span
${SU(1,1)/ U(1)} \ti SO(6,22)\Big/ \Big(SO(6) \ti SO(22)\Big)$,
where the $SU(1,1)/U(1)$ factor corresponds to the complex scalar
in the gravitational multiplet ($T$ for IIA, $U$ for IIB) \cite{fk}.

Type II string theory on $K3 \times T^2$ at the $T^4/Z_2$ orbifold
point is described by the following partition function:
\bea
\tn{22}: \;\;
Z &=&
{1 \over \t_2\,  \vert\eta \vert^{24}}
 {1\over 2}\sum_{a,b=0}^1 (-1)^{a+b+ab}
 \vartheta^2{a\atopwithdelims[]b}
{1\over 2}\sum_{\ba,\bb=0}^1 (-1)^{\ba+\bb+\mu \ba \bb}
 \thb^2{\ba\atopwithdelims[]\bb}
\cr
&& \ti \, {1\over 2}\sum_{h,g=0}^1
\vartheta{a+h\atopwithdelims[]b+g}
\vartheta{a-h\atopwithdelims[]b-g}
\thb{\ba+h\atopwithdelims[]\bb+g}
\thb{\ba-h\atopwithdelims[]\bb-g}
\Gamma_{4,4}\hg \Ga_{2,2} (T,U)\, ,
\label{pf1}
\eea
where we use the same $\Ga_{4,4}\hg$ blocks as in (\ref{z44}).

\subsection{Helicity supertraces and $R^2$ corrections\label{mdgr}}

The helicity supertrace $B_4$ entering in the threshold
(\ref{db4}) can be readily computed from (\ref{pf1}) using the methods
of Appendix B, with the result:
\be
B_4 = 36~\Ga_{2,2}\, .
\label{b41}
\ee
It is easy to check the $\t_2 \to \infty$ limit, where
only short BPS massless states contribute, with the result:
\be
\left. B_4 \right\vert_{\rm massless} =   1 \ti 3 + 22 \ti \frac{3}{2}
= 36\, ,
\ee
where we used the contributions in Eq.  (\ref{b4r}) for the
supergravity and vector multiplets.
The expression in Eq. (\ref{b41})
further
shows that the rest of the contributions to $B_4$ come from the tower
of massive short BPS multiplets whose vertex operators are those of the
massless states plus momenta and windings of the two-torus.
The matching condition implies that we should have $\vec m  \vec
n=0$
for these states and they are in $N=4$ supermultiplets similar to the
massless
ones.
This result is expected, since we know that a left-moving state
breaks
half of the  two left-moving supersymmetries.
Thus states that are ground states both on the left and right (plus
momentum
of the two-torus) are expected to break half out of the total of
four supersymmetries in agreement with the helicity
supertrace.

Using Eq. (\ref{gres6}), not
much more work is required to extract the $B_6$ supertrace
\be
B_6  = 90~\Ga_{2,2}\, ,
\label{b61}
\ee
whose $\t_2\rightarrow \infty$ limit again agrees with the
massless (short BPS) spectrum since  $ 1 \ti {195 \over 4} + 22 \ti
\frac{15}{8} = 90$.
However, although we know that intermediate multiplets, corresponding
to states that are ground states on the left only, but with arbitrary
oscillator excitations  on the right (or reversed)
could contribute to $B_6$, they turn out to cancel
as a consequence of identity (\ref{hdeid}).
We therefore conclude that intermediate BPS multiplets
come in combinations that can always be paired
into long massive multiplets and thus do not contribute to $B_6$.
Their multiplicities and mass formulae are therefore not protected from
quantum corrections.
This example indicates that one has to be careful when invoking
non-renormalization theorems for BPS states.
Only BPS states having non-zero ``index" are protected from quantum
corrections.

We now insert $B_4$ into Eq.  (\ref{db4}) and use the
fundamental-domain integral (\ref{dkli}) to obtain the $R^2$
thresholds:
\bs
\be
\mbox{type IIA:}\; \;  \D_{\rm gr} (T) =  -36 \log \left(
T_2 \left|
\et(T)\right|^4 \right)
 + {\rm const.}\, ,
\label{789} \ee
\be
\mbox{type IIB:}\; \;  \D_{\rm gr} (U) = -36  \log \left(
U_2 \left|
\et(U)\right|^4 \right)
 + {\rm const.}\, ,
\ee
\es
where the constant is undetermined in our scheme.
The above result is  in agreement with \cite{hm1}.
Note that the one-loop thresholds are respectively invariant under
$SL(2,Z)_T$ and
$SL(2,Z)_U$, as they should.
Moreover, since only
the twisted sectors $(h,g)\ne(0,0)$ of $T^4/Z_2$
contribute to $B_4$, $\D_{\rm gr}$ as well as the other thresholds
are independent of the untwisted moduli of $K3$, and therefore
of all $K3$ moduli. Consequently, the result obtained at the orbifold
point
$T^4/Z_2$ is valid everywhere in the moduli space of $K3$.

\subsection{Decompactification limit of CP-even couplings: a puzzle
\label{puz}}

It is important to confront this result to our six-dimensional result
(\ref{638}),
which should be retrieved in the
decompactification limit of the two-torus,
$T_2=\sqrt{G}
\to \infty$:
\be
\tn{22} : \;\;
\D_{\rm gr} (T) \limit{\longrightarrow}{T_2 \to \infty}
-\! 36 \log T_2 + 12\p T_2
+{O}\left({\re}^{-T_2}\right)
\label{lv}
\ee
in the type IIA situation.
This agrees with Eq.  (\ref{638}) provided we set
\be
\N_6={4\p^2 \over 3}\, .
\ee
On the other hand, taking the large-volume limit in the type IIB
theory does not affect the $U$-dependent threshold. However, only terms
of
order $T_2$ (the volume of the torus) can be seen in the
decompactification limit, so this agrees with the vanishing
of $R^2$ coupling in six dimensions~(\ref{639}).

We can repeat the same discussion for the four-dimensional $\na H \na
H$
threshold, which has the same behaviour up to $T\leftrightarrow U$
interchange, and
predict
that the six-dimensional coupling $\na H \na H$ should occur only in
type IIB and not in type IIA, in contrast to $R^2$.
This is in disagreement with our six-dimensional result,
which showed that cancellation between ${\bar{e}{\rm-}e}$ and
${\bar{o}{\rm-}o}$
spin structures had to occur in the same way for both $R^2$ and
$\na H \na H$. Note that we could also have performed the
three-graviton--two-form scattering calculation directly in four
dimensions,
finding the same result for $\bar{e}{\rm-}e$ as in six dimensions, but
a
vanishing
$\bar{o}{\rm-}o$ contribution. We would have concluded that $R^2$ and
$\na H \na H$ have to occur with the same $(T,U)$-dependent coupling,
in both types IIA and IIB. This shows that the three-particle amplitude
has to be interpreted with great care.

\subsection{CP-odd couplings and  holomorphic anomalies}

Moving on to the CP-odd couplings and focusing on the IIA case for
definiteness,
Eq. (\ref{tfo}) yields
\be
\partial_T\Theta_{\rm gr}=-18 i \partial_T\log \left(T_2\left|
\et(T)\right|^4\right)\sp
\partial_{\bT}\Theta_{\rm gr}=18 i \partial_{\bT}\log\left(T_2\left|
\et(T)\right|^4\right)\, .
\label{640}
\ee
Would the non-harmonic $T_2$ term be absent, those two equations
could be easily integrated and would give
\be
\Theta_{\rm gr}(T) = 18 \Im \log \eta^4(T)\, .
\ee
However, in the presence of the $T_2$ term the notation
$\pa_T \Theta$ and $\pa_{\bar{T}} \Theta$
for CP-odd couplings between two gravitons
and one modulus no longer makes sense.
This non-integrability of CP-odd couplings
has already been encountered before \cite{agnt}. This
problem can be evaded
simply by rewriting the CP-odd coupling as
\be
{\cal I}^{\rm CP-odd}_{\rm gr}=\int\O \wedge  (Z_T \rd T + Z_{\bT} \rd
\bT)\, ,
\label{641}
\ee
where $\O$ is the gravitational Chern--Simons three-form, such that
$\rd \O=R\w R$.
In the special case $Z_T=\partial_{T}\Theta(T,\bT)$,
$Z_{\bT}=\partial_{\bT}\Theta(T,\bT)$, one retrieves by partial
integration the usual integrable CP-odd coupling. In the case at hand,
\be
Z_T=-18 i \partial_T\log \left(T_2\left|
\et(T)\right|^4\right)\sp
Z_{\bT}=18 i \partial_{\bT}\log\left(T_2\left|
\et(T)\right|^4\right)\, .
\label{6422}
\ee
We can take advantage of the special structure of Eq.  (\ref{6422}) and
rewrite Eq.  (\ref{641}) as
\be
{\cal I}^{\rm CP-odd}_{\rm as}=18  \p\int
\left({\Im}\left(\log\eta^4(T)\right)R \w R-
{1\over T_2} \, \Omega \w \rd T_1 \right)\, .
\label{643p}
\ee
In the decompactification limit $T_2\to \infty$, only the first
term survives and we obtain
\be
{\cal I}^{\rm CP-odd}_{\rm gr}=18 \int\left({\pi\over 3}\,
T_1 \, R\w R
+{O}(1/ T_2)\right)\, .
\label{644}
\ee
This reproduces the six-dimensional type IIA result (\ref{638}):
\bea
{\cal I}_{\, \rm IIA}^{\rm six \ dim}
&=& 16\p^3\N_6 \int \rd^6 x \sqrt{-g_6}\,  \e^{\m\n\k\l\r\s}_{\phl}
B_{\m\n}^{\phl}
  R_{\k\l\a\b}^{\phl} R_{\r\s}^{\hphantom{\r\s}\a\b}
\cr
&\stackrel{\rm four \ dim}{\longrightarrow}&
 6\p \int \rd^4 x \sqrt{-g_4}\,  T_2 \e^{IJ}B_{IJ} \e^{\k\l\r\s}_{\phl}
R_{\k\l\a\b}^{\phl} R_{\r\s}^{\hphantom{\r\s}\a\b}
\eea
since $\e^{IJ} B_{IJ}=2 T_1/T_2$ and $\N_6=4\p^2/3$.

Exactly the same feature arises for the $\Theta_{\rm as},\Theta_{\rm
gr-as}$
and $\Theta_{\rm dil-as}$
cases, for which, in the type IIA case, the correct coupling should
instead
be written as
\bs
\label{643}
\be
{\cal I}^{\rm CP-odd}_{\rm as}=18  \p\int \left(\Im
\left( \log\eta^4(T)\right) \na H \w \na H-
{1\over T_2} \, H \w \na H \w \rd T_1 \right)
\label{6434}
\ee
\be
{\cal I}^{\rm CP-odd}_{\rm gr-as}=36  \p\int \left(\Im
\left(\log\eta^4(U)\right)R\w \na H
-{1\over T_2}\,  R\w H \w \rd T_1 \right)
\label{6435}
\ee
\be
{\cal I}^{\rm CP-odd}_{\rm dil-as}=24  \p\int \left(\Im
\left(\log\eta^4(U)\right)\na\na\Fi\w \na H
-{1\over T_2} \, \na\na\Fi \w H \w \rd T_1 \right)\, .
\label{6436}
\ee
\es
Note also that
$\na H\na H$ correctly decompactifies to the $B\w\na H \w \na H$ of
six-dimensional type IIA theory in just the same way as $R \w R$,
while in type IIB $R\w \na H$ gives the correct
$H\w H \w R = -3 B \w R \w \na H$ six-dimensional coupling.
The $\na\na\Fi\w \na H$ coupling cannot be checked here since we did
not
consider six-dimensional dilaton scattering.

\subsection{From type II to heterotic string}

Coming now to duality, it is well known
that heterotic on $T^4$ -- type IIA on $K3$ duality in six dimensions
implies,
after compactification, the duality of the corresponding
four-dimensional
theories under exchange of $S$ and $T$, where $S$ is the
axion--dilaton multiplet, sitting in the gravitational multiplet
on the heterotic side.

For definitess we recall the partition
function of heterotic string on $T^6$:
\be
\he{22} : \;\;
Z={1 \over \t_2 \, \et^{12} \bar{\et}^{24}}
\frac{1}{2}
\sum_{a,b=0}^1 (-1)^{a+b+ab} \th^4 \ar{a}{b}
\Ga_{6,22} (G,B,A) \, ,
\ee
where $\Ga_{6,22} (G,B,A)$ depends on the six-dimensional metric $G$,
the
antisymmetric tensor $B$ and the Wilson lines $A$. At generic points of
the moduli space (i.e. with gauge group broken to $U(1)$ factors), the
massless
bosonic spectrum is
\be
\mbox{1 supergravity multiplet} \sp \mbox{22 vector multiplets}\, ,
\ee
in agreement with (\ref{ordsp}), as expected by duality.
Contrary to the type II case, the heterotic string theory possesses
a tree-level $R\wedge R$ coupling required for anomaly cancellation
through the Green--Schwarz mechanism, together with an $R^2$ coupling
required for supersymmetry.
The world-sheet fermions now have 10 zero-modes, so that the one-loop
three-particle amplitude vanishes (in even spin structure, one would
need four fermionic contractions to have a non-vanishing result
after spin-structure summation). In particular, we conclude that
there is no one-loop correction  to tree-level $R^2$ coupling.

Following \cite{hm1}, we can therefore translate the type IIA result
(\ref{789}) for the heterotic string on $T^6$:
\be
\he{22} : \;\;   \D_{\rm gr}  (S) = - 36 \log \left( S_2 \left|\et (S)
\right|^4 \right)
\limit{\longrightarrow}{S_2 \ra \infty}
-\! 36 \log S_2 +12 \p S_2 +O \left({\re}^{-S_2}\right)\, .
\label{hsd}
\ee
The ${S_2 \ra \infty}$ heterotic weak-coupling limit exhibits
the tree-level $R^2$ coupling together with a non-perturbatively
seen logarithmic divergence. The latter was omitted in Ref. \cite{hm1},
where only the Wilsonian effective action was investigated,
but is also present in other instances~\cite{kp}.
The full threshold is manifestly invariant under $SL(2,Z)_S$,
and could in fact be inferred from $SL(2,Z)_S$ completion
of the tree-level result. The exponentially suppressed terms
in Eq.  (\ref{hsd}) were identified in \cite{hm1} with the instanton
contributions of the neutral heterotic NS 5-brane wrapped on $T^6$,
the only instanton configuration that can possibly occur in
four-dimensional heterotic string.

The same mapping can be executed for the CP-odd $R\w R$ coupling
from Eq.  (\ref{643}):
\be
{\cal I}^{\rm CP-odd}_{\rm gr}=18  \int \left(\Im
\left(\log\eta^4(S)\right)
R\w R-{1\over S_2}\, \O \w  \rd S_1 \right)\, .
\label{643h}
\ee
There, however, in addition to the tree-level term and instead
of the logarithmic divergence, we find a coupling between
the axion and the gravitational Chern--Simons form.
Dualizing the axion into a two-form and keeping track
of the powers of the heterotic coupling $S_2$, this translates
into a {\it one-loop} coupling $H_{\m\n\r} \Omega^{\m\n\r}$
between one two-form and two gravitons, precluded by
a one-loop heterotic calculation. Happily enough,
the Chern--Simons form is co-closed, so that this coupling is a
total derivative.

\boldmath
\section{Reduced-rank $N=4$ models and breaking of $S$-duality
\label{rrm}}
\unboldmath

Although the most studied $N=4$ dual string pair is the standard
heterotic on $T^6$ -- type IIA on $K3\times T^2$ pair with generic
gauge group $U(1)^{28}$, more
exotic
models with a lower gauge-group rank do exist. Since all $N=4$ matter
multiplets have to transform into the adjoint representation
of the gauge group, their expectation values cannot break it to a group
with
lower rank, and those theories therefore have to live in disconnected
moduli spaces.

On the type II side,
such models can be easily obtained by
compactifying the six-dimensional IIA on $K3$ theory at orbifold points
of $K3$ using a generalized Scherk--Schwarz mechanism
\cite{ss1,ss2,kr}  to give a (moduli-dependent)
mass \cite{decoa, kk, decol} to part of the vector multiplets
originating from the
twisted
sectors of $K3$. This can be implemented by orbifolding the
IIA on $K3\ti T^2$ theory by a translation on the torus accompanied by
an action on the twisted sector.

On the heterotic side, such models have been constructed in Ref.
\cite{Chaudhuri} with fermionic characters, but it is difficult
to identify them with models dual to the above type II, since that
would
require
identifying the point in heterotic moduli space corresponding to
the orbifold points of $K3$. Nevertheless, if one trusts
six-dimensional
heterotic--type IIA duality, such heterotic duals are guaranteed to
exist.

The construction on the type II side makes it clear that $T$-duality
is {\it broken} to a subgroup by the precise translation vector on
the $\Ga_{2,2}$ lattice, which translates in heterotic variables into
a {\it breaking} of $S$-duality.
This breaking modifies the non-perturbative instanton corrections in
lower-rank heterotic or type II theories discussed below.
In the following, we shall examine the four-derivative perturbative
gravitational corrections in various type II models, and translate
them in terms of non-perturbative effects on the heterotic side.
\vskip 0.3cm

\subsection{The $\itbs$--$\itfs $  $U$-dual type II pair}

Here we consider a variation of the type II over
$T^4/Z_2\ti T^2$ compactification described above (see model $\tn{22}$,
(\ref{pf1})).
The $Z_2$ will now
act both as a twist on the $T^4$
and as a shift on the two-torus. This model is a spontaneously broken
$N=8 \ra  4$ theory with ($2,2$) supersymmetry, as will become clear
shortly, and will be denoted by $\tbs (w)$.
The partition function reads:
\bea
\tbs(w) : \;\; Z  &=&
{1 \over \t_2\,  \vert\eta \vert^{24}}
 {1\over 2}\sum_{a,b=0}^1 (-1)^{a+b+ab}
 \vartheta^2{a\atopwithdelims[]b}
{1\over 2}\sum_{\ba,\bb=0}^1 (-1)^{\ba+\bb+\mu \ba \bb}
 \thb^2{\ba\atopwithdelims[]\bb}
\cr
&& \ti \, {1\over 2}\sum_{h,g=0}^1
\vartheta{a+h\atopwithdelims[]b+g}
\vartheta{a-h\atopwithdelims[]b-g}
\thb{\ba+h\atopwithdelims[]\bb+g}
\thb{\ba-h\atopwithdelims[]\bb-g}
\Gamma_{4,4}^{\phl} \ar{h}{g}
\Ga_{2,2}^w \ar{h }{ g}\, ,
\label{rr2}
\eea
where $\Gamma_{4,4} \ar{h}{g}$ are the twisted ($4,4$) lattice sums
(see Eq. (\ref{z44})) and
$\Ga_{2,2}^w \ar{h}{g}$ are the shifted ($2,2$) lattice sums given in
Appendix C. Modular invariance requires the shift vector $w$
to satisfy $w^2 = 0$.
The 16 twisted vector multiplets from the $T^4 /Z_2 \ti T^2$
model
now acquire a mass of the order of the inverse radii of $T^2$,
so that the massless spectrum becomes:
\be
\mbox{1 supergravity multiplet} \sp \mbox{6 vector multiplets}\, .
\label{sp3}
\ee
The scalars of the 6 vector multiplets parametrize
$SO(6,6)\Big/\Big(SO(6)\ti
SO(6)\Big)$,
while the complex scalar in the gravitational multiplet corresponds
to the $T$ modulus (resp. $U$) in type IIA (resp. B) theories.

It is now straightforward to compute helicity supertraces
directly from Eq. (\ref{rr2})
and using (\ref{gres4}), (\ref{gres6}).
They read\footnote{The primed summation over $(h,g)$ stands
for $(h,g) \in
\{(0,1),
(1,0), (1,1)\} $.}
\bs
\label{b4all}
\be
B_4^{\phl} = 12   \sump  \Ga_{2,2}^w \ar{h}{g} \sim 12\, ,
\label{b43}
\ee
\be
B_6^{\phl} = 15  \sump
\left(2 + \Re H \ar{h}{g}\right) \Ga_{2,2}^w \ar{h}{g}
\sim 60\, ,
\ee
\es
where $H\ar{h}{g}$ are given in Eq. (\ref{hde}).
In Eq. (\ref{b4all}), we have indicated after the $\sim$ sign
the contributions of the massless states, obtained by using the fact
that
only the
$\Ga^w_{2,2} \ar{0}{1}$ block contains massless states, as well as  the
leading behaviour $H\ar{0}{1}= 2 + O (q)$.
As a check, we observe the correct values for the contributions of the
massless states,
\bs
\be
 B_4 \ve_{\rm massless} =
 1 \ti 3  + 6 \ti \frac{3}{2} = 12\, ,
\label{b4c}\ee
\be
 B_6\ve_{\rm massless} =
 1 \ti \frac{195}{4}   + 6 \ti \frac{15}{8} = 60\, ,
\label{b6c}\ee
\label{b46c} \es
where we used the elementary contributions (\ref{b4r}) and (\ref{b6r}).
Moreover, we observe that in contrast to the ordinary type IIA theory
on $K3\ti T^2$
(model (\ref{pf1})), the intermediate multiplets do contribute to
$B_6$.

Inserting the result (\ref{b43}) in Eq. (\ref{db4}) that we recall here
\be
\pa_\f \D_{\rm gr} (T,U) = {1 + \vep\s_\f \over 2}
\int_{\cal F} {\rd^2 \t\over\t_2}\pa_\f B_4\, ,
\label{db4p}
\ee
allows us to determine the
gravitational thresholds in terms of $B_4$.
Fundamental-domain integrals involving $\Ga_{2,2} \ar{h}{g}$
are computed in Appendix C, and yield, for the type IIA
case,
\be
\tbs (w) : \;\;
 \D_{\rm gr} (T) =-12 \log \left(T_2 \left|\th_i (T) \right|^4\right) +
{\rm const.}\, ,
\label{b404}
\ee
where  $i=2,3,4$, depending on the shift vector $w$ (see Appendix C).
An important consequence is that the  resulting corrections break
the $SL(2,Z)_T$ duality group to a $\Gamma(2)_T$ subgroup. The precise
subgroup depends on $i$ as indicated in Appendix C.

This model was argued in \cite{sv} to be $U$-dual to a ($4,0$)
supersymmetric
type II model, to which we now turn. This model is obtained as a
$Z_2$ orbifold of type II on $T^6$, where the $Z_2$ acts as
$(-1)^{F_{\rm L}}$
together with a translation on $T^6$.
Again, this model exhibits
spontaneously broken $N=8 \to 4$ supersymmetry and we will denote it
by $\tfs (w)$.
The resulting partition function reads:
\bea
\tfs (w) : \;\; Z&=&
{1 \over \t_2\,  \vert\eta \vert^{24}}
 {1\over 2}\sum_{a,b=0}^1 (-1)^{a+b+ab}
 \vartheta^4{a\atopwithdelims[]b}
{1\over 2}\sum_{\ba,\bb=0}^1 (-1)^{\ba+\bb+\mu \ba \bb}
 \thb^4{\ba\atopwithdelims[]\bb}
\cr
&& \ti \, {1\over 2}\sum_{h,g=0}^1
(-1)^{a g + b h  +  g  h}
\Gamma_{4,4}^{\phl}
\Ga_{2,2}^w \ar{h }{ g}\, .
\label{pf4}
\eea
To compute the massless spectrum, we first recall that for $N=8$ type
II,
obtained by compactifying on $T^6$, the spectrum is as follows:
NS--NS gives $G_{\m \n}$, $B_{\m \n}$, $\Phi$ and 12 vectors as ($6,6$)
and $6\ti 6 = 36$ scalars; R--R gives 16 vectors as $(0,16)$ and 32
scalars.
Because of the $(-1)^{F_{\rm L}}$ orbifold, the R--R sector is
projected out so we
are left with the NS--NS states only, which combine into the following
four-dimensional $N=4$ multiplets:
\be
\mbox{1 supergravity multiplet} \sp \mbox{6 vector multiplets}\, ,
\ee
in agreement with the massless spectrum (\ref{sp3}) of the dual theory.
The complex scalar in the gravitational multiplet now corresponds
to the axion--dilaton field.

For completeness, the helicity supertraces for this model can be
computed using (\ref{pv2}) and (\ref{pv4}):
\bs
\be
B_4^{\phl} =   {3 \over 4 }\, \Ga_{4,4}^{\phl}
\sump H_4^{\phl} \ar{h}{g}  \Ga^w_{2,2} \ar{h}{g}
\sim 12 \, ,
\ee
\be
B_6^{\phl} = \frac{15}{8}\, \Ga_{4,4}^{\phl}
\sump \left( H_4^{\phl} \ar{h}{g} + H_6^{\phl}
\ar{h}{g} \right)
\Ga^w_{2,2}
\ar{h}{g}
\sim 60\, ,
\ee
\label{b464} \es
where
\be
H_4 \ar{h}{g} = {\re}^{i\pi hg}{\th^4 \ar{1 - h}{1- g} \over {\et}^{12}
}
\sp
H_6 \ar{h}{g}  =
 \cases{
 \frac{1}{2}\frac{\th_3^8 - \th_4^8}{\et^{12}}  \sp (h,g) = (0,1)\cr
 \frac{1}{2}\frac{\th_2^8 - \th_3^8}{\et^{12}}  \sp (h,g) = (1,0) \cr
 \frac{1}{2}\frac{\th_4^8 - \th_2^8}{\et^{12}}  \sp (h,g) = (1,1)\, ,
\cr
}
\ee
from which we see that $B_4$  again receives contributions only  from
massless
and massive short BPS multiplets, while $B_6$ also gets contributions
from
intermediate ones.

However, for ($4,0$) supersymmetric models, a four-graviton
scattering calculation shows that
the one-loop corrections to $R^2$ terms do not involve helicity
supertraces. Instead, the one-loop corrections simply vanish, and
the only contributions to $R^2$ couplings, as argued in
the introduction, are non-perturbative. Now,
$U$-duality can  be invoked to obtain the non-perturbative
($4,0$) result from the one-loop result (\ref{b404}) of the ($2,2$)
dual,
by identifying the $T$-modulus of the ($2,2$) theory with the
$S$-modulus of the $($4,0$)$ theory. There is however an important
subtlety
involved in identifying the lattice shifts on both sides.
We recall that in  the full non-perturbative spectrum, states have not
only
electrical charges $m_i,n^i$ under the Kaluza--Klein gauge fields
of $T^2$, but also have magnetic charges $\tm_i,\tnn^i$.
Under $S \lra T$ interchange, electric and magnetic
charges are mapped to each other according to \cite{kk}
\be
(m_i,n^i,\tm_i,\tnn^i) \to
(m_i,\epsilon^{ij}\tm_j,-\epsilon_{ij}\tnn^j,\tnn^i)\, .
\ee
In particular, a $(-1)^{n^1}$ projection on states with even electric
winding
$n^1$ on the ($2,2$) side translates into a $(-1)^{\tm_2}$
projection on the ($4,0$) side, of no effect in perturbation theory.
A $(-1)^{m_1}$ projection on the other hand in the ($2,2$) theory
translates into a perturbative
$(-1)^{m_1}$ in the dual ($4,0$) theory. These two projections
have a geometrical interpretation of doubling one radius of $T^2$,
in contrast to the $(-1)^{n^1}$ one.
However, ($2,2$) perturbative modular invariance
requires at the same time half-integer $n^1$ charges in the twisted
sector. This implies also half-integer $\tm_2$ charges in the twisted
sector of the dual ($2,2$) theory, which should presumably be
accompanied
by a $(-1)^{\tnn^2}$ under some ``non-perturbative modular invariance''
requirement. This in turn would imply that the
correct projection on the ($2,2$)
side is $(-1)^{m_1+\tnn^2}$, which reduces to $(-1)^{m_1}$ in the
perturbative
spectrum. This ambiguity does not affect the perturbative evaluation of
thresholds. As for non-perturbative corrections, the relevant
instantons
are a subset of the original ones, which have been shown to
not contribute to $R^2$ couplings.  Restricting to a projection
on the electrical momenta only (cases I, II, III in Table C.1), we find
from Eq. (\ref{b404}) the result:
\be
\label{hfu}
\tfs (w_{\I, \II, \III}) : \;\;
\D_{\rm gr} (S) =-12 \log \left(S_2 \left|\th_4 (S) \right|^4\right) +
{\rm const.}
\ee
This exhibits the expected feature \cite{vw2} that the $S$-duality
symmetry is
broken
to a $\Ga(2)_S$ subgroup of $SL(2,Z)_S$, namely the subgroup
that leaves $\th_4 (S)$ invariant.
The two theories are  weakly coupled
in the regime $T_2,S_2 \to \infty$. The $T_2\to \infty$
decompactification
limit of shifted ($2,2$) lattice sums was investigated in
\cite{decoa}, with the result:
\be
\Ga_{2,2}^{w_{\I, \II, \III}} \hg\,
\limit{\longrightarrow}{T_2\to\infty}\,
{T_2 \over \t_2}\, \delta_{h,0}\, \delta_{g,0}
\ee
up to exponentially suppressed corrections.
This selects the untwisted unprojected sector of the two models
(\ref{rr2}), (\ref{pf4}), thereby restoring $N=8$ supersymmetry for
both
of them, in agreement with $U$-duality conjecture.
Expanding Eq.  (\ref{hfu}) in the weak ($4,0$) coupling limit, we find
\be
\label{hful}
\tfs (w_{\I, \II, \III}) : \;\;
\D_{\rm gr} (S) \limit{\longrightarrow}{S_2\to\infty}
-\! 12 \log S_2 + O\left({\re}^{-S_2}\right)\, .
\ee
The result exhibits the correct vanishing of perturbative
$O\left(S_2^n\right)$
corrections, together with the already encountered non-perturbative
logarithmic divergence.

Let us now turn to the strong-coupling behaviour of the ($4,0$)
ground state.
The $S_2 \to 0$ limit of ($4,0$) is
mapped under duality to the $T_2\to 0$ limit
of the ($2,2$) ground state, for which we can again use the results of
Ref. \cite{decoa}:
\be
\label{g220}
\Ga_{2,2}^{w_{\I,\II,\III}} \hg\,
\limit{\longrightarrow}{T_2\to 0}\,
{1\over \t_2 \, T_2} \; \forall h,g
\ee
up to exponentially suppressed corrections. The orbifold action does
not
affect the $T^2$ part any longer, thereby yielding the standard type II
on
$K3\ti T^2$ model of Section \ref{m1} at small radius.
This is strictly true only in the perturbative regime of type II,
because
of
the non-perturbative ambiguities mentioned before.
This is further mapped to the $\he{22}$ model at large coupling
$S_2 \to 0$ and large radius $T_2\to\infty$.
We therefore conclude that the ($4,0$) model
and the standard heterotic model on $T^6$ are equivalent in
the strong-coupling large-radius limit\footnote{One may ask whether the
two limits commute. The correct
prescription is to first take $T_2 \to \infty$ and then only $S_2\to 0$
in ($4,0$) variables, since we needed the ($2,2$) dual to be weakly
coupled before we could conclude anything about its small-radius
limit.}.
This can also be checked on the explicit $R^2$ coupling
\be
\label{hful2}
\tfs (w_{\I, \II, \III}) : \;\;
\D_{\rm gr} (S) \limit{\longrightarrow}{S_2\to 0}
-\! 12 \log S_2 + 12\p S_2 + O\left({\re}^{-S_2}\right)\, ,
\ee
which reproduces the correct heterotic on $T^4$ tree-level coupling
(\ref{hsd}). This set of relations is depicted
on the upper and rear faces of the cube in Fig. 1.

\fig{8cm}{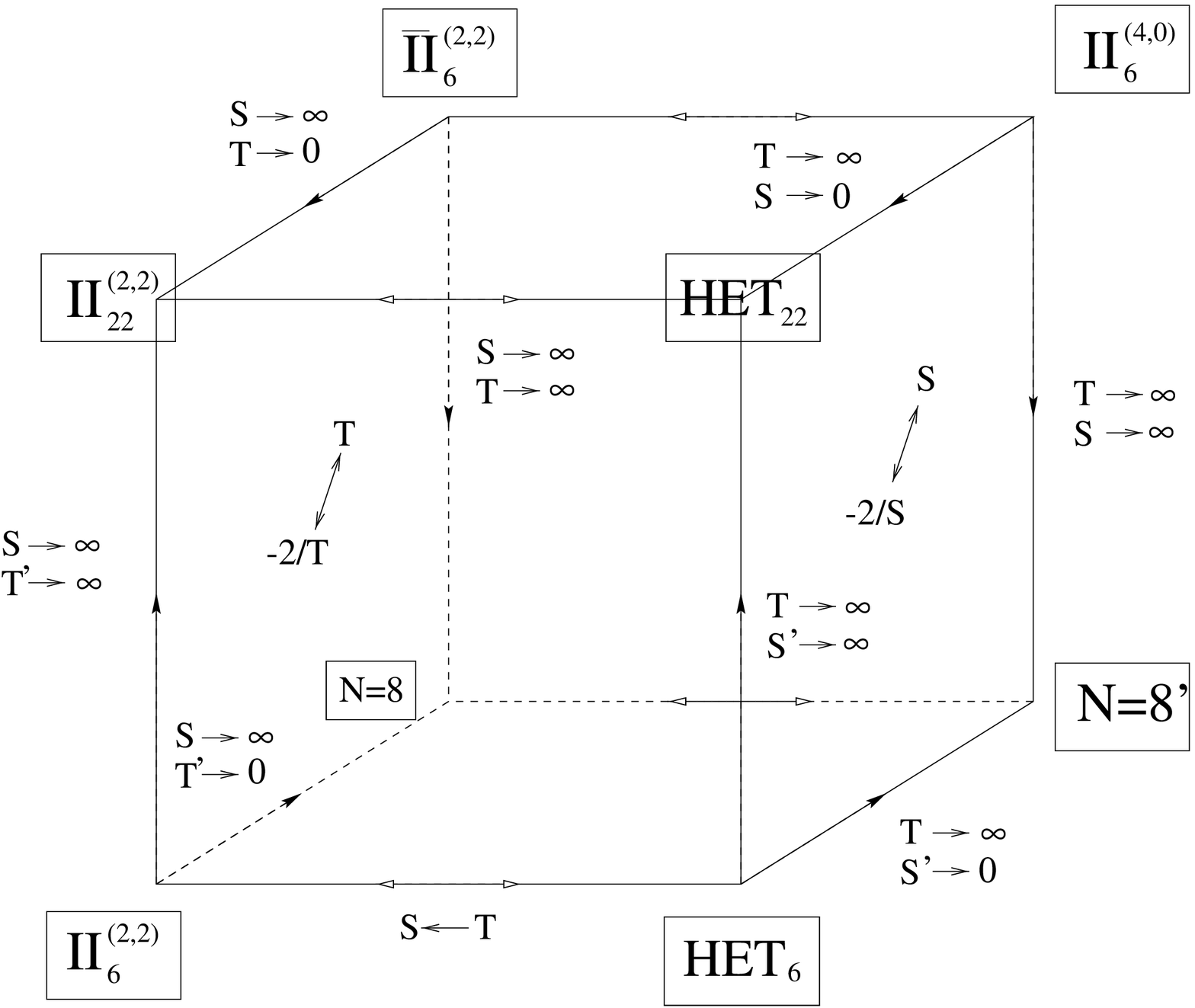}{The cube of duality, decompactification and
strong/weak
coupling relations}{f1}

\subsection{$\itn{6}$ free orbifold of type II on $K3$  and
its heterotic dual $\ihe{6} $ \label{m2}}

We now turn to another example of $N=4$ four-dimensional duality,
which this time descends from six-dimensional string--string duality
by a freely-acting orbifold, namely a half-lattice shift on $T^2$
together with a minus sign on the twisted sector of $K3$.
The adiabatic argument~\cite{vw1} guarantees
that the heterotic model obtained by translating this action in
heterotic string on $T^4$ is still dual to the type II orbifold.
To be explicit, the resulting partition function for this type II
model,
denoted by $\tn{6} (w) $,  is given by
\bea
\tn{6}(w) : \;\; Z &=&
{1 \over \t_2\,  \vert\eta \vert^{24}}
 {1\over 2}\sum_{a,b=0}^1 (-1)^{a+b+ab}
 \vartheta^2{a\atopwithdelims[]b}
{1\over 2}\sum_{\ba,\bb=0}^1 (-1)^{\ba+\bb+\mu \ba \bb}
 \thb^2{\ba\atopwithdelims[]\bb}\cr
&& \ti \, {1\over 2}\sum_{h,g=0}^1{1\over 2}\sum_{h',g'=0}^1
(-1)^{ h g' +  g  h'  }
\vartheta{a+h\atopwithdelims[]b+g}
\vartheta{a-h\atopwithdelims[]b-g}
\thb{\ba+h\atopwithdelims[]\bb+g}
\thb{\ba-h\atopwithdelims[]\bb-g}\cr
&& \ti \,
\Gamma_{4,4}^{\phl} \ar{h}{g}
\Ga_{2,2}^w \ar{h' }{ g'}\, .
\label{rr1}
\eea
Again, the shift vector $w$ has to  satisfy $w^2 = 0$ for
modular invariance.
The $(h,g)$ projections are associated with the $T^4/Z_2$ orbifold,
while
the
freely-acting transformations correspond to the $(h',g')$
projection.

The massless spectrum is most easily obtained from the results at the
beginning of
Section~\ref{m1}, by noting that the $(-1)^h$ orbifold projects out
the twisted states, so that we are left with the following untwisted
four-dimensional
$N=4$ multiplets:
\be
\mbox{1 supergravity multiplet} \sp \mbox{6 vector multiplets}\, .
\label{sp2} \ee
The relevant helicity supertraces are (we use again the results
(\ref{gres4})
and (\ref{gres6}))
\bs
\be
B_4^{\phl} = 6 \left( 3 \Ga_{2,2}^{\phl} -\sump
\Ga_{2,2}^w \ar{h}{g} \right) \sim 12\, ,
\label{b44} \ee
\be
B_6^{\phl} =  15 \left( 3 \Ga_{2,2}^{\ph{w}} - \sump
\left( 1 -\Re H \ar{h}{g} \right) \Ga_{2,2}^w \ar{h}{g} \right)
\sim 60 \, ,
\label{b46 b4}\ee
 \label{b4}
\es
where functions
$H\ar{h}{g}$
are given in Eq.  (\ref{hde}).
We deduce the type IIA gravitational thresholds:
\be
\tn{6} (w) : \;\;
 \D_{\rm gr} (T)
= -12 \log \left( T_2 {\left|\eta(T)\right|^6 \over
\left|\th_i(T)\right|^2} \right)
+{\rm const.}\, ,
\label{d1e}
\ee
where $i=2,3,4$, depending on the shift vector $w$ (see Table C.1).
As advocated in the previous section, we shall restrict our
discussion to shift vectors leading to $i=4$, for which
the resulting $T$-duality group is $\Ga^+(2)_T$.

We now want to discuss the heterotic dual for this model.
 From six-dimensional string duality, the $Z_2$ symmetry
    acting as $-1$ on all twisted states of $K3$ at the orbifold
point has to have an equivalent in the dual heterotic string
for the corresponding values of the $SO(6,22)$ heterotic moduli.
At present, there remains a puzzle as to what these values are
\cite{asp}.
Nevertheless, this symmetry can in principle be used to
construct a freely-acting orbifold of heterotic string on $T^4$,
and the adiabatic argument guarantees that the resulting model
will be dual to the present $\tn{6} (w)$ model.
Henceforth we shall refer to this model as $\he{6} (w)$.
The heterotic coupling is given by the area
of the type II torus,
which, owing to the free action, is $T/2$. We therefore deduce
the non-perturbative threshold for $\he{6}(w)$:
\bea
\he{6} (w_{\I,\II,\III}) : \;\;
\D_{\rm gr} (S)\!\!\!\!\!\!\!\!
 &&= -12 \log \left( 2S_2 {\left|\eta(2S)\right|^6 \over
\left|\th_4(2S)\right|^2} \right)
+{\rm const.}
\cr
&&\limit{\longrightarrow}{S_2\to \infty}
-\! 12 \log S_2 + 12\pi S_2 + O \left({\re}^{-S_2} \right)\, .
\label{d1eh6}
\eea
In particular, we observe that the tree-level contribution matches
the one of the $\he{22} $ model (\ref{hsd}), as it should, since the
tree-level effective action is universal for all heterotic ground
states. 
The cases  corresponding to
$i=2,3$ in the threshold (\ref{d1e}) are obtained by applying
$T$-duality on the type II side, yielding
$T \to -1/2S, T \to 2S-1$, respectively.

The (large-radius) weak-coupling limit of $\he{6}(w)$ is mapped to
the (weak-coupling) large-radius
limit of $\tn{6}(w)$, which by the same techniques as in the previous
section turns out to be the standard $\tn{22}$ model.
The latter being dual to the standard $\he{22}$, we conclude that
$\he{6}(w)$ and $\he{22}$ are the same in the (large-radius)
perturbative regime. The relation between the quartet of
theories that we have been discussing can be seen on the front side
of the cube in Fig. 1.

The (large-radius)
strong-coupling limit of $\he{6}(w)$ can be discussed in the same
way as for the $\tfs(w)$ model: it corresponds to the (weak-coupling)
small-radius limit of $\tn{6}(w)$, which from the partition
function (\ref{rr1}) and
from Eq.  (\ref{g220}) appears to restore $N=8$ supersymmetry. In fact,
$\tn{6}(w)$ and $\tbs(w)$ are identical under transformation
of the moduli, thanks to the relation (\ref{con})
\be
\frac{1}{2}\sum_{h',g'=0}^1
(-1)^{hg'+gh'}
\Ga_{2,2}^w \ar{ h'}{  g' }(T',U')=\Ga_{2,2}^w \ar{ h}{g}(T,U) \, .
\ee
The precise mapping $(T,U)\to(T',U')$ is shown in Table C.1.
for the various lattice shifts, $T \to -2/T$ for the cases
$\I, \II, \III$ at
hand, leading to $i=4$ in the above formula (\ref{d1e}).
The $N=8$ (weak-coupling)
large-radius limit of $\tn{6}(w)$ therefore
coincides with the $N=8$ (weak-coupling) small-radius limit of
$\tbs(w)$,
and is dual to the $N=8$ (large-radius) strong-coupling limit of
$\he{6}(w)$.
Furthermore, this implies that $\he{6}(w)$ and $\tfs(w)$ are mapped
to each other under $S \to -2/S$.

The various relations among the octet of theories that has been
discussed
in this and the previous section are summarized by the duality cube in
Fig. 1. In this figure, the horizontal connections correspond to $S
\leftrightarrow T$
duality and the  various connections on the sides of the cube are
limits.

\subsection{$\itn{14}$ free orbifold of type II on $K3$  and
its heterotic dual $\ihe{14}$ \label{m3}}

We now turn to another example of $N=4$ four-dimensional duality,
which this time descends from six-dimensional string--string duality.
We now wish to construct models with an intermediate gauge-group
rank.
To achieve that we need to project out part of the 16
twisted states of $T^4/Z_2$. This can be done by using a $Z_2$ subgroup
of the  $(D_4)^4$
discrete
symmetry
of the orbifold $T^4/Z_2$ \cite{verlinde}, generated by
\bs
\be
D: \; \; |+\rangle \leftrightarrow |-\rangle\sp
    |m,n\rangle \rightarrow (-1)^m |m,n\rangle
\ee
\be
\tilde{D}: \; \; |+\rangle \rightarrow -|+\rangle\sp
    |-\rangle \rightarrow  |-\rangle\sp
    |m,n\rangle \rightarrow (-1)^n |m,n\rangle
\ee
\label{Dsh}
\es
on each circle, where $|\pm\rangle$ denote the two twisted states
and $|m,n\rangle$ the untwisted momentum-winding states
corresponding to the chosen circle. The operation $D$ can be
interpreted as the remnant of a $Z_2$ translation on the original
circle, carrying one fixed point onto the other.

As a first step we will examine the possibility of projecting out one
half
of the twisted states and obtain an $SO(6,14)$ model. Starting from the
$T^4/Z_2 \ti T^2$ orbifold blocks,
${\Gamma_{4,4}\hg  \Gamma_{2,2}\over \vert \et \vert^{12}}$, we mod
out a
further $Z_2$, which acts as a shift on the two-torus, and as the
$D$-operation described above on the $T^4/Z_2$. The ($6,6$) conformal
blocks entering the partition function (\ref{gp})
now read:
\be
\tn{14} (w): \;\;
Z_{6,6} \ar{h}{g} = {1\over 2}\sum_{h',g'=0}^1
{\Gamma_{4,4}^{\ph{w}}\ar{h;h'}{g;g'}
\Gamma_{2,2}^{w}\ar{h'}{g'}\over \vert \et \vert^{12}}
\sp \forall h,g \, .
\label{orb614}
\ee
In this expression, $(h,g)$ refer to the original twist while
$(h',g')$ refer to the $D$-shift. According to the definition of the
latter
(see Eq.  (\ref{Dsh})), the ($4,4$) orbifold blocks possess the
following
properties:
for $(h,g)\neq (0,0)$,
$\Gamma_{4,4}\ar{h;0}{g;0}=
\Gamma_{4,4}\ar{h;h}{g;g}=
\Gamma_{4,4}\ar{h}{g}$ (ordinary twist); $\Gamma_{4,4}\ar{0;h}{0;g}$
is a ($4,4$) lattice sum with one shifted momentum (or winding if
$\tilde
D$
is used instead of $D$), analogous to the ($2,2$) constructions of
Appendix C;
finally, ($4,4$) orbifold blocks with $(h,g)\neq (0,0)$, $(h',g')\neq
(0,0)$ and $h\neq h'$ or $g\neq g'$ vanish because the trace is
performed over the original twisted states with the insertion of an
operator under which half of the states have eigenvalue $+1$ and the
others $-1$.

We can now proceed to the computation of the helicity supertraces. For
$(h,g)\neq (0,0)$ our orbifold blocks are of the form (\ref{fact66}).
We therefore use the results (\ref{gres4}) and (\ref{gres6}) and find:
\bs\be
B_4 = 6  \left( 3\Gamma_{2,2}^{\ph{w}} + \sump
\Gamma_{2,2}^w \hg
\right)\sim 24
\, ,
\ee
\be
B_6=15
\left( 3 \Ga_{2,2}^{\ph{w}} + \sump
\left(1+ {1\over 2} \Re H \ar{h}{g}\right)\Ga_{2,2}^w \ar{h}{g}\right)
\sim 75
\, .
\ee
\es
We note again that the infrared behaviours of $B_4$ and $B_6$ are in
agreement with the massless content of the model, namely 1
supergravity multiplet and 14 vector multiplets (of which 8 are
twisted).

The gravitational threshold corrections follow from Eq.  (\ref{db4p}),
\be
\tn{14} (w) : \;\;
\D_{\rm gr} (T) =-24 \log
\left(T_2 \left\vert\th_i (T) \right\vert
\left\vert\et(T)\right\vert^3
\right) + {\rm const.}\, ;
\label{thr614}
\ee
here the index $i$ depends on the choice of shift vector $w$
(see Table C.1).

As in Subsection \ref{m2}, this model is guaranteed to have a heterotic
dual
obtained by translating the $(D_4)^4$ action
on the heterotic side, at the corresponding
point in moduli space. This symmetry is likely to be non-perturbative
again.
However, in this case it is possible to construct a perturbative
heterotic dual
with the correct rank 14, which we will denote by $\he{14}(w)$.
We consider the decomposition of the
$\Gamma_{6,22}$ lattice according to
\be
\Gamma_{6,22}= \Gamma_{5,5}\oplus \Gamma_{1,1} \oplus \Gamma_{0,8}
\oplus \Gamma_{0,8}\, ,
\ee
where the last two terms give $E_{8} \times E_{8}$. The operation
that reduces the rank acts as an exchange of the two $\Gamma_{0,8}$
lattices
coupled with a translation in $\Gamma_{1,1}$, thereby reducing
$E_8 \ti E_8$ to its diagonal level-2 subgroup\footnote{A rank-14
heterotic model has also been constructed in
\cite{Chaudhuri}.}.

Again, the
heterotic non-perturbative threshold is obtained by exchanging $T$ with
$2S$ 
(for lattice shift corresponding to $i=4$ in Eq.\ (\ref{thr614})), 
and reads:
\be
\he{14}(w_{\I, \II, \III}): \; \;
\D_{\rm gr}  (S) =-24 \log
\left(2 S _2 \left\vert\th_4 (2S ) \right\vert
\left\vert\et(2 S )\right\vert^3
\right) + {\rm const.}
\ee
The above expression exhibits the correct tree-level heterotic
contribution and the breaking of $S$-duality by instanton effects.

\subsection{$\itn{10}$ free orbifold of type II on $K3$  and
its heterotic dual $\ihe{10}$ \label{m4}}

The method presented in the previous section can be slightly modified
so that the original twisted sector of the $T^4/Z_2$ is left with one
quarter of the states only. The model obtained in this way will have
rank 16 and
${SO(6,10) \Big/\Big( SO(6)\ti SO(10)\Big)}$ moduli space.

Starting from the orbifold blocks (\ref{orb614}) of the $SO(6,14)$
model, we perform an extra $Z_2$, which acts on the ($4,4$) part as a
$D$-operation along another circle (see Eq.  (\ref{Dsh})), while it
amounts
to a further shift on the
($2,2$) with respect to some momentum-winding direction. In other
words,
we perform a $Z_2 \ti Z_2$ on the original $T^4/Z_2 \times T^2$
construction. The result for the ($6,6$) blocks is
\be
\tn{10} (w) : \;\;
Z_{6,6}\ar{h}{g}= {1\over 2}\sum_{h_1,g_1=0}^1 {1\over
2}\sum_{h_2,g_2=0}^1
{\Gamma_{4,4}^{\ph{w}}\ar{h;h_1,h_2}{g;g_1,g_2}
\Gamma_{2,2}^{w_1,w_2}\ar{h_1,h_2}{g_1,g_2}\over \vert \et \vert^{12}}
\sp \forall  h,g
\, ,
\label{orb610}
\ee
where $\Gamma_{2,2}^{w_1,w_2}\ar{h_1,h_2}{g_1,g_2}$ are the $Z_2\ti
Z_2$
freely-acting constructions explained in Appendix C  and
$\Gamma_{4,4}\ar{h;h_1,h_2}{g;g_1,g_2}$
are orbifold blocks whose non-vanishing components are the following:
for $(h,g)\neq (0,0)$
$\Gamma_{4,4}\ar{h;0,0}{g;0,0}=\Gamma_{4,4}\ar{h;h,0}{g;g,0}
=\Gamma_{4,4}\ar{h;0,h}{g;0,g}=\Gamma_{4,4}\ar{h;h,h}{g;g,g}
=\Gamma_{4,4}\ar{h}{g}$ (ordinary twist);
$\Gamma_{4,4}\ar{0;h_1,h_2}{0;g_1,g_2}$, which is an ordinary ($4,4$)
shifted lattice sum corresponding to a freely-acting $Z_2 \ti Z_2$,
analogous to the ones studied in Appendix C for the ($2,2$) lattices.
The precise structure of the latter plays no role for the computation
of helicity supertraces, since only the $(h,g)\neq (0,0)$
blocks contribute to gravitational thresholds.
By using the results (\ref{lat22}), these blocks are
recast as:
\be
\tn{10} (w) : \;\;
Z_{6,6} \ar{h}{g}= {1\over 4}{\Gamma_{4,4}\ar{h}{g}\over \vert \et
\vert^8}
\left(
{\Gamma_{2,2}^{\ph{w}}\over \vert \et \vert^4}+
\sum_{w\in \{ w_1,w_2,w_{12} \}}
{\Gamma_{2,2}^w \ar{h}{g}\over \vert \et \vert^4}
\right)
\sp {\rm for} \ (h,g)\neq (0,0)\, .
\label{orb610t}
\ee
This expression, combined with Eqs.  (\ref{fact66}), (\ref{gres4}) and
(\ref{gres6}), therefore leads to the following helicity supertraces:
\be
B_4 =
3 \left( 3\Gamma_{2,2}^{\ph{w}} +
\sum_{w\in \{ w_1,w_2,w_{12} \}}
\sump \Gamma_{2,2}^w \hg
\right)\sim 18
\, ,
\ee
\be
B_6 = 15
\left( 3 \Ga_{2,2}^{\ph{w}} +
\sum_{w\in \{ w_1,w_2,w_{12} \}}
\sump
\left(1+ {1\over 2} \Re H \ar{h}{g}\right)\Ga_{2,2}^w \ar{h}{g}\right)
\sim {135\over 2}
\, .
\ee
The leading infrared behaviours reflect the presence of 10 vector
multiplets,
6 untwisted and 4 twisted, as expected by construction. The
gravitational thresholds are determined as usual:
\be
\tn{10} \left(w_{({\rm i}),({\rm ii}),({\rm iii})}\right) : \;\;
\D_{\rm gr} (T) =-18\log
\left(T_2 \left\vert\th_i (T) \right\vert^2
\left\vert\et(T)\right\vert^2
\right) + {\rm const.}\, ,
\label{thr610}
\ee
where $i=4,2,3$ respectively for the  $Z_2\ti Z_2$ shifted models
(\romannumeral1), (\romannumeral2) and (\romannumeral3) in Table
C.2. Models (\romannumeral4), (\romannumeral5) and (\romannumeral6)
lead, on the other hand, to the result:
\be
\tn{10} \left(w_{({\rm iv}),({\rm v}),({\rm vi})}\right)  : \;\;
 \D_{\rm gr} (T) =-18\log
\left(T_2 \,
\left\vert\et(T)\right\vert^4
\right) + {\rm const.}
\label{thr610p}
\ee
It is remarkable that this threshold is invariant under the full
$SL(2,Z)_T$ duality, but one should refrain from concluding that the
$SL(2,Z)_T$ symmetry is restored, since the breaking may appear
in quantities other than $R^2$ thresholds.

In order to construct the heterotic dual with
rank 10, we consider the $SO(8) \times SO(8)$ decomposition of
each $E_{8}$ and the decomposition of $\Gamma_{6,6}^{\phl}$ into
$\Gamma_{4,4}^{\phl} \oplus \Gamma_{1,1}^{(1)} \oplus
\Gamma_{1,1}^{(2)}$.
This lattice has an enhanced
$SO(8) \times SO(8)' \times  SO(8)'' \times SO(8)'''$
symmetry point\footnote{In four dimensions it is possible to build a
model with such
gauge group by
switching on appropriate discrete Wilson lines, which act by breaking
$E_{8} \times E_{8}$ and shifting the mass of the spinors, preventing
the
reconstruction of $E_{8}$.}, from which we can switch on two discrete
Wilson
lines, which act independently with exchange and shift as for rank 14.
We then perform two $Z_2 $ orbifolds, the first
one exchanging $SO(8) \times SO(8)'$ with $SO(8)'' \times SO(8)'''$
while shifting the $\Gamma_{1,1}^{(1)}$, and the
second one exchanging $SO(8) \times SO(8)''$ with $SO(8)'' \times
SO(8)'''$
while shifting the $\Gamma_{1,1}^{(2)}$. The remaining gauge symmetry
is
 $SO(8)$ at level 4.
Again, identifiying the precise heterotic dual would require knowing
the
point
in heterotic moduli space corresponding to the $K3$ orbifold point, a
piece of
information that is lacking at present \cite{asp}.

Finally, for a lattice shift corresponding to
$i=4$ in the threshold (\ref{thr610}), we find that the
duality maps $T$ to $4S$. The other cases $i=2,3$ are
obtained by applying $T$-duality on the type II side, yielding
$T \to -1/4S, T \to 4S-1$, respectively.
We therefore conclude that the exact gravitational threshold
in heterotic variables reads:
\be
\he{10}\left(w_{({\rm i}),({\rm ii}),({\rm iii})}\right):\; \;
\D_{\rm gr} (S) =-18\log
\left(4S_2 \left\vert\th_4 (4S) \right\vert^2
\vert\et(4S)\vert^2
\right) + {\rm const.}
\ee
These results indeed are in agreement with the fact that the heterotic
dilaton
should correspond to the volume form of the base of the $K3$-fibration,
which in
this case is $T^2/(Z_2 \times Z_2)$.

On the other hand, in the case of models (\romannumeral4),
(\romannumeral5), (\romannumeral6) leading to (\ref{thr610p}),
the
correct
tree-level term on the heterotic side is only obtained by
substituting $T \to 2S$, in apparent
contradiction
with the fact that we have a $Z_2 \times Z_2$ orbifold.
This is due to the particular
translation on $T^2$ used to obtain this models: one $Z_2$ acts as a
translation on the electric momenta, which are mapped under
type II--heterotic duality to the electric momenta on the heterotic
side \cite{kk}.
The second $Z_2$ acts instead on the electric windings, which are
mapped
to the magnetic momenta on the heterotic side, so it is not visible in
the
heterotic perturbative theory: from the heterotic point of view there
is
only one $Z_2$. The correct map is therefore $T \rightarrow 2S$,
and we obtain the threshold:
\be
\he{10} \left(w_{({\rm iv}),({\rm v}),({\rm vi})}\right)  : \;\;
\D_{\rm gr} (S) =-18\log
\left(2S_2 \,
\left\vert\et(2S)\right\vert^4
\right) + {\rm const.}
\ee

\section{Conclusions}

We have considered the threshold corrections to low-energy $R^2$ and
other four-derivative couplings in heterotic and type II ground states
with 16
unbroken supercharges. In particular, we have discussed the ordinary
$K3$ compactification and a family of type II vacua that have
spontaneously broken
$N=8 \to 4$ supersymmetry and~4~massive gravitinos in the
perturbative spectrum.
Those are special cases of more general models with spontaneously
broken supersymmetry studied in \cite{decoa, kk, decol}.

We have argued that there are no perturbative or non-perturbative
corrections to the $R^2$ couplings in heterotic ground states in
dimension higher than four.
In four dimensions, instanton corrections are expected from the
heterotic
Euclidean 5-brane, and they depend on the $S$ field only.
In type II ground states with ($2,2$) supersymmetry we have argued that
there are no non-perturbative corrections to the $R^2$ couplings in
four dimensions or more.
The full result arises from one loop.
We have first analysed this threshold in six dimensions, which provides
a guide
on what to expect in lower dimensions.
We have subsequently evaluated this one-loop threshold for several
($2,2$) four-dimensional models
with various numbers of massless vector multiplets. All such
ground states have heterotic duals, and the type II result translates
into 5-brane
instanton corrections from the heterotic point of view.
Most reduced-rank models have an Olive--Montonen duality group that is
a
subgroup
of $SL(2,Z)_S$, namely $\Gamma(2)_S$, which is reflected in the
behaviour of the non-perturbative corrections.

The above non-perturbative results should provide a guideline
towards the determination of the rules for calculating instanton
corrections in string theory. Several steps in this direction
were recently taken \cite{hm1,kp,bk2,apt,sup,ov,gg,gv,a,tr}.
The ultimate goal is to be able to handle non-perturbative effects with
less
supersymmetry or in its absence.

We have also analysed the CP-odd $R^2$ four-dimensional couplings, and
resolved an apparent puzzle: the type II result implies, via duality, a
CP-even coupling at one loop on the heterotic side between the
antisymmetric tensor and the gravitational Chern--Simons form.
We have shown that this is compatible with heterotic perturbation
theory since
such a coupling is invisible in on-shell amplitudes.

Finally we have considered type II dual pairs with 16 supercharges
and ($2,2$) or ($4,0$) supersymmetry.
The situation there is analogous to the type II--heterotic case.
By using some additional perturbative relationships, we find quartets
of dual
models, one of which is a heterotic ground state, with $N=4$ in four
dimensions,
and which, at strong coupling, exhibits enhanced $N=8$ supersymmetry!
The interpretation of this ground state is a spontaneously broken
$N=8 \to 4$ theory, with 4 solitonic massive gravitinos that become
massless at strong coupling, enhancing the supersymmetry to $N=8$.
We believe this possibility to be valuable for constructing
interesting models with less supersymmetry and an $N=8$ high-energy
behaviour.

\vskip 0.3cm
\centerline{\bf Acknowledgements}

\noindent
We would like to thank I. Antoniadis and R. Woodard for helpful
discussions.
N.A.O. acknowledges the Niels Bohr Institute for hospitality.
The work of C.K. was supported by the TMR contract ERB-4061-PL-95-0789,
and that of E.K. and  P.M.P. by the contract TMR-ERBFMRXCT96-0090.
\vskip 0.3cm

\setcounter{section}{0}
\setcounter{equation}{0}
\renewcommand{\theequation}{A.\arabic{equation}}
\section*{\normalsize{\centerline{\bf Appendix A: Kinematics and
on-shell field theory vertices}}}

\noindent
Throughout this paper, we use a $d$-dimensional metric
$g_{\m\n}$ with signature
$(+,-,-,\ldots)$. We evaluate the leading fourth order in momenta
scattering amplitudes of gravitational particles in six dimensions,
together with moduli in four dimensions. Particles are characterized
by their light-like momentum $p_i$ and (except for the moduli)
their  transverse (i.e. $p_i\e_i~=~0$)
polarization tensors $\e_i$. The latter are symmetric for gravitons
$h$, antisymmetric
for antisymmetric tensors $b$ and pure trace for dilatons $\Phi$. By
the
latter
we mean a polarization $\e_{\m\n} = (g_{\m\n}- p_\m k_\n - k_\m p_\n)$,
where $k$ is an auxiliary vector such that $k \cdot p=1$. We let
$\r_i=\pm 1$ according to whether $\e_i$ is symmetric ($h,\Phi$) or
antisymmetric ($b$). All amplitudes exhibit the gauge invariance
$\e_i \rightarrow \e_i +  p_i \otimes \zeta_i + \r_i \, \zeta_i \otimes
p_i$,
where $\zeta_i$ is the transverse (i.e. $p_i \cdot \zeta_i=0$ )
infinitesimal gauge transformation parameter in
momentum space (different for each particle).
These gauge symmetries correspond to general covariance for gravitons,
gauge invariance for antisymmetric tensors, and $k$-arbitrariness for
dilatons. Therefore, $k$ drops out of all amplitudes involving
dilatons,
and can safely be set to zero so long as one imposes the correct
$\Tr \e=2$ for the dilaton polarization tensor (as is obvious in
light-cone gauge).

Whenever possible, we omit Lorentz indices and implicitly
contract indices from left to right, for example
\bs
\label{levi2}
\be
p_1^{\phl} \e_2^{\phl} \e_1^{\phl} p_2^{\phl} \equiv
p_1^{\hphantom{1}\m}
\e^{\phl}_{2\m\n} \e_1^{\hphantom{1}\n\r} p^{\phl}_{2 \r} \, ,
\ee
\be
p_1^{\phl} \w p_2^{\phl} \w p_1^{\phl}\, \e_2^{\phl} \w p_2^{\phl}\,
\e_1^{\phl} \w \e_3^{\phl}\equiv
\e_{\l\m\n\r\s\t}^{\phl}
p_1^{\hphantom{1}\l}
p_2^{\hphantom{2}\m}
p^{\phl}_{1\k}
\e_2^{\hphantom{2}\k \n}
p^{\phl}_{2\x}
\e_1^{\hphantom{1}\x \r}
\e_3^{\hphantom{3}\s\t}\, ,
\ee
\es
where we define the CP-odd antisymmetric Levi--Civita tensor $\e$ such
that $\e_{0123}=+\sqrt{-g}$ and $\e_{012345}=+\sqrt{-g}$ in
four and six dimensions, respectively.
Our convention for $n$-forms is such that
$A= A_{\m\n\ldots\r} \rd X^{\m} \w \rd X^{\n} \w \ldots \w \rd X^{\r}$.
The exterior derivative acts as
$\rd A = \pa_\a A_{\m\ldots\r} \rd X^\a \w \rd X^{\m}\w  \ldots\w \rd
X^{\r}$.

First quantized string
perturbation theory forces us to restrict to on-shell amplitudes, and
we systematically impose, in the three-particle scattering case:
\be
p_i \cdot p_j = p_i \, \e_i = p_1 + p_2 + p_3 =0\, .
\ee
This drastically reduces the number of independent kinematic
structures,
for instance
\be
p_1 \e_2 p_3 = - p_1 \e_2 p_1 = - p_3 \e_2 p_3 =
p_3 \e_2 p_1 = \r_2\,  p_1 \e_2 p_3\, .
\ee
In several instances we reduce the product of two CP-odd Levi--Civita
tensors in CP-even terms using the Minkowskian identity
\be
\label{twoeps}
\e^{\a_1\a_2\ldots\a_d} \e^{\b_1\b_2\ldots\b_d}
=- \sum_{{\cal S}_d} \t(\s) \left( g^{\a_1 \b_{\s(1)}} g^{\a_2
\b_{\s(2)}}
\ldots g^{\a_d \b_{\s(d)}} \right)\, ,
\ee
where the sum runs over the $d!$ permutations $\s$ of $d$ elements
with signature $\t(\s)=\pm 1$.
When some indices are already contracted on the left-hand side,
one can significantly reduce the number of terms in the sum by using
the Minkowskian\footnote{The minus sign has to be omitted in Euclidean
space.}
identities:
\be
\eqalign{
\e^{\a_1\a_2\ldots\a_d} \e^{\b_1\b_2\ldots\b_d} &g_{\a_d\b_d}
=- 1! \sum_{{\cal S}_{d-1}} \t(\s) \left( g^{\a_1 \b_{\s(1)}} g^{\a_2
\b_{\s(2)}}
\ldots g^{\a_{d-1} \b_{\s(d-1)}} \right)
\cr
\e^{\a_1\a_2\ldots\a_d} \e^{\b_1\b_2\ldots\b_d}
g_{\a_{d-1}\b_{d-1}}&g_{\a_d\b_d}
=- 2! \sum_{{\cal S}_{d-2}} \t(\s) \left( g^{\a_1 \b_{\s(1)}} g^{\a_2
\b_{\s(2)}}
\ldots g^{\a_{d-2} \b_{\s(d-2)}} \right)
\cr
\vdots&\cr
\e^{\a_1\a_2\ldots\a_d} \e^{\b_1\b_2\ldots\b_d}
g_{\a_1\b_1}\ldots &g_{\a_d\b_d}
= - d!}
\ee
The four-derivative low-energy effective action is obtained by finding
Lorentz invariants that will induce the same interactions of the
massless
spectrum as those given by the string amplitude. Three-particle
interactions
have the simplification that no field-theory subtraction is required,
and the field-theory vertex has to match the precise string amplitude.
This is also the case in the four-particle interactions we are
considering.

The field theory vertices are obtained by expanding the Lorentz
invariant around flat backgrounds $g_{\m\n}=\eta_{\m\n}+ h_{\m\n},
B_{\m\n}=0+b_{\m\n}, \Phi=\Phi_0 + \d \Phi$, going to momentum space
variables
\be
h_{\m\n} (x)
= {\int { \rd^4 p \over (2\pi)^4 }} h_{\m\n}(p) {\re}^{i p x}\,
\ee
and imposing on-shell conditions.

To order $h^2$, the Riemann tensor with covariant indices becomes
\bea
R_{\a\b\c\d}^{\phl} &=& \Bigg( {1\over 2} h_{\a\d,\b\c}^{\phl}
  +{1\over 8} \bigg( h_{\a\d,\m}^{\phl}\,
h_{\b\c}^{\hphantom{\b\c},\m}
  +  \left( h^{\m}_{\hphantom{\c}\a,\d} + h^{\m}_{\hphantom{\b}\d,\a}-
2 h_{\a\d,\m}^{\phl} \right)
     \left( h^{\m}_{\hphantom{\c}\b,\c} + h^{\m}_{\hphantom{\b}\c,\b}
\right)
  \bigg)
  \Bigg) \cr
  &&- (\a\leftrightarrow\b) - (\c\leftrightarrow\d)
  + \Big((\a,\b)\leftrightarrow (\c,\d)\Big)
\label{lap}
\eea
so that
\bea
\!\!\!\!\!\!\!\!\int \rd^6 x \sqrt{-g} R_{\m\n\r\s}
R^{\m\n\r\s}&&\!\!\!\!\!\!
\stackrel{\rm on \ shell}{=} \int { \rd^6 p_1 \, \rd^6 p_2  \over
(2\pi)^6 }
  \d^{(6)}(p_1+p_2)
  \Big(p_1 h(p_2) p_1\Big) \Big(p_2 h(p_1) p_2\Big) \cr
+&& \!\!\!\!\!\!\!\!
  \int { \rd^6 p_1 \, \rd^6 p_2 \, \rd^6 p_3 \over (2\pi)^{12} }
  \d^{(6)}(p_1+p_2+p_3)
  \Big(p_1 h(p_2) p_3\Big) \Big(p_2 h(p_1) h(p_3) p_2\Big)\, .
\label{R2full}
\eea
The first term vanishes on shell because of momentum conservation,
but becomes relevant for two-graviton--one-modulus scattering
when the coefficient of $R^2$ is moduli-dependent
as in Eq.  (\ref{4Dvert}).
The second term induces a three-graviton amplitude:
\be
\sum_{\rm 6 \ perm} (p_1 \e_2 p_3) (p_2 \e_1 \e_3 p_2)\, ,
\ee
reproducing Eq.  (\ref{617}).
The same kind of manipulations yield the other vertices in
Eqs.~(\ref{R2vert})
and (\ref{4Dvert}).

\boldmath
\setcounter{section}{0}
\setcounter{equation}{0}
\renewcommand{\theequation}{B.\arabic{equation}}
\section*{\normalsize{\centerline{\bf Appendix B: Helicity supertraces
and
$\th$-function
identities}}}
\unboldmath

\noindent
Helicity supertraces are defined as
\be
B_{2n}\equiv \Str\l^{2n}\, ,
\label{helsup}
\ee
where $\l$ stands for the physical four-dimensional helicity.
In models with $N=4$ supersymmetry, $B_2$ vanishes (this is responsible
for the vanishing of the one-loop corrections
to two-derivative terms in the effective action), $B_4$ receives
contributions from short representations only, while
$B_6$ receives also contributions from
intermediate ones. This property can be proved by computing supertraces
for individual supermultiplets:
\bs
\be
B_4 ({\rm supergravity}) = 3
\sp
B_4 ({\rm vector}) = \frac{3}{2}\, ,
\label{b4a}
\ee
\be
B_4 \left(S^j\right) =\frac{3}{2}(2j+1)  (-1)^{2 j}
\sp
B_4 \left(I^j\right) = 0
\sp
B_4 \left(L^j\right) = 0\, ;
\label{b4b b4r}
\ee
\label{b4r}
\es
\bs
\be
B_6 ({\rm supergravity}) = { 195 \over 4}
\sp
B_6 ({\rm vector}) = \frac{15}{8}\, ,
\label{b6a}\ee
\be
B_6 \left(S^j\right) =\frac{15}{8}(2j+1)^3  (-1)^{2 j}
\sp
B_6 \left(I^j\right) =\frac{45}{4}(2j+1)  (-1)^{2 j +1 }
\sp
B_6 \left(L^j\right) = 0\, ,
\label{b6b b6r}
\ee
\label{b6r}\es
where $S^j, I^j , L^j$ are the short, intermediate and long
representations, respectively.

In the framework of string theory, the physical four-dimensional
helicity is $\l=\l_{\rm L}+\l_{\rm R}$,
where $\l_{\rm L,R}$ are the contributions to the helicity from the
left- (right-) movers.
We introduce the helicity-generating function as
\be
Z(v,\bar v)=\Tr'
q^{L_0-{c\over 24}}\,
\bar q^{\bar L_0 - {\bar c \over 24}}\,
{\re}^{2\pi i\left(v\l_{\rm L}-\bar v \l_{\rm R}\right)}
\, ,
\label{1}\ee
where the prime over the trace excludes the zero-modes related to the
space-time coordinates (consequently
$Z(v,\bar v)\vert_{v=\bar v=0}= \t_2 Z$).
At the perturbative level,
helicity supertraces are obtained by taking appropriate
derivatives of (\ref{1}), using
\be
\l_{\rm L} = {1 \over 2 \p i}\pa_v \sp
 \l_{\rm R} = -{1 \over 2 \p i}\pa_{\bv}\, .
\ee

In this paper we are mostly interested in $N=4$ type II
four-dimensional models of the $Z_2$-orbifold type, for which the
partition function (\ref{gp}) results into
a helicity-generating function\footnote{We use the short-hand notation
$\th \ar{a}{b} (v)$
for $\th \ar{a}{b} (v\vert \t)$.}
\begin{eqnarray}
Z(v,\bv)&=&
{1 \over \vert\eta \vert^{12}}
 {1\over 2}\sum_{a,b=0}^1 (-1)^{a+b+ab}
 \vartheta{a\atopwithdelims[]b}(v)\,
 \vartheta{a\atopwithdelims[]b}
{1\over 2}\sum_{\ba,\bb=0}^1 (-1)^{\ba+\bb+\mu \ba \bb}
\thb{\ba\atopwithdelims[]\bb}(\bv)\,
\thb{\ba\atopwithdelims[]\bb}
\cr
&&\ti \,  {1\over 2}\sum_{h,g=0}^1
\vartheta{a+h\atopwithdelims[]b+g}
\vartheta{a-h\atopwithdelims[]b-g}
\thb{\ba+h\atopwithdelims[]\bb+g}
\thb{\ba-h\atopwithdelims[]\bb-g}
 Z_{6,6} {h\atopwithdelims[]g}
\xi(v)\,
\bar \xi(\bv)\, ,
\label{hgp}
\end{eqnarray}
where
\be
\xi(v)=\prod_{n=1}^{\infty}{(1-q^n)^2\over \left(1-q^n {\re}^{2\pi
iv}\right)\left(1-q^n {\re}^{-2\pi iv}\right)}=
{\sin\pi v\over \pi}{\th_1'(0)\over
\th_1(v)}
\label{bhf} \ee
counts the helicity contributions of the space-time bosonic
oscillators.

Owing to the ($2,2$) supersymmetry of our models\footnote{In situations
where $N=4$ supersymmetry is realized as ($4,0$) (see e.g. model
$\II^{(4,0)}_6$ with vacuum amplitude given in (\ref{pf4})), formulas
(\ref{gb46}) get modified as follows:
$$
B_4^{\ph{2}}  = \left\langle \l_{\rm L}^4 \right\rangle
= {1\over 16 \,\p^4} \left.\pa_v^4 \, Z(v,\bv)
\right\ve_{v = \bv =0}\, ,
$$
$$
B_6^{\ph{2}}  =  \left\langle\l_{\rm L}^6  \right\rangle  + 15
\left\langle \l_{\rm L}^4 \, \l_{\rm R}^2\right\rangle
=-{1\over 64 \, \p^6}
\left.\left(\pa_v^6 + 15 \pa_v^4 \, \pa_{\bv}^2 \right) Z(v,\bv)
\right\ve_{v =\bv =0 }\, .
$$}, the first non-trivial
helicity supertraces can be computed by using the following
formulas:
\bs
\label{gb46}
\be
B_4^{\ph{2}}  = \left\langle \left(\l_{\rm L}^{\ph{2}} +
\l_{\rm R}^{\ph{2}}\right)^4 \right\rangle
= 6 \left\langle \l_{\rm L}^2  \l_{\rm R}^2  \right\rangle
={  6\over 16 \p^4} \left.\pa_v^2 \pa_{\bv}^2  Z(v,\bv)
\right\ve_{v = \bv =0}\, ,
\label{gb41} \ee
\be
B_6^{\ph{2}}  = \left\langle
\left(\l_{\rm L}^{\ph{2}} + \l_{\rm R}^{\ph{2}}\right)^6 \right\rangle
= 15   \left\langle \l_{\rm L}^4  \l_{\rm R}^2  + \l_{\rm R}^4  \l_{\rm
L}^2  \right\rangle
=-{15 \over 64  \p^6} \left.\left(
\pa_v^4  \pa_{\bv}^2 + \pa_v^2  \pa_{\bv}^4
\right) Z(v,\bv)
\right\ve_{v =\bv =0 }\, .
\label{gb61}
\ee
\label{gb}
\es
In the rest of this appendix, we collect some of the identities
involving
$\th$-functions, which are useful for these
computations.

Our conventions for the $\th$-functions
are\be
\th \ar{a}{b} (v)=
 \sum_{p \in Z}  {\re}^{\pi i \t \left(p+{a \over 2}\right)^2 +
2\p i \left(v + {b \over 2} \right)\left(p + {a \over 2} \right) }
\ee
so that
\be
\th_1= \th \ar{1}{1} \sp
\th_2= \th \ar{1}{0} \sp
\th_3= \th \ar{0}{0} \sp
\th_4= \th \ar{0}{1}\, .
\ee
We also recall that
\be
\pa_v^2 \th \ss= 4 \pi i \pa_\t^{\ph{2}} \th \ss
\label{thd}
\ee
and
\be
\th_1'(0) = - 2 \p  \et^3 = - \p  \th_2  \th_3  \th_4\, .
\label{thp}
\ee

A very useful identity is the Riemann identity:
\begin{eqnarray}
\frac{1}{2}\sum_{a,b=0}^1 (-1)^{a + b + \mu ab}&&\!\!\!\!\!\!\!
 \th \ss (v)\,
 \th \ss(0)\,
 \th \ar{a +h}{ b +g }(0)\,
\th \ar{a -h}{ b-g}(0)\cr
=&&\!\!\!\!\!\!\!
 \th\ar{1 }{ 1} \left({v\over 2}\right)\,
 \th\ar{1 }{ 1} \left({v\over 2}\right)\,
 \th \ar{1 +h }{ 1+g } \left({v\over 2}\right)\,
\th \ar {1 -h}{ 1-g} \left({v\over 2}\right) \, .
\label{RiemI}
\end{eqnarray}
Taking the second derivative of Eq. (\ref{RiemI})
with respect to $v$ at $v=0$ and using (\ref{thd}) and~(\ref{thp})
leads to
\be
 \sum_{a,b=0}^1
(-1)^{a+b+\mu ab} \th^2 \ss
\th \ar{a +h}{ b +g }
\th \ar{a -h}{ b -g }
i  \pa_\t \log { \th \ss \over
\et }
= \p  \et^6 \th \ar{1 +h}{ 1 +g }
\th \ar{1 -h}{ 1 -g }\, .
\label{613}
\ee

Finally, we present  some properties involving the bosonic helicity
factor $\x (v)\equiv \x(-v)$ defined in (\ref{bhf}):
\be
\x (0) = 1 \sp
\x ' (0) = 0
\sp
\x ''(0) = { \p^2 \over 3} \left(E_2 -1 \right)
\sp E_2 = {12 \over  \p i} \pa_\t \log \et\, ,
\ee
as well as the following relations:
\be
\left.\pa_v^2 \left( \th \ss (v) \, \x (v) \right)  \right\ve_{v=0}
= \left(
 4 \p i \pa_\t \log {\th \ss \over\et}
-
\frac{\p^2}{3} \right) \th \ss\, ,
\label{pv2}
\ee
\begin{eqnarray}
&&\left.\pa_v^4
\left(\th^2 \ar{1 }{ 1 }  \left({v \over 2}\right)\,
\th \ar{1 +h }{ 1 +g }\left({v \over 2}\right) \,
\th \ar{1 -h}{ 1-g}\left({v \over 2}\right) \,
\x (v) \right) \right\ve_{v=0}\cr
&&= 2 \p^2 \et^6  \left( 12  \p i  \pa_\t
\log {\th \ar{1-h}{1-g}\over \et} - 2 \p^2\right)
\th \ar{1 +h }{ 1 +g }
\th \ar{1 -h}{ 1-g} \, .
\label{pv4}
\end{eqnarray}

We will now focus on a specific class of ($2,2$) models that appear in
the
text,
which share the following property: the corresponding orbifold blocks
are of the form
\be
Z_{6,6}\hg=
{\Gamma_{4,4}\hg \over \vert \et \vert^8}
Z_{2,2}\hg
\sp {\rm for} \;\; (h,g)\neq(0,0) \, .
\label{fact66}
\ee
Here $\Gamma_{4,4}\hg$ are the ordinary $Z_2$-twisted ($4,4$)
lattice sums (see (\ref{z44})), whereas
$Z_{2,2}\hg$
are generic blocks. For these models, the above identities
(\ref{pv2}) and (\ref{pv4}) can be used together with the definitions
(\ref{gb})
and the helicity-generating function (\ref{hgp}) to obtain finally:
\be
B_4=12\vert \et \vert^4 \sump
Z_{2,2}\hg
\label{gres4}
\ee
and
\be
B_6=30\vert \et \vert^4 \sump
\left(
1+ {1\over 2} \Re H\hg
\right)Z_{2,2}\hg\, ,
\label{gres6}
\ee
where the functions $H\hg$
are given by
\be
H \ar{h}{g} =  {12 \over  \p i} \pa_\t \log {\th \ar{1-h}{1-g} \over
\et}
= \cases{\hphantom{-} \th_3^4 + \th_4^4  \sp (h,g)=(0,1)\cr
          -  \th_2^4 - \th_3^4  \sp (h,g)=(1,0)\cr
\hphantom{-} \th_2^4 - \th_4^4  \sp (h,g)=(1,1)\, .\cr }
\label{hde}
\ee
Notice also the property
\be
\label{hdeid}
\sump H\hg =0\, .
\ee

\setcounter{section}{0}
\setcounter{equation}{0}
\renewcommand{\theequation}{C.\arabic{equation}}
\section*{\normalsize{\centerline{\bf Appendix C: $\Ga_{2,2}$
lattice sums and fundamental-domain integrals}}}

\noindent
In this appendix we give our notation and conventions for the usual
($2,2$) and shifted ($2,2$) lattice sums used in the text. We also
give the explicit results for the relevant fundamental-domain integrals
of these lattice sums.

The ($2,2$) lattice sum is given by
\begin{eqnarray}
\Gamma_{2,2} (T,U)  &=&
\sum_{\{p_{\rm L},p_{\rm R}\} \in \Ga_{2,2}}
q^{{1\over 2}p_{\rm L}^2} \,\bq^{{1\over 2}p_{\rm R}^2}\cr
&=& \frac{T_2}{\t_2} \sum_{A \in GL(2,Z)}
\exp \left( - 2\pi i  T \det A -{\p T_2 \over \t_2 U_2}
\left\vert
\left(\matrix{1 & U}\right)
A
\left( \matrix{\t \cr 1} \right)
\right\vert^2 \right)\, ,
\label{a2a}
\end{eqnarray}
where
\be
p_{\rm L}^2={\left\vert U m_1-m_2+T n^1+T U n^2\right\vert^2\over 2 T_2
U_2}
\sp
p_{\rm L}^2 - p_{\rm R}^2=  2\vec m \vec n
\label{g22}
\ee
($\vec m \vec n$ stands for $m_I n^I$). In terms of the background
fields $G_{IJ}$ and
$B_{IJ}$, the left and right momenta can be written as
\be
p_{\rm L}^I = {1\over \sqrt{2}}(N+n)_{\ph{L}}^I \sp
p_{\rm R}^I = {1\over \sqrt{2}}(N-n)_{\ph{L}}^I\, ,
\ee
where
\be
N^I = G^{IJ} \left(m_J - B_{JK} n^K \right)\, ,
\ee
so that
\be
p_{\rm L}^2 = p_{\rm L}^I  G^{\ph{L}}_{IJ}  p_{\rm L}^J
\sp
p_{\rm R}^2 = p_{\rm R}^I  G^{\ph{L}}_{IJ}  p_{\rm R}^J\, .
\label{mde}\ee
The matrix identities (\ref{mid}), (\ref{sde}) follow,
after
some algebra,
using the parametrization of $G_{IJ}$ and $B_{IJ}$ in terms of the
moduli $T$ and $U$\footnote{When $T_1=U_1=0$, the usual parametrization
is
$T_2=R_1R_2$,   $U_2=R_2/R_1$, where $R_i$ are the radii of
compactification.}:
\be
G = \frac{T_2}{U_2} \left( \matrix{ 1 & U_1 \cr
U_1 & \vert U\vert^2 \cr} \right)
\sp
B = T_1 \left( \matrix{ 0 & -1 \cr
1 & 0 \cr} \right)\, .
\ee
The relation (\ref{did}) follows  from the
definition of the lattice sum and the identity,
\be
{\pa p_{\rm L}^2 \over \pa E_{IJ}} = {\pa p_{\rm R}^2 \over \pa E_{IJ}}
=
{1\over 2}(N+n)_{\ph{L}}^I (N-n)_{\ph{L}}^J  = p_{\rm L}^I p_{\rm R}^J
\sp E
\equiv G +B\, ,
\label{latid}
\ee
which may be derived from (\ref{mde}).
Finally, the relevant fundamental-domain integral is \cite{dkl}
\be
\ifd \left( \Ga_{2,2} (T,U) - 1 \right)
= -\log \left(T_2 \left| \et (T) \right|^4  U_2
\left| \et (U) \right|^4 \right) -
 \log  { 8 \p   {\rm e}^{1 -\gamma}  \over 3 \sqrt{3} }\, .
\label{dkli} \ee
The subtraction of the massless-states contribution in this integral
is necessary for regularizing the logarithmic divergence,
and results in a non-harmonic dependence on $T,U$.

The $Z_2$-shifted ($2,2$) lattice sums are
\be
\Gamma_{2,2}^w (T,U) \hg  =
\sum_{\{p_{\rm L},p_{\rm R}\} \in \Ga_{2,2}+w{h\over 2}}
{\re}^{-\p i g \ell \cdot w }
q^{{1\over 2}p_{\rm L}^2} \bq^{{1\over 2}p_{\rm R}^2}\, ,
\label{b2}
\ee
where the shifts $h$ and projections $g$ take the values 0 or
1. Here, $w$ denotes the shift vector with components
$\left(a_1,a_2,b^1,b^2\right)$, and
$\ell \equiv \left(m_1,m_2,n^1,n^2\right)$. We have also introduced
the inner product\footnote{For
$w_1=\left({\vec a}_1,{\vec b}_1\right)$ and
$w_2=\left({\vec a}_2,{\vec b}_2\right)$,
the inner product is defined as
$w_1\cdot w_2= {\vec a}_1 {\vec b}_2+{\vec a}_2 {\vec b}_1$.}
\be
\ell \cdot w = \vec m \vec b + \vec a \vec n
\sp
w^2 = 2  \vec a \vec b \, ,
\label{inpr}
\ee
so that $a_I$ generates a winding shift in the $I$ direction, whereas
$b^I$ shifts the $I$th momentum. The vector $\ell$ is associated with
the $\Ga_{2,2}$ lattice and therefore the vector associated with
the shifted lattice will be
\be
p\equiv \ell+w{h\over 2}\, .
\label{b3}
\ee
With these conventions, left and right momenta read:
\bs
\be
p_{\rm L}^2=
{\left\vert
U \left( m_{1}+a_1 {h \over 2}\right)-
\left( m_{2}+a_2 {h \over 2}\right)+
T \left( n^{1}+b^1 {h \over 2}\right)+
T U \left( n^{2}+b^2 {h \over 2}\right)
\right\vert^2\over 2 T_2 U_2}\, ,
\ee
\be
p_{\rm L}^2 - p_{\rm R}^2= 2
\left( m_{I}+a_I  {h \over 2}\right)
\left( n^{I}+b^I  {h \over 2}\right)\, .
\ee
\es

It is easy to check the periodicity properties ($h,g$ integers)
\be
Z_{2,2}^{w} \ar{h}{g}=
Z_{2,2}^{w} \ar{h+2}{g}=
Z_{2,2}^{w} \ar{h}{g+2}=
Z_{2,2}^{w} \ar{-h}{-g}
\label{b9}
\ee
as well as the modular transformations that expression
\be
Z_{2,2}^{w} \ar{h}{g}=
{\Gamma_{2,2}^{w} \ar{h}{g}\over \vert \et \vert^4}
\ee
obeys:
\bs
\be
\tau\to\tau+1 : \; \; Z_{2,2}^{w} \ar{h}{g}
\to
{\re}^{\p i{w^2\over 2}{h^2 \over 2}} Z_{2,2}^{w}
\ar{h}{h+g}
\label{b8a}
\ee
\be
\tau\to-{1\over \tau} :\; \; Z_{2,2}^{w} \ar{h}{g} \to
{\re}^{-\pi i {w^2\over 2}{hg}} Z_{2,2}^{w} \ar{g}{-h} \, .
\label{b8b b8}
\ee
\label{b8}
\es
The relevant parameter for these transformations is
\be
\l \equiv {w^2\over 2}=\vec a \vec b \, .
\label{b12}
\ee
{}From expressions (\ref{b2})  we learn that the
integers $a_I$ and $b^I$ are defined modulo 2, in the sense that adding
2 to any one of them amounts at most to a change of sign in
$Z_{2,2}^{w}{1\atopwithdelims[]1}$. Such a
modification is necessarily compensated by an appropriate one in
the rest of the partition function in order to ensure modular
invariance; we are thus left with the same model. On the other
hand, adding 2 to $a_I$ or $b^I$ translates into adding a multiple of 2
to $\l$. Therefore,
although $\l$ can be any integer, only $\l=0$ and $\l=1$ correspond to
truly different situations.

We now
would like to discuss the issue of target-space duality in these
models, where the $Z_2$ orbifold acts as
a translation in one complex plane. The moduli dependence of the
two-torus shifted sectors
(see Eq.  (\ref{b2})) reduces in general  the
duality group
to some
subgroup\footnote{The subgroups
of $SL(2,Z)$ that will actually appear in the sequel are
$\Gamma^{\pm}(2)$ and $\Gamma(2)$. If
$\left(\matrix{a&b \cr c&d}\right)$
represents an element of the modular group,
$\Gamma^+(2)$ is defined by $a,d$ odd and $b$ even, while for
$\Gamma^-(2)$ we have $a,d$ odd and $c$ even. Their intersection is
$\Gamma(2)$.}
of
$SL(2,Z)_T^{\vphantom U} \times SL(2,Z)^{\vphantom T}_U \times Z_2^{T
\leftrightarrow U}$. Transformations that do not belong to this
subgroup map a model $w$ to some other model $w '$
leaving invariant, however, $\l={w^2 \over 2}={w'^2 \over 2}$. This
means
in particular that for a given model,
decompactification limits that are related
by transformations that do not belong to the actual duality group are
no longer equivalent.

To be more specific, by using expression (\ref{b2}), we can
determine the transformation properties of
$\Gamma_{2,2}^{w}{h\atopwithdelims[]g}$ under the full group
$SL(2,Z)_T^{\vphantom U} \times SL(2,Z)^{\vphantom T}_U \times Z_2^{T
\leftrightarrow U}$:
\bs
\be
SL(2,Z)_T :\ \
\left(\matrix{a_1 \cr a_2 \cr b^1 \cr b^2 \cr}\right)
\to
\left(
\matrix{
d&{\hphantom{-}}0&0&b\cr 0&{\hphantom{-}}d&-b{\hphantom{-}}&0\cr
0&-c&a&0\cr c&{\hphantom{-}}0&0&a\cr}
\right)
\left(
\matrix{a_1 \cr a_2 \cr  b^1 \cr b^2\cr}
\right)\ ,\ \ ad-bc=1\, ,
\label{b13a}\ee
\be
SL(2,Z)_U :\ \
\left(\matrix{a_1 \cr a_2 \cr b^1 \cr b^2 \cr}\right)
\to
\left(
\matrix{
{\hphantom{-}}a'&-c'{\hphantom{-}}&0&0\cr -b'&d'&0&0\cr
{\hphantom{-}}0&0&d'&b'\cr {\hphantom{-}}0&0&c'&a'\cr}
\right)
\left(
\matrix{a_1 \cr a_2 \cr  b^1 \cr b^2\cr}
\right)\ ,\ \ a'd'-b'c'=1
\label{b13b}\ee
and
\be
Z_2^{T \leftrightarrow U} :\ \
\left(
\matrix{a_1 \cr a_2 \cr b^1 \cr b^2 \cr}
\right)
\to
\left(
\matrix{0&0&1&0\cr 0&1&0&0\cr 1&0&0&0\cr 0&0&0&1\cr}
\right)
\left(\matrix{a_1 \cr a_2 \cr b^1 \cr b^2 \cr}\right)\, .
\label{b14}
\ee
\es

\noindent
Thus, we can determine the duality group for a given model by demanding
that the components of the vectors $\vec a$ and $\vec b$ remain
invariant modulo 2. For example, in the $\l = 0$ situation defined
by
$\vec a=(0,0)$ and $\vec b=(1,0)$, the target-space duality group turns
out to be
$\Gamma ^+ (2)_T \times \Gamma ^- (2)_U$, whereas for the case with $\l
= 1$ and $\vec a=(1,0)$, $\vec b=(1,0)$, we find
$\Gamma  (2)_T \times \Gamma  (2)_U \times Z_2^{T \leftrightarrow U}$.

At this point, we would like to mention a remarkable identity,
which plays a role in the
computation of fundamental-domain integrals, as well as in the
identification of several type~II
constructions (see Subsection \ref{m2}).
Starting from the definition (\ref{b2}) of shifted lattice sums, one
checks easily that
\be
\frac{1}{2}\sum_{h',g'=0}^1
(-1)^{hg'+gh'}
\Ga_{2,2}^w \ar{ h'}{  g' }(T',U')=\Ga_{2,2}^w \ar{ h}{g}(T,U) \;
\label{con}
\ee
for any shift vector such that $w^2=0$.
The precise relation between $(T',U')$ and $(T,U)$ depends on
the specific shift vector $w$, and is presented in the Table C.1 for
all distinct $\l=0$ situations.

\begin{center}
\begin{tabular}{| c || c | c || c |  c || c | c | }
\hline
{\rm case}&$\vec{a}$ & $\vec{b}$ & $T'$
                           & $U'$       & $i$ & $j$         \\ \hline
I   &$(0,0) $  & $(1,0)$   & $-{2\over T}   $
                           & $-{1\over 2U}  $ & 4 & 2    \\ \hline
II  &$(0,0) $  & $(0,1)$   & $-{2\over T}   $
                           & $-{2\over U}   $  & 4 & 4  \\ \hline
III &$(0,0) $  & $(1,1)$   & $-{2\over T}   $
                           & ${1+U\over 1-U}$  & 4 & 3   \\ \hline
IV  &$(1,0) $  & $(0,0)$   & $-{1\over 2T}  $
                           & $-{2\over U}   $  & 2 & 4 \\ \hline
V   &$(0,1) $  & $(0,0)$   & $-{1\over 2T}  $
                           & $-{1\over 2U}  $  &2 &  2  \\ \hline
VI  &$(1,1) $  & $(0,0)$   & $-{1\over 2T}  $
                           & ${1+U\over 1-U}$  & 2 & 3 \\ \hline
VII &$(1,0) $  & $(0,1)$   & ${1+T\over 1-T}$
                           & $-{2\over U}   $  & 3 & 4 \\ \hline
VIII&$(0,1) $  & $(1,0)$   & ${1+T\over 1-T}$
                           & $-{1\over 2U}  $  & 3 & 2 \\ \hline
IX  &$(1,-1)$  & $(1,1)$   & ${1+T\over 1-T}$
                           & ${1+U\over 1-U}$  & 3 & 3 \\ \hline
\end{tabular}
\end{center}
\centerline{Table C.1: The nine physically distinct models with
$\l=0$.}

After Poisson resummation in $m_1,m_2$, the shifted lattice sum
(\ref{b2})
takes the alternative form
\bea
\Gamma_{2,2}^w (T,U) \hg
= \frac{T_2}{\t_2}\sum_{A}
&&\!\!\!\!\!\!\!\! \exp -\pi i \left({w^2\over 4}{hg}
-\vec a\left( g\vec n-h\vec m \right) \right)\cr
&&\!\!\!\!\!\!\!\! \exp\left(- 2\pi i  T \det A -{\p T_2 \over \t_2
U_2}
\left\vert
\left(\matrix{1 & U}\right)
A
\left( \matrix{\t \cr 1} \right)
\right\vert^2\right) \, ,
\label{b6}
\eea
where the summation is performed over the set of matrices of the form
\be
A = \pmatrix{n^{1}+b^1 {h \over 2}&
m_{1}+b^1 {g \over 2}\cr
n^{2}+b^2 {h \over 2}&
m_{2}+b^2 {g \over 2}\cr}\, .
\label{b7}
\ee
Modular-invariant combinations of blocks
$\Gamma_{2,2}^w (T,U) \hg  $
can be integrated over the fundamental domain by decomposing
the set of matrices $A$ with respect to
orbits of the modular group.
In this paper, we are mainly interested in the case
$\l=0$\footnote{Heterotic
constructions with $\l = 1 $ can be found in \cite{decol}.}, for which
the relevant integrals can be obtained from (\ref{dkli})
by using (\ref{con}) together with
\be
\th_2(\t)= 2 {\eta^2(2\t) \over \eta(\t)}
\sp
\th_4(\t)= {\eta^2\left({\t\over 2}\right) \over \eta(\t)}
\sp
\th_3(\t)=
{2 {\re}^{-{i \pi \over 3}} \over 1-\t}
{\eta^2\left({1+ \t \over 1-\t}\right) \over \eta(\t)}\, .
\ee
As a result,
\be
\ifd\left(\sump\Gamma^{w}_{2,2}\hg (T,U) -1\right)=
-\log
\left(
T_2\left|\vartheta_i(T) \right|^4
U_2\left|\vartheta_j(U) \right|^4
\right)
- \log {\pi  {\re}^{1-\gamma}\over 6\sqrt{3}}\, ,
\label{b15}
\ee
where the relation between the shift vector $w=(\vec{a},\vec{b})$  and
the pairs  $(i,j)$ is taken from Table~C.1.

In the construction of reduced-rank models of Section \ref{rrm}, we
introduce
shifted ($2,2$) lattices where the free action is of the type $Z_2\ti
Z_2$. Each of the $Z_2$'s acts according to the above analysis on a
given set of momenta and windings. Consistency of the $Z_2\ti Z_2$
action demands that the intersection of these two sets be empty. In
other words, the corresponding shift vectors
$w_1$ and $w_2$ must satisfy $w_1 \cdot w_2=0$. Notice that the union
of these sets corresponds to the action of the diagonal $Z_2$. The
lattice sum will be denoted
$
\Gamma_{2,2}^{w_1,w_2}\ar{h_1,h_2}{g_1,g_2}
$
and we have in particular
\be
\Gamma_{2,2}^{w_1,w_2}\ar{h,0}{g,0}=
\Gamma_{2,2}^{w_1}\ar{h}{g}
\sp
\Gamma_{2,2}^{w_1,w_2}\ar{0,h}{0,g}=
\Gamma_{2,2}^{w_2}\ar{h}{g}
\sp
\Gamma_{2,2}^{w_1,w_2}\ar{h,h}{g,g}=
\Gamma_{2,2}^{w_{12}}\ar{h}{g}
\, ,
\label{lat22}
\ee
where $w_{12}\equiv w_1 + w_2$ reflects the action of the diagonal
$Z_2$.

As an example, consider the situation where $w_1=(0,0,1,0)$,
$w_2=(0,0,0,1)$ and therefore $w_{12}=(0,0,1,1)$. In that case, the
first
(resp. second) $Z_2$ shifts the momenta of the first (resp. second)
plane
(insertion of $(-1)^{m_1}$ (resp. $(-1)^{m_2}$)), while the diagonal
$Z_2$
amounts to inserting $(-1)^{m_1+m_2}$. The lattice sum now reads:
\begin{eqnarray}
\Gamma_{2,2}{h_1,h_2\atopwithdelims[]g_1,g_2}&=&
\sum_{\vec m \vec n \in Z}
(-1)^{m_1  g_1+m_2  g_2}
\exp
\Bigg(
2\pi i\bar\t\left(
m_{1}\left( n^{1}+{h_1 \over 2}\right)+m_{2}\left( n^{2}+{h_2 \over
2}\right)
\right)\cr &&-\,
{\pi \t_2 \over T_2 U_2}
\left|T \left( n^{1}+{h_1 \over 2}\right) + TU\left( n^{2}+{h_2 \over
2}\right)+Um_1-m_2\right|^2
\Bigg)\, ,
\label{lat22ex}
\end{eqnarray}
from which
Eqs.  (\ref{lat22}) are immediatly checked.

In the framework of Subsection \ref{m4}, the requirement of modular
invariance
implies that $w_1^2=w_2^2=w_{12}^2=0$. This reduces the number of
distinct possibilities to the six listed in Table C.2.
The first of these corresponds to the example whose lattice sum is
given in Eq.~(\ref{lat22ex}).

\begin{center}
\begin{tabular}{|c||c|c|}
\hline
{\rm case}       & $w_1$         & $w_2$         \\ \hline
(\romannumeral1) & $(0,0,1,0) $  & $(0,0,0,1) $  \\ \hline
(\romannumeral2) & $(1,0,0,0) $  & $(0,1,0,0) $  \\ \hline
(\romannumeral3) & $(1,0,0,1) $  & $(0,-1,1,0)$  \\ \hline
(\romannumeral4) & $(1,0,0,0) $  & $(0,0,0,1) $  \\ \hline
(\romannumeral5) & $(0,0,1,0) $  & $(0,1,0,0) $  \\ \hline
(\romannumeral6) & $(0,0,1,1) $  & $(1,-1,0,0)$  \\ \hline

\end{tabular}
\end{center}
\centerline{Table C.2: The six physically distinct models with
$w_i^{\vphantom{1}} \cdot w_j^{\vphantom{1}} =0 \; \forall i,j = 1,2$.}

\setcounter{section}{0}
\setcounter{equation}{0}
\renewcommand{\theequation}{D.\arabic{equation}}
\renewcommand{\thesection}{D}
\section*{\normalsize{\centerline{\bf Appendix D: Details of string
amplitude calculations}}}

\noindent
In this section we compute in great detail the stringy scattering
amplitude (\ref{64}) of three gravitons (or two-forms) for type II
superstring on $K3$, and
subsequently the scattering amplitude (\ref{tpa}) of two
gravitons (or two-forms or dilatons)
with moduli of $T^2$ for type II on~$K3\ti T^2$.

\subsection*{\normalsize{\bf D.1 String amplitude toolbox}}

For these computations we use the following
contraction formulae:
\bs
\be
\label{101}
\left\langle X^\m (\bz,z) X^\n (0) \right\rangle
= g^{\m\n} \Delta(\bz,z)
\equiv - g^{\m\n} \log \left( {\re}^{-2\p {z_2^2 \over \t_2} }
                        \left| {\th_1(z)\over \th'_1(0)}\right|^2
\right)
\ee
\be
\label{1010}
\left\langle \bpa X^\m (\bz,z) \pa X^\n (0) \right\rangle
= -g^{\m\n} \bpa\pa \Delta(\bz,z)
= -{\p \over \t_2} g^{\m\n}
\ee
\bea
\left\langle
\pa_1^{\phl} X^\m_{\phl} (\bz_1,z_1)\,  p_2^{\phl} \cdot X( \bz_2,z_2 )
\right\rangle
\left\langle
\pa_2^{\phl} X^\l_{\phl} (\bz_2,z_2)\,  p_1^{\phl} \cdot X( \bz_1,z_1 )
\right\rangle
= \cr
=- p_2^{\hphantom{1}\m} p_1^{\hphantom{1}\l} \left\langle
 \pa_1^{\phl} X (\bz_1,z_1) X (\bz_2,z_2) \right\rangle^2
\label{105}
\eea
\be
\label{106}
\left\langle \bpa X^I_{\phl}(\bz,z) \pa X^J_{\phl}(0) \right\rangle =
p_{\rm R}^I p_{\rm L}^J
  + G^{IJ}_{\phl} {\pi \over \t_2}  - G^{IJ}_{\phl}
\pa\bpa\Delta(\bz,z) \equiv
p_{\rm R}^I p_{\rm L}^J
\ee
\be
\label{108}
\left\langle \ps(z_1)^\m \ps(z_2)^\n \right\rangle \ar{a}{b}
= g^{\m\n} \langle \ps(z_1) \ps(z_2) \rangle \ar{a}{b}
= g^{\m\n} {\th \ss(z_{12}) \th_1'(0) \over \th \ss(0) \th_1(z_{12}) }
\ee
\be
\label{109}
\left\langle  p_1^{\phl} \cdot \ps (z_1) \, \ps^\l(z_2)_{\phl}
\right\rangle
\left\langle \ps^\m_{\phl} (z_1)\,   p_2^{\phl} \cdot \ps (z_2)
\right\rangle
= p_1^{\hphantom{1}\l} p_2^{\hphantom{1}\m} \langle \ps(z_1) \ps(z_2)
\rangle^2\, ,
\label{610b}
\ee
\es
where the Greek space-time indices run from 0 to 5 (resp. 3) in the
six- (resp. four-) dimensional case, and the indices $I,J$ run on the
two compactified
directions of $T^2$. As in Appendix C, $p_{\rm L,R}$ denote the left-
and right-moving momenta
of $T^2$.

A few remarks about these equations are in order. Equation (\ref{101})
gives the propagator of a non-compact boson on the space of
non-zero-modes.
Equation (\ref{1010}) omits a delta function singularity, which has to
be subtracted for tree-level factorization. The first term in
(\ref{106}) is
the contribution of the winding zero-modes of a compact boson written
in Hamiltonian representation, to be added
to the non-zero-mode contribution (\ref{1010}).
Equation (\ref{108}) holds only for even spin structures where the
world-sheet fermions do not have any zero-modes. In the odd spin
structure,
there is one zero-mode for each space-time or $T^2$ fermion, a total of
six
in both the six- and four-dimensional cases. These zero-modes have to
be saturated in
order
to give a non-vanishing result, and we normalize them as
\bs
\be
\left\langle
\ps^\m  \ps^\n   \ps^\k  \ps^\l  \ps^\r \ps^\s
\right\rangle
= \e^{\m \n \k \l \r \s}
\; \;  \mbox{in six dimensions,}
\label{610c}
\ee
\be
\left\langle
\ps^\m  \ps^\n   \ps^\k  \ps^\l
\ps^I \ps^J \right\rangle
= \e^{\m \n \k \l} \e^{IJ}
\; \;  \mbox{in four dimensions,}
\ee
\es
where $\e^{12} = -\e^{21} = 1/\sqrt{G}$.
Saturation of the zero-modes at the same time induces the replacement
\be
\th^2 \ar{1}{1}  \ra \left( \th' \ar{1}{1}(0) \right)^2 = 4\pi^2\eta^6
\label{611}
\ee
in the partition function.

We also need the integrated propagators on the torus:
\bs
\be
\int {\rd^2 z \over \t_2  }
 \Big( \pa \Delta (\bz,z) \Big) ^2
=- 4 \p i \pa_\t \log \left(\et (\t) \t_2^{1/2}\right) \, ,
\ee
\be
\int {\rd^2 z \over \t_2  }  \langle \ps (z) \ps (0) \rangle^2 \ss
=  4 \p i \pa_\t \log \left (\th \ss (\t) \t_2^{1/2} \right)\, ,
\label{612b id1} \ee
\label{id1}
\es
where $ (a,b)$ is an even spin structure. We normalize the measure
of integration on vertex positions as
$\int {\rd^2 z \over \t_2  }=1$ .
Expressions analogous to (\ref{610b},c) and (\ref{612b id1}) for the
left
side follow by complex conjugation. Useful Riemann identities for
the summing of spin structures are assembled in Appendix B.

\subsection*{\normalsize{\bf D.2 Three-graviton scattering in six
dimensions}}

Here we wish to evaluate the amplitude (\ref{64}) and derive the
corresponding
four-derivative terms in the effective action. We need to distinguish
according to the spin structures on both sides.

\ni {\sl A.  CP-even $\bar{e}{\rm-}e$}.
In this sector we need to compute the
correlation function
\be
\eqalign{
A^{\bar{e}{\rm-}e}=
\Big\langle &
\left(\bpa_1 X^\m (\bz_1,z_1) + i p_1 \cdot \bps (\bz_1) \bps^\m
(\bz_1) \right)
\Big(\pa_1 X^\n (\bz_1,z_1) + i p_1 \cdot \ps (z_1) \ps^\n (z_1)
\Big)
{\rm e}^{i p_1 \cdot X (\bz_1,z_1) }
\cr
\ti &
\left(\bpa_2 X^\k (\bz_2,z_2) + i p_2 \cdot \bps (\bz_2) \bps^\k
(\bz_2) \right)
\Big(\pa_2 X^\l (\bz_2,z_2) + i p_2 \cdot \ps (z_2) \ps^\l (z_2)
\Big)
{\rm e}^{i p_2 \cdot X (\bz_2,z_2) }
\cr
\ti &
\left(\bpa_3 X^\r (\bz_3,z_3) + i p_3 \cdot \bps (\bz_3) \bps^\r
(\bz_3) \right)
\Big(\pa_3 X^\s (\bz_3,z_3) + i p_3 \cdot \ps (z_3) \ps^\s (z_3)
\Big)
{\rm e}^{i p_3 \cdot X (\bz_3,z_3) }
\Big\rangle\, .
\cr
} \label{bee}
\ee
The Riemann identity (\ref{RiemI}) shows that at least two pairs of
fermions
must
be contracted together on both sides, since contributions with less
fermionic
contractions vanish after a sum on even spin structures.
Each fermion pair comes with one
power of momentum; we therefore need precisely two such
contractions\footnote{It is known
that those singularities arising when two vertices come together
can yield poles ${O}(1/(p_i p_j))$ that can cancel against
six-derivative terms to yield ${O}(p^4)$ terms \cite{minahan}.
We evaluated these contributions and found a precise cancellation
of the corresponding terms, in agreement with the expectation that
these terms reproduce field-theory subtractions that are
absent in the case at hand.}.
The two pairs of fermions have to be chosen in two different vertices
on both sides, since the polarizations are traceless:
\bea
A^{\bar{e}{\rm-}e}_{\rm four \ deriv}&=& (i)^4
\left\langle \bpa_1 X^\m \pa_3 X^\s  \right\rangle
\left\langle  p_2 \cdot \bps (\bz_2) \bps^\r(\bz_3) \right\rangle
\left\langle  p_3 \cdot \bps (\bz_3) \bps^\k(\bz_2) \right\rangle
\cr
&&\ti \,
\left\langle  p_1 \cdot \ps (z_1) \ps^\l(z_2)   \right\rangle
\left\langle  p_2 \cdot \ps (z_2) \bps^\n(z_1)  \right\rangle
+ \mbox{perm.}
\eea
Making use of Eq.
(\ref{610b}) and of the Riemann identities
in Appendix B, it can be shown that the integrand, after summation over
spin structures, no longer depends on the position of
the vertices, so that we can apply Eq.  (\ref{id1}) to obtain:
\be
{\cal I}^{\bar{e}{\rm-}e} =
\T^{\bar{e}{\rm-}e}  \int_{\cal F} {\rd^2 \t \over \t_2^2} {\t_2^3
\over
\p^3} Z_{\II}
 {\p \over \t_2}
 (4 \pi i) \pa_{\t} \log \left({\th \ss\over \et }\right)
 (-4 \pi i) \pa_{\bar{\t}} \log \left({\thb \bss \over
\bar{\et}}\right)\, .
\ee
Here $Z_{\II}$ stands for the unintegrated partition function in
Eq.  (\ref{61}).
We also defined the kinematic structure
\be
\T^{\bar{e}{\rm-}e}
= (p_1 \e_2 p_3) (p_2 \e_1 \e_3 p_2) + \mbox{5 perm.}
\label{617}
\ee
It can easily be shown that
\be
\T^{\bar{e}{\rm-}e}   = \r_1 \r_2 \r_3 \T^{\bar{e}{\rm-}e}\, ,
\ee
so that the amplitude is non-vanishing only for three gravitons or for
two
antisymmetric tensors and one graviton.
Identity (\ref{613}) shows that the untwisted sector
$(h,g)=(0,0)$ does not contribute, and we use the explicit
expression (\ref{z44}) for the twisted $\Ga_{4,4}\ar{h}{g}$ to
obtain:
\be
{\cal I}^{\bar{e}{\rm-}e}_{\rm reg} =
\T^{\bar{e}{\rm-}e}   \int_{\cal F} {\rd^2 \t \over \t_2^2}
{\t_2^2  \over \p^2}
\frac{1}{8} {1 \over \t_2^2} \ti 4 \pi^2 \ti  4 \pi^2
 \ti 16  \ti 3
= 32 \p^3 \T^{\bar{e}{\rm-}e}\, ,
\ee
where we also used the standard modular-invariant integral
$ \int_{\cal F} \rd^2 \t /\t_2^2 = \p/3$.

Comparing Eq.  (\ref{617}) with Eq.  (\ref{R2vert}) we find that the
three-graviton--
and one-graviton--two-two-forms in ${\bar{e}{\rm-}e}$ spin structure
can be
described by the following vertex in the effective action:
\be
{\cal I}_{\rm eff}^{\bar{e}{\rm-}e}
= 32 \p^3  \int \rd^6 x \sqrt{-g} \left( R^2 + {1\over 6 } \na H \na H
\right)\, .
\label{621}
\ee

\ni{\sl B. CP-even $\bar{o}{\rm-}o$.}
In this sector the correlation function we have to compute is modified
to
\bea
A^{\bar{o}{\rm-}o}\!\!\!\!&=&\!\!\!\!
\Big\langle
\left(\bpa_1 X^\m (\bz_1,z_1) + i p_1 \cdot \bps (\bz_1) \bps^\m
(\bz_1) \right)
\Big(\pa_1 X^\n (\bz_1,z_1) + i p_1 \cdot \ps (z_1) \ps^\n (z_1)\Big)
{\rm e}^{i p_1 \cdot X (\bz_1,z_1)}
\cr
&&\ti \,
\left(\bpa_2 X^\k (\bz_2,z_2) + i p_2 \cdot \bps (\bz_2) \bps^\k
(\bz_2) \right)
\Big(\pa_2 X^\l (\bz_2,z_2) + i p_2 \cdot \ps (z_2) \ps^\l (z_2)\Big)
{\rm e}^{i p_2 \cdot X (\bz_2,z_2)}
\cr
&&\ti \,
 \bps^\r  (\bz_3)
 \ps^\s (z_3)
{\rm e}^{i p_3 \cdot X (\bz_3,z_3) }
 \pa X^\a (0) \ps_\a (0)
 \bpa X^\b (0) \bps_\b (0)
\Big\rangle\, ,
\eea
where we have used for the third vertex operator the $-1$-picture on
both
the left and the right sides, and inserted the left- and right-moving
supercurrents. All fermions have to be contracted in order to saturate
zero-modes, and the remaining $\bpa X^\a \pa X^\b$ from the two
supercurrents does not yield any singular contribution because of the
antisymmetry of Levi--Civita tensors.

We are therefore left with the following term
\bea
A^{\bar{o}{\rm-}o}_{\rm four \ deriv}&=& (i)^4
p_{1\a_1} p_{2\a_2}  \e^{\a_1 \m \a_2 \k \r \a}
p_{1\b_1} p_{2\b_2}  \e^{\b_1 \n \b_2 \l \s \b}
{\p \over \t_2} g_{\a \b}
\eea
obtained by using Eq.
(\ref{610c}). Then using Eq.  (\ref{611}) on both
the left and the right sides, the integrated three-point amplitude
becomes,
after some algebra:
\be
{\cal I}^{\bar{o}{\rm-}o}
= \T^{\bar{o}{\rm-}o} \int_{\cal F} {\rd^2 \t \over \t_2^2} {\t_2^3
\over \p^3}
\frac{1}{8} {1 \over \t_2^2} \ti (-4 \pi^2) \ti
(-1)^\m  4 \pi^2 \ti
{ \pi \over  \t_2} \ti 16   \ti 3
= - \vep  32 \p^3  \T^{\bar{o}{\rm-}o}\, ,
\ee
where we have defined the tensor structure
\be
\T^{\bar{o}{\rm-}o} = \e_{1\m \n} \e_{2\k \l} \e_{3\r \s}
p_{1\a_1} p_{2\a_2}  \e^{\a_1 \a_2 \m \k \r \a}
p_{1\b_1} p_{2\b_2}  \e^{\b_1 \b_2 \n \l \s \b}
g_{\a \b}\, .
\label{ots} \ee
Expanding the product of the two CP-odd Levi--Civita tensors
in terms of the metric in Eq.~(\ref{levi2}) and comparing Eqs.
(\ref{617}) and
(\ref{ots}),
we can show that, without any assumption on the symmetry properties
of the polarization tensors,
\be
\T^{\bar{o}{\rm-}o} = - (p_1 \e_2 p_3) (p_2 \e_1 \e_3 p_2) + \mbox{5
perm.}
\ee
Therefore, the ${\bar{o}{\rm-}o}$ spin structure yields exactly
the same interactions as the $\bar{e}{\rm-}e$ one, but with a sign
depending on
whether we are in type IIA or IIB.

Hence, we record for the corresponding term in the effective action
\be
{\cal I}_{\rm eff}^{\bar{o}{\rm-}o}
 =  32 \p^3 \vep \int \rd^6 x \sqrt{-g} \left(  R^2 + {1\over 6}  \na H
\na H \right)\, .
\label{628} \ee

\ni {\sl C. CP-odd.}
We first work out the result in the sector $\bar{e}{\rm-}o$, in which
we need to compute the correlator
\bea
A^{\bar{e}{\rm-}o}\!\!\!\!&=&\!\!\!\!
\Big\langle
\left(\bpa_1 X^\m (\bz_1,z_1) + i p_1 \cdot \bps (\bz_1) \bps^\m
(\bz_1) \right)
\Big(\pa_1 X^\n (\bz_1,z_1) + i p_1 \cdot \ps (z_1) \ps^\n (z_1)\Big)
{\rm e}^{i p_1 \cdot X (\bz_1,z_1) }
\cr
&&\ti \,
\left(\bpa_2 X^\k (\bz_2,z_2) + i p_2 \cdot \bps (\bz_2) \bps^\k
(\bz_2) \right)
\Big(\pa_2 X^\l (\bz_2,z_2) + i p_2 \cdot \ps (z_2) \ps^\l (z_2)\Big)
{\rm e}^{i p_2 \cdot X (\bz_2,z_2) }
\cr
&&\ti \,
\left(\bpa_3 X^\r (\bz_3,z_3) + i p_3 \cdot \bps (\bz_3) \bps^\r
(\bz_3)
\right)\ps^\s (z_3)
{\rm e}^{i p_3 \cdot X (\bz_3,z_3) }
 \pa X^\g (0) \ps_\g (0)
\Big\rangle\, ,
\eea
where we have used the $-1$-picture on the right for the third vertex
operator and inserted the right-moving supercurrent (\ref{68}). Again,
no contact terms are involved and
the relevant four-derivative term is
\bea
A^{\bar{e}{\rm-}o}_{\rm four \ deriv} &=& \!\!\! (i)^4
\bigg(
\left\langle  p_1 \cdot \bps (\bz_1) \bps^\k(\bz_2) \right\rangle
\left\langle \bps^\m (\bz_1)  p_2 \cdot \bps (\bz_2) \right\rangle
\left\langle \bpa X^\r \pa X^\g \right\rangle
\cr
& + &  \!\!\!
\left\langle  p_1 \cdot \bps (\bz_1) \bps^\r(\bz_3) \right\rangle
\left\langle \bps^\m (\bz_1)  p_3 \cdot \bps (\bz_3)  \right\rangle
\left\langle \bpa X^\k \pa X^\g  \right\rangle
\cr
& + & \!\!\!
\left\langle  p_2 \cdot \bps (\bz_2) \bps^\r(\bz_3) \right\rangle
\left\langle \bps^\k (\bz_2)  p_3 \cdot \bps (\bz_3)  \right\rangle
\left\langle \bpa X^\m \pa X^\g  \right\rangle
\bigg)  p_{1\a} p_{1\b} \e^{\a \n \b \l \s \d} g_{\g\d}
\cr
& = &  \!\!\!
\frac{\p}{\t_2}
\left(\e^{\a \b\n  \l \s \r }_{\phl}
 p_1^{\hphantom{1}\k} p_2^{\hphantom{1}\m}
+ \e^{\a \b\n  \l \s \k }_{\phl}
p_1^{\hphantom{1}\r} p_3^{\hphantom{1}\m}
+ \e^{\a \b \n  \l \s \m }_{\phl}
p_2^{\hphantom{1}\r} p_3^{\hphantom{1}\k} \right)
p_{1\a}^{\phl} p_{2\b}^{\phl}
\cr
&\ti & \!\!\!
\left\langle \bps (\bz) \bps (0) \right\rangle^2\, ,
\label{630b}
\eea
where in the second step we used Eq.
(\ref{610b}) and
the fact that each of the three fermion correlators
will contribute the same amount thanks to translation invariance.

Using the result (\ref{630b}) in (\ref{64}) along with the
partition function (\ref{61}), the integrated correlator (\ref{612b
id1})
and the replacement (\ref{611}) on the right side, we obtain
for the integrated three-point function in this sector
\be
{\cal I}^{\bar{e}{\rm-}o}
= \T^{\bar{e}{\rm-}o} \int_{\cal F} {\rd^2 \t \over \t_2^2} {\t_2^3
\over \p^3}
\frac{1}{8} {1 \over \t_2^2} \ti (-4 \pi^2) \ti 4 \pi^2 \ti
{ \pi \over  \t_2} \ti 16   \ti 3
=  -32 \p^3  \T^{\bar{e}{\rm-}o}\, ,
\ee
where we have defined the tensor structure
\bea
\T^{\bar{e}{\rm-}o} & = & \e_{1\m \n}^{\phl} \e_{2\k \l}^{\phl}
 \e_{3\r \s}^{\phl}
\left( \e^{\a\b \n \l \s \r }_{\phl}
p_1^{\hphantom{1}\k} p_2^{\hphantom{1}\m}
+ \e^{\a \b \n \l \s \k }_{\phl}
p_1^{\hphantom{1}\r} p_3^{\hphantom{1}\m}
+ \e^{\a \b \n \l \s \m }_{\phl}
p_2^{\hphantom{1}\r} p_3^{\hphantom{1}\k}\right)
p_{1\a}
p_{2\b} \cr
& = & {1\over 2}\,
p_1 \w p_2 \w p_1 \, \e_2 \w  p_2 \, \e_1 \w  \e_3 + \mbox{perm.}
\label{567}
\eea
Following the same steps, and using Eq.  (\ref{66}), it is not
difficult to
show that in the other CP-odd sector the result is
\be
{\cal I}^{\bar{o}{\rm-}e} = - \vep  \r_1 \r_2 \r_3 {\cal
I}^{\bar{e}{\rm-}o}\, ,
\label{632} \ee
so that the total result for the CP-odd part of the three-point
function
(\ref{64})
is
\be
{\cal I}^{\rm CP-odd} = - (1- \vep \r_1 \r_2 \r_3) 32 \p^3
\T^{\bar{e}{\rm-}o}\, .
\ee
This implies that we need, for the non-zero couplings:
\bs
\be
\mbox{type IIA}: \;\; \r_1 \r_2 \r_3 = -1\, ,
\ee
\be
\mbox{type IIB}: \;\; \r_1 \r_2 \r_3 =  1 \, ,
\ee
\es
so that in type IIA we need an odd number of antisymmetric tensors
and in type IIB an even number. Note also that one cannot construct
any CP-odd four-derivative on-shell coupling between three gravitons.

Comparing Eq.  (\ref{567}) with (\ref{R2vert}), we conclude that the
one-loop correction to CP-odd four-derivative gravitational couplings
is
\bs
\be
{\cal I}_{\rm eff,\ IIA}^{\rm CP-odd} =
 32\p^3 \int \rd^4 x \sqrt{-g}\,
{1\over 2}\, ( B\w R \w R + B\w \na H \w \na H )\, ,
\ee
\be
{\cal I}_{\rm eff, \ IIB}^{\rm CP-odd} =
 - 32\p^3 \int \rd^4 x \sqrt{-g}
\left({1 \over 6}  H \w H \w R \right)\, .
\ee
\label{637app}
\es

\boldmath
\subsection*{\normalsize{\bf D.3 Two-graviton--$N$-moduli scattering
in four dimensions}}
\unboldmath

Here we evaluate the (leading) four-momentum piece of the $(N+2)$-point
amplitude in Eq.~(\ref{tpa}). We first define a set of signs specifying
the nature of the moduli:
\be
\chi_\f = \cases{ \hphantom{-}1\sp  \f = T,U \cr  -1  \sp\f = \bT,\bU
\cr
}\quad
  \s_\f = \cases{ \hphantom{-}1\sp  \f = T,\bT \cr  -1  \sp\f = U,\bU
\, .\cr }
\ee
With these notations, the selection rules read:
\be
\eqalign{
v(\f_i)_{IJ} G^{JK} v(\f_j)_{LK} =& 0  \sp \mbox{if} \quad \s_i = \s_j
\, ,
\cr
v(\f_i)_{JI} G^{JK} v(\f_j)_{KL} =& 0  \sp \mbox{if} \quad
     \s_i \chi_i= \s_j \chi_j\, .
}
\ee
Let us first focus on the $\bar{e}{\rm-}e$ case, and first on the left
side.
If the moduli are chiral, they all have the same vertex
$\pa X + i p \cdot \ps \Ps$ on the left side; therefore, they can
only contribute through the zero-mode $p_{\rm L}$ of $\pa X$. If on the
other
hand modulus $i$ is chiral and modulus $j$ is antichiral, there can
a priori be a contraction $\pa X (\bz_i,z_i) \pa \tX (\bz_j, z_j)$,
but this will be a total derivative with respect to  $z_i$,
unless there is also
a contraction $\bpa X (\bz_i,z_i) \bpa X (\bz_j, z_j)$ on the right
side.
But this can only occur if $\f_i$ and $\f_j$ have also opposite
vertices
on the right side, that is $\f_i=\bar{\f_j}$, a case that we excluded.
Therefore,  only the zero-modes $p^I_{\rm R} p^J_{\rm L}$ of $\bpa X^I
\pa X^J$
contribute. Moreover, we must contract the fermionic parts of the
graviton--two-form vertices together, since other contractions vanish
after the sum over even spin structures, thereby providing four powers
of momenta. All in all,
\be
A^{\bar{e}{\rm-}e} =
\left(p_1^{\hphantom{1}\k} p_2^{\hphantom{1}\m} - p_1^{\phl} \cdot
p_2^{\phl} \, g^{\m\k}_{\phl}\right)
\left(p_1^{\hphantom{1}\l} p_2^{\hphantom{1}\n} - p_1^{\phl} \cdot
p_2^{\phl} \, g^{\n\l}_{\phl}\right)
\Big\langle \ps \ps \Big\rangle^2
\left\langle \bps \bps \right\rangle^2
\prod_{j=3}^{N+2} v_{IJ}^{\phl} (\f_j)  p_{\rm R}^I p_{\rm L}^J\, .
\ee
A similar reasoning applies when one of the spin structures is odd
and shows that the 2 fermionic zero-modes on $T^2$ have to come
from the vertex in the $-1$-picture together with the $T^2$ piece
of the supercurrent, while all other vertices contribute through
the bosonic zero-modes. The space-time fermionic zero-modes are
then provided by the graviton or two-form vertex operators. We find:
\be
\eqalign{
&A^{\bar{o}{\rm-}o} =
\e^{\k \m \a \b} \e^{\l \n \r \s}
p_{1\a} p_{2\b} p_{1\r} p_{2\s}
\Big( G \e v(\f) \e G\Big)_{IJ} \,  p_{\rm R}^I \, p_{\rm L}^J \cr
&A^{\bar{e}{\rm-}o} =
\left(p_1^{\hphantom{1}\k} p_2^{\hphantom{1}\m} - p_1^{\phl} \cdot
p_2^{\phl}\,  g^{\m\k}_{\phl}\right) \e^{\l \n \a \b}_{\phl}
p_{1\a}^{\phl} p_{2\b}^{\phl}
\left(\langle \bps \bps \rangle^2 - \langle \bpa X X \rangle^2 \right)
\Big(  v(\f) \e G\Big)_{IJ} \,  p_{\rm R}^I\,  p_{\rm L}^J \cr
&A^{\bar{o}{\rm-}e} =
\e^{\k \m \a \b}_{\phl}
p_{1\a}^{\phl} p_{2\b}^{\phl}  \left(p_1^{\hphantom{1}\l}
p_2^{\hphantom{1}^\n} - p_1^{\phl} \cdot p_2^{\phl}\,
g^{\n\l}_{\phl}\right)
\left(\left\langle \ps \ps \right\rangle^2 - \left\langle \pa X X
\right\rangle^2 \right)
\Big( G \e v(\f) \Big)_{IJ} \,  p_{\rm R}^I \,  p_{\rm L}^J \, .\cr}
\ee

The Riemann identity (\ref{RiemI}) allows us to carry out the spin
structure
summation
and shows that the integrand is in fact independent of the position
of the vertices. In the odd spin structure, the saturation of
zero-modes
induces the replacement (\ref{611}).

We can simplify the kinematic structures by making use of
Eq.  (\ref{twoeps}),
and rewrite them as (recall that $p_1 \cdot p_2$
is not restricted to vanish anymore):
\be
\eqalign{
\T^{\bar{e}{\rm-}e} =&\,
(p_1^{\phl} \e_2^{\phl} p_1^{\phl})(p_2^{\phl} \e_1^{\phl} p_2^{\phl})
-(p_1^{\phl} p_2^{\phl})\bigg(\! \left(p_2^{\phl} \e_1^T \e_2^{\phl}
p_1^{\phl} \right) + \left(p_2^{\phl} \e_1^{\phl} \e_2^T p_1^{\phl}
\right)\!\bigg)
+ (p_1^{\phl} p_2^{\phl})^2 \left(\e_1^T \e_2^{\phl}\right)
\cr\T^{\bar{o}- o} =&
- (p_1 \e_2 p_1)(p_2 \e_1 p_2)
+(p_1 p_2)\Big( (p_2 \e_1 \e_2 p_1 ) + (p_2 \e_1 \e_2 p_1 )\Big)\cr
&+ (p_1 p_2)^2 \Big((\e_1)(\e_2)-(\e_1 \e_2)\Big)
\cr
\T^{\bar{e} - o} =&\,
p_1^{\phl} \, \e_2^{\phl}\w p_2^{\phl} \,  \e_1^{\phl}\w  p_1^{\phl}\w
p_2^{\phl} + (p_1^{\phl} p_2^{\phl})  \, p_1^{\phl} \w p_2^{\phl} \w
\e_1^T\e_2^{\phl}
\cr
\T^{\bar{o}{\rm-}e} =&\,
\e_2^{\phl}  \, p_1^{\phl} \w \e_1^{\phl} \,  p_2^{\phl} \w
p_1^{\phl}\w  p_2^{\phl} + (p_1^{\phl} p_2^{\phl})  \, p_1^{\phl} \w
p_2^{\phl} \w
\e_1^{\phl} \, \e_2^T\, .
\cr }
\label{tes}
\ee
If particle $1$ is a dilaton, this can be further reduced to
\be
\T^{\bar{e}- e}=\T^{\bar{o}{\rm-}o}=(p_1 p_2)^2 (\e_2)
\sp
\T^{\bar{e} -o}=-\T^{\bar{o}{\rm-}e}= -(p_1 p_2)\,  p_1 \w p_2 \w
\e_2\, ,
\ee
so that one dilaton only couples to another dilaton (CP-even)
or to an antisymmetric tensor (CP-odd) at this order.
One also notes that the $\bar{o}{\rm-}o$ contribution is opposite
to the $\bar{e}{\rm-}e$ contribution in the two-graviton case, equal in
the
$b^2$ and $\Phi^2$ cases.

We can then make use of the identities
\bs
\be
v(\f) \e G = i\chi_\f  v (\f)
\sp
\chi_\f = \cases{ \hphantom{-}1 \sp \f = T,U \cr  -1 \sp \f = \bT,\bU
\cr }
\label{mid}
\ee
\be
 G \e v (\f) = i \s_\f \chi_\f v (\f)
\sp
\s_\f = \cases{\hphantom{-} 1 \sp \f = T,\bT \cr  -1 \sp \f = U,\bU \cr
}
\label{sde}
\ee
\be
\sum_{p_{\rm L}, p_{\rm R}} v_{IJ}^{\phl} (\f) p_{\rm R}^I p_{\rm L}^J
q^{{1\over 2}p_{\rm L}^2} \bq^{{1\over 2}p_{\rm R}^2}
= \frac{1}{\pi \t_2} \pa_\f \Ga_{2,2}^{\phl}
\label{did}
\ee
\es
and find the general result:
\be
{\cal I}_\f^{\bi j }= \T^{\bi j}_{\phl} \k^{\bi j}_{\phl}
\int_{\cal F} {\rd^2 \t \over \t_2^2} \left({\t_2 \over \pi}\right)^3
{1\over \p \t_2}
\pa_\f^{\phl} Z^{\bi j}_{\phl}\, ,
\ee
where $i,j$ run over the even and odd spin structures.
The quantities $\k^{\bi j}$ and $Z^{\bi j}$ are defined in
Eqs.~(\ref{kad}) and (\ref{zij}).  In the last equation, $\pa_\f$
stands for the product
of derivatives with respect to the moduli $\f_i$. For $N> 1$,
these derivatives are actually promoted to modular covariant
derivatives
due to reducible diagrams that we disregarded. The signs $\k^{\bi j}$
make reference to the modulus in the $-1$-ghost-picture.
The various coefficients in Eq. (\ref{gfo})
are then obtained by comparing the kinematical factors
(\ref{tes}) to the vertices (\ref{4Dvert})
and taking proper account of symmetry weights.

\end{document}